\documentclass[fleqn,usenatbib,useAMS]{mnras}

\usepackage{graphicx}	
\usepackage{amsmath}	
\usepackage{amssymb}	
\usepackage{multicol}        
\usepackage{bm}		
\usepackage{pdflscape}	
\usepackage{float}
\usepackage{threeparttable} 
\usepackage{caption}
\usepackage[title]{appendix}

\usepackage[T1]{fontenc}
\usepackage{ae,aecompl}

\usepackage{newtxtext,newtxmath}

\title[Locating gamma-ray emission in FSRQs]{Locating the gamma-ray emission region in the brightest \textit{Fermi}-LAT Flat Spectrum Radio Quasars}

\author[A. Acharyya et al.]{
Atreya Acharyya,$^{1}$\thanks{E-mail: atreya.acharyya@durham.ac.uk (AA)}
Paula M. Chadwick,$^{1}$
Anthony M. Brown$^{1}$
\\
$^{1}$Department of Physics, Durham University, Durham, DH1 3LE, UK
}

\date{Accepted 2020 October 30. Received 2020 October 27; in original form 2020 July 3}

\pubyear{2020}

\begin{document}
\label{firstpage}
\pagerange{\pageref{firstpage}--\pageref{lastpage}}
\maketitle

\begin{abstract}
We present a temporal and spectral analysis of the gamma-ray flux from nine of the brightest flat spectrum radio quasars (FSRQs) detected with the \textit{Fermi} Large Area Telescope (LAT) during its first eight years of operation, with the aim of constraining the location of the emission region. 
Using the increased photon statistics provided from the two brightest flares of each source, we find evidence of sub-hour variability from B2 1520+31, PKS 1502+106 and PKS 1424-41, with the remaining sources showing variability on timescales of a few hours.
These indicate gamma-ray emission from extremely compact regions in the jet, potentially compatible with emission from within the broad line region (BLR).
The flare spectra show evidence of a spectral cut-off in 7 of the 18 flares studied, further supporting the argument for BLR emission in these sources.
An investigation into the energy dependence of cooling timescales finds evidence for both BLR origin and emission from within the molecular torus (MT). 
However, Monte Carlo simulations show that the very high energy (VHE, $E_{\gamma} \geq 20$ GeV) emission from all sources except 3C 279, 3C 454.3 and 4C 21.35 is incompatible with a BLR origin.
The combined findings of all the approaches used suggest that the gamma-ray emission in the brightest FSRQs originates in multiple compact emission regions throughout the jet, within both the BLR and the MT.

\end{abstract}





\section{Introduction}
Flat Spectrum Radio Quasars (FSRQs) constitute a subclass of blazars, Active Galactic Nuclei (AGN) with their jets closely aligned to our line of sight. Being closely oriented with our line of sight means that the emission from these objects is highly Doppler-boosted, making them some of the brightest objects in the gamma-ray sky. However, unlike BL Lacertae objects (BL Lacs),  FSRQs are characterised by strong, broad emission lines (\cite{RN20}).  
The close orientation of the jet to the line-of-sight renders the resolution of structures within the jet difficult, and consequently uncovering the location and origin of the emission remains one of the most active areas of research. In this respect, the Large Area Telescope (LAT) on board the \textit{Fermi} satellite (\cite{Fermi_LAT}) has been particularly important. This pair-conversion telescope, launched in June 2008, is sensitive to photon energies between 20 MeV-2 TeV and has the ability to scan the entire gamma-ray sky every three hours. 

Localizing the gamma-ray emission is an indirect process and a variety of different methods have been used previously. The emission is assumed to be coming from compact regions, supported by the rapid flux variability found in these objects. Timescales of the order of a few hours have been detected in several FSRQs, for example 3C 454.3 (\cite{abdo3c454}), PKS 1510-089 (\cite{RN11}, \cite{Saito_2013}) and 4C 21.35 (\cite{RN41}). There is also evidence of timescales as short as a few minutes, as has been reported in the cases of CTA 102 (\cite{2018shuklacta102}) and 3C 279 (\cite{Ackermann_2016}). 

Assuming constant jet geometry, the size of the emission region, r, can be used to infer the distance from the supermassive black hole (SMBH), R, using $\text{r}=\psi \text{R}$, where $\psi$ is the semi-aperture opening angle of the jet (\cite{Ghisellini_2009}, \cite{2009Dermer}). 
This relation has been used to constrain the location of the emission region to be close to the base of the jet.
For example, in a study of 3C 454.3, 3C 273 and 4C 21.35 undertaken by \cite{Foschini_2011} using $\sim$ 2 years of \textit{Fermi}-LAT observations, the emission was constrained to be from within the broad line region (BLR) under the assumption that the full width of the jet is responsible for the emission.

Further arguments towards BLR origin are based on evidence of a spectral cut-off at GeV energies. This has been interpreted as a consequence of photon-photon pair production of gamma-rays with the Helium Lyman recombination continuum within the photon-rich BLR environment (\cite{RN2}, \cite{Stern_2014}). However, this interpretation has been questioned by \cite{RN51} who found the location of the cut-off inconsistent with the absorption model proposed.
A cut-off in the spectrum can also be the consequence of a break in the energy distribution of the emitting electrons (\cite{Dermer_2015}).

Other studies suggest the emission originates farther out, on parsec scale distances from the SMBH, and thus within the Molecular Torus (MT) region. 
Some of these studies use multi-wavelength observations of a single source, which have revealed that gamma-ray flares are often accompanied by flares at optical or radio wavelengths which are known to be resolved to parsec scale distances from the SMBH (\cite{Marcher}, \cite{RN26}, \cite{Jorstad_2013}). For instance, \cite{Marcher} studying optical, radio and gamma-ray flares in PKS 1510-089 found a single emission feature to be a superluminal knot outside the BLR.
While emission at parsec scales would appear to contradict the short-term variability timescales observed in these objects, the two can be reconciled by assuming localised emission in turbulent cells (\cite{RN18}, \cite{giannios2}).

The observation of very high energy (VHE) photons ($E_{\gamma} \geq 20$ GeV) also supports the theory of emission from outside the BLR (\cite{Donea_2003}, \cite{RN28}). VHE photons would be expected to be severely attenuated by interactions with the photons in the BLR and their detection is difficult to explain if the emission were to originate in regions near the central engine. At the time of writing, 8 FSRQs have been detected at $E_{\gamma} \geq 100$ GeV, of which 3C 279 (\cite{Errando_final}, \cite{RN29}), PKS 1510-089 (\cite{PKS1510_Magic}, \cite{HESSPKS1510_discovery}) and 4C 21.35 (\cite{4c2135_atel2}, \cite{4c2135_magic}) are included in this study \footnote[1]{\color{blue}http://tevcat.uchicago.edu/ (accessed on 11/06/20)}. In addition, \cite{RN3},  studying high energy flares from a sample of FSRQs  using multi-wavelength SED modelling, found the emission to be located significantly outside the BLR.

A possible solution to accommodate both the short variability timescales and VHE photons observed is to abandon the one-zone emission model and invoke the presence of multiple emission regions.
Multi-zone emission models have been proposed to interpret the VHE observations of misaligned AGN (for example (\cite{lenain_m87}, \cite{BrownNGC1275})) as well as the multi-wavelength spectral distribution of blazars (for example PKS 1510-089 (\cite{nale_multi})).
Furthermore, it has been suggested that these multiple simultaneously active emission regions lie at various points throughout the jet, both in the BLR and MT (for example PKS 1510-089 (\cite{RN11})).
In a study of the absorption of VHE photons in the BLR field of FSRQs, \cite{RN16} suggest that the opacity constraints derived can be circumvented by resorting to multi-zone models.
In such a model, the GeV and VHE emission would not be produced co-spatially, with the latter being emitted at a scale of several parsecs from the central engine.

The detection of 4C 21.35 with the MAGIC imaging atmospheric Cherenkov telescope (IACT) (\cite{4c2135_magic}) has been explained by invoking the presence of axion-like particles (ALPs; \cite{Tavecchio_2012}).
ALPs (\cite{axion_we}) are light, neutral bosons and have been predicted by the extension of the standard model in particle physics. 
Gamma-rays produced inside the BLR are assumed to oscillate into ALPs, which do not interact with BLR photons and are therefore not absorbed until they are converted back into photons in magnetic fields outside the BLR (\cite{Galanti_2019}). This leads to a considerable fraction of VHE photons escaping absorption inside the BLR. 
Multiple experiments are in operation to confirm the presence of ALPs (see \cite{Graham_2015} for a review).

This work investigates the gamma-ray emission from a sample of nine bright FSRQs (see Section \ref{subsec:2.1}) detected with the \textit{Fermi}-LAT during the first eight years of observations.
In particular, we identify periods of high flux with the aim of using the increased photon statistics to constrain the characteristics and location of the emission region under the assumption of a leptonic model for the origin of the gamma-rays.
This is followed by a study of the VHE ($E_{\gamma} \geq 20$ GeV) photon emission from each source.
More specifically, we want to address the following issues:\\
a) identify the shortest variability timescales for the two brightest flare periods in each source and understand the implications on the size and location of the emission region;\\
b) investigate further evidence for either BLR or MT emission in the flare spectra such as a possible spectral cut-off and evidence for energy dependent cooling timescales;\\
c) determine whether the VHE emission observed with the \textit{Fermi}-LAT for the sample is compatible with BLR origin and what the findings tell us about the nature of the emission region(s);\\
d) assess whether there is an overarching trend in the results obtained for the sources and consider how they compare with other studies in the literature.

In Section \ref{sec:2} we define our sample of FSRQs and the data analysis routines used in this study. The gamma-ray lightcurves for each source are shown in Section \ref{sec:3}, where we also present our definition of flare periods. In Section \ref{sec:4} we describe the methods used for constraining the size and location of the emission region for the two brightest flares observed for each source. 
In Section \ref{sec:5} we discuss the VHE emission for the sample and compare observations with Monte Carlo simulations to ascertain if this is compatible with BLR origin. In Section \ref{subsec:6.1} we discuss the findings of all our methods for both flares from each individual source. This includes a comparison with other studies of the same source in the literature as well as the interpretation of our results in the context of the nature of the  emission region. A brief discussion of the implications of the results is given in Section \ref{subsec:6.9}. We summarise our conclusions and suggest ideas for future investigations in Section \ref{sec:7}.

\section{Source Selection and Data Reduction}
\label{sec:2}

\begin{table*}
	\centering
	\caption{List of FSRQs selected for this study along with their right ascensions (RA) and declinations (DEC) in degrees  (\protect\cite{2019arXiv190210045T}) and redshifts (z; references given below). Also shown are the results of the eight year likelihood analysis in the energy range 100 MeV-300 GeV. All sources, with the exception of 3C 454.3, were found to be best modelled by a log parabola (see equation \ref{eq:2}) with the spectral parameters being the spectral index ($\alpha$), spectral curvature ($\beta$) and the pivot energy ($E_{0}$). 3C 454.3 was found to be best modelled by a power law with a super exponential cut-off (see equation \ref{eq:3}) having the spectral parameters index1 ($\gamma$), index2 ($\alpha$), pivot energy ($E_{0}$) and the cut-off energy ($E_{\text{cut}}$). The final two columns list the observed eight year averaged flux and the TS values (see equation \ref{eq:1}) of each source obtained from the likelihood analysis.}
	\label{tab:sample_table}
	\resizebox{\linewidth}{!}{
	\begin{tabular}{lcccccccr} 
		\hline
		Source &  RA     &  DEC  & z &  $E_{0}$     &$\alpha$ &$\beta$ &  Flux    & TS  \\
		   & [deg] & [deg] & &[MeV] & & & [$10^{-7}$ photons $\text{cm}^{-2}\text{s}^{-1}$] &  \\
		\hline
		 CTA 102 & 338.15 & 11.73 & 1.0320 $\pm$ $0.0030^{[1]}$ & 414.1  & 2.32 $\pm$ 0.01 &  0.078 $\pm$ 0.005 & 4.19 $\pm$ 0.04 &75211\\
		 B2 1520+31 & 230.55 & 31.74 & 1.4886 $\pm$ $0.0002^{[2]}$  & 593.4  & 2.40 $\pm$ 0.01 & 0.059 $\pm$ 0.006 & 3.10 $\pm$ 0.03 &63775\\
		 PKS 1510-089 & 228.22 & -9.11 & 0.3600 $\pm$ $0.0020^{[3]}$ & 743.5  & 2.39 $\pm$ 0.01 & 0.045 $\pm$ 0.003 & 8.71 $\pm$ 0.07 &180884\\
	     PKS 1502+106 & 226.10 & 10.49 & 1.8381 $\pm$ $0.0015^{[4]}$ & 496.7  & 2.18 $\pm$ 0.01 & 0.075 $\pm$ 0.005  & 3.04 $\pm$ 0.03 &66529\\
		 PKS 1424-41 & 216.99 & -42.11 & 1.5220 $\pm$ $0.0002^{[5]}$ & 677.7  & 2.12 $\pm$ 0.01 & 0.069 $\pm$ 0.003 & 4.87 $\pm$ 0.04 &122369\\
		 3C 279 & 194.04 & -5.79 & 0.5362 $\pm$ $0.0004^{[6]}$  & 442.1  & 2.32 $\pm$ 0.01  & 0.049 $\pm$ 0.004 & 5.27 $\pm$ 0.04 &107214\\
		 4C 21.35 & 186.23 & 21.38 & 0.4320 $\pm$ $0.0010^{[7]}$  & 393.7  & 2.31 $\pm$ 0.01 & 0.031 $\pm$ 0.004 & 4.09 $\pm$ 0.03 &88689\\
		 PKS 0454-234 & 74.26 & -23.41 & 1.0030 $\pm$ $\text{NA}^{[8]}$  & 477.7  & 2.12 $\pm$ 0.01 & 0.069 $\pm$ 0.005 & 2.87 $\pm$ 0.03 &72177\\
		 \end{tabular}}
		 \resizebox{\linewidth}{!}{
		 \begin{tabular}{lccccccccr}
		 \hline
		Source & R.A   &  DEC    & z &  $E_{0}$   &$\gamma$ &$\alpha$ & $E_{\text{cut}}$  &  Flux   & TS   \\
		 & [deg]   &[deg] &  & [MeV]    & & &   [MeV]  &  [$ 10^{-7}$ photons $\text{cm}^{-2}\text{s}^{-1}$]  &   \\
		\hline
		3C 454.3 & 343.50 & 16.15 &0.8590 $\pm$ $0.0001^{[9]}$  & 413.3  & 1.75 $\pm$ 0.01  & 0.283 $\pm$ 0.005 & 47.91 $\pm$3.92 & 15.63 $\pm$ 0.38  &871437\\
    	\hline
    \end{tabular}}
    \begin{tablenotes}
      \small
      \item  Redshift references : [1]: \cite{2016AJ....152...25M}, [2]: \cite{2014A&A...563A..54P}, [3]: \cite{Thompson_1990}, [4]: \cite{2010MNRAS.405.2302H}, [5]: \cite{2018yCat.1345....0G}, [6]: \cite{1996ApJS..104...37M}, [7]: \cite{1987ApJ...323..108O}, [8]: \cite{1989A&AS...80..103S}, [9] \cite{2002LEDA.........0P}.
    \end{tablenotes}
\end{table*}

\begin{figure*}
    \resizebox{0.85 \textwidth}{!}{
    \includegraphics{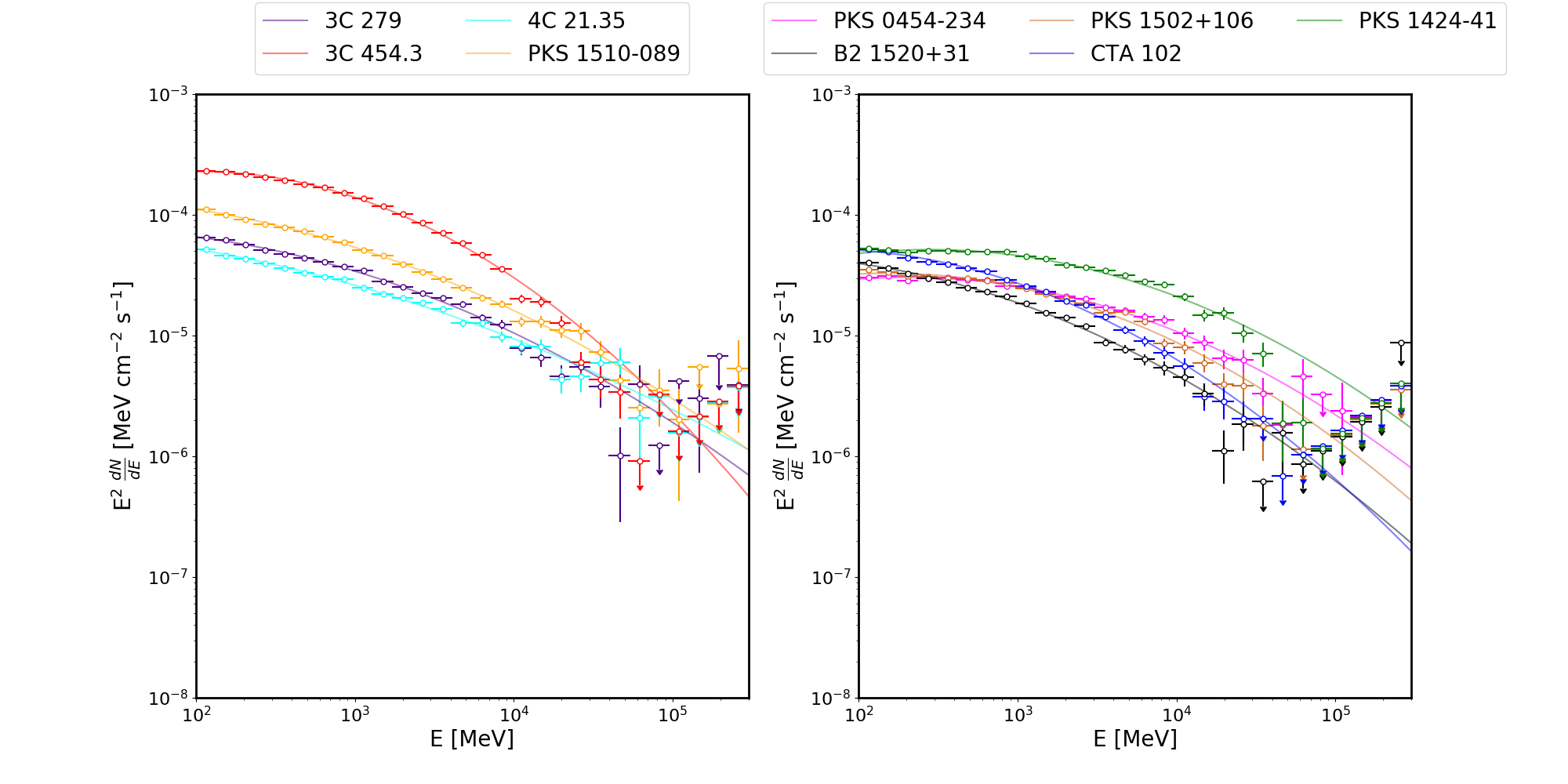}}
    \caption{The eight year averaged \textit{Fermi}-LAT spectra obtained for the sample of bright FSRQs in the energy range 100 MeV-300 GeV, shown in two plots for clarity. The data points are shown as circles along with the corresponding uncertainties. The curves represent the best fits to the spectra with the spectral parameters for each source tabulated in Table \ref{tab:sample_table}. 
    The data are binned into eight energy bins per decade, with individual bins having a TS < 10 considered as upper limits.}
    \label{fig: Figure 1.}
\end{figure*}

\subsection{Source Selection}
\label{subsec:2.1}

The main goal of this investigation is to locate the origin of the gamma-ray emission in FSRQs. This led to a three step process in the identification of suitable sources, primarily governed by having sufficient photon statistics to allow for a detailed study of the gamma-ray emission.
The first step involved surveying the \textit{Fermi}-LAT 8 year catalog of detected sources (\cite{2019arXiv190210045T}) for point sources identified as FSRQs and ordering these by the detection significance of each identification. 

It was also desirable to choose from these bright FSRQs sources having flaring episodes with averaged daily fluxes $\geq 10^{-6}$ cm$^{-2}$ s$^{-1}$ within uncertainties of $1\sigma$ above 100 MeV\footnote[2]{as reported in the \textit{Fermi}-LAT list of monitored sources. See  \color{blue}https://fermi.gsfc.nasa.gov/ssc/data/access/lat/msl\char`_lc/}.
Finally, it was essential that all the identified sources had known redshifts as this is important for interpretation. 
The final sample of nine sources chosen for this study is shown in Table \ref{tab:sample_table}.

\subsection{Data reduction}

Throughout the analysis we use the \textit{Fermi} Science Tools version $11-05-03$ \footnote[3]{\color{blue}http://fermi.gsfc.nasa.gov/ssc/data/analysis/software} and \textit{FERMIPY} version 0.18.0 \footnote[4]{\color{blue}http://fermipy.readthedocs.io} (\cite{wood2017fermipy}) in conjunction with the latest \textit{PASS 8} instrument response functions (IRFs; \cite{atwood2013pass}). 
We select all `Source' class photons from both the front and back of the detector observed between modified Julian dates (MJD) 54682.66 and 57604.66.
This corresponds to midnight on the August 4, 2008 until midnight on August 4, 2016. 

We consider the energy range 100 MeV-300 GeV and a region of interest (RoI) with radius 15$^{\circ}$ centred on each source.  
Furthermore, we selected only photon events within a maximum zenith angle of 90$^{\circ}$ to reduce contamination from background photons from the Earth's limb, produced by the interaction of cosmic rays with the upper atmosphere.
The initial model for each analysis consisted of all sources within 20$^{\circ}$ of the RoI centre with the spatial positions of each source given by the  RA and DEC obtained from the 4FGL catalog (\cite{2019arXiv190210045T}).
Also included in the model were the most recent templates for isotropic and Galactic diffuse emission\footnote[5]{The isotropic diffuse emission model used in the analysis was iso\char`_P8R3\char`_SOURCE\char`_V2\char`_v1.txt. The galactic diffuse emission applied was  gll\char`_iem\char`_v07.fits. For more information see \color{blue}{https://fermi.gsfc.nasa.gov/ssc/data/access/lat/BackgroundModels.html}.}.

The analysis began with an initial automatic optimisation of the RoI by iteratively fitting the sources. This ensures all parameters are close to their global likelihood maxima.
The spectral normalisation of all modelled sources within the RoI were left free as were the normalisation factor of both the isotropic and Galactic diffuse emission templates.
Furthermore, the spectral shape parameters of all sources within 5$^{\circ}$ of the centre of the RoI were left free to vary while those of other sources were fixed to the values reported in the 4FGL catalog (\cite{2019arXiv190210045T}). 

The \textit{gtfindsrc} routine was then applied to search for any additional point sources present in our model and not included in the 4FGL catalog. 
No significant additional point sources were detected indicating that all sources in the model had been accounted for.
A binned likelihood analysis was performed to obtain the spectral parameters best describing the data during the eight year observation period.
We chose a spatial binning of 0.1$^{\circ}$ pixel$^{-1}$ and eight energy bins per decade.

The significance of the gamma-ray emission from each source was evaluated using the maximum likelihood test statistic (TS). The TS is defined as the log likelihood ratio between the maximized likelihoods with and without an additional source, $L_{1}$ and $L_{0}$ respectively (\cite{RN7}):
\begin{equation}
    \hspace*{3cm}
    \centering
   TS=-2 \text{ln} \left(\frac{L_{1}}{L_{0}}\right)
	\label{eq:1}
\end{equation}

During the likelihood analysis, eight sources in the sample were found to be best modelled by a log parabola:
\begin{equation}
    \hspace*{3cm}
    \centering
   \frac{dN}{dE}=N_{0} \left(\frac{E}{E_0}\right)^{-\alpha - \beta \text{ln} \left(\frac{E}{E_0}\right)}
	\label{eq:2}
\end{equation}

where $N_{0}$ is the normalisation (in units of photons $\text{cm}^{-2}\text{s}^{-1} \text{MeV}^{-1}$), $E_{0}$ is the pivot energy in MeV, $\alpha$ the spectral index and $\beta$ the curvature. 

The brightest source in the sample, 3C 454.3, was found to be best modelled by a power law with super exponential cut-off:

\begin{equation}
    \hspace*{2.5cm}
    \centering
   \frac{dN}{dE}=N_{0} \left(\frac{E}{E_0}\right)^{-\gamma} \exp \left(-\frac{E}{E_{\text{cut}}}\right)^{\alpha}
	\label{eq:3}
\end{equation}

where $\gamma$ and $\alpha$ are index1 and index2 respectively and $E_{\text{cut}}$ is the cut-off energy in MeV. 

The resulting eight year averaged spectra of all sources are shown in Figure \ref{fig: Figure 1.} with the spectral parameters obtained from the fit tabulated in Table \ref{tab:sample_table} along with the observed time-averaged flux and TS values of each source.

\section{Gamma-ray Lightcurves}
\label{sec:3}

\begin{figure*}
    \vspace*{-2cm}
    \centering
    \resizebox{\textwidth}{!}{
    \includegraphics{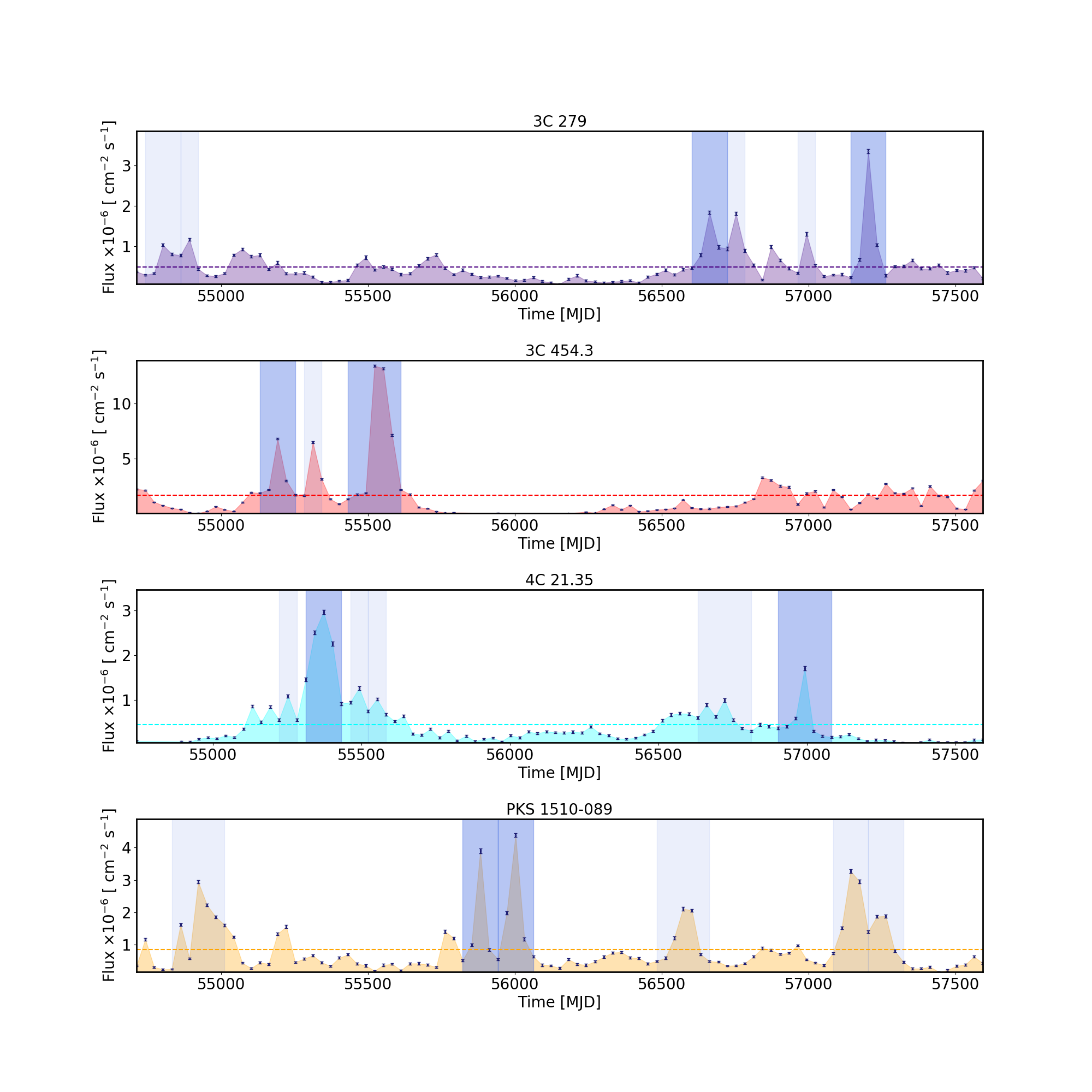}}
    \caption{\textbf{a.} The eight year gamma-ray lightcurves for 3C 279, 3C 454.3, 4C 21.35 and PKS 1510-089  between August 4, 2008 (MJD 54682.66) and August 4, 2016 (MJD 57604.66) binned in monthly periods. The errors are purely statistical and only data points with TS $\geq$ 10 are shown. The horizontal lines indicate the average flux of each source during the entire period. The blue shaded regions indicate periods of flaring activity, with the dark blue shaded regions being the time intervals studied in this investigation.}\label{fig: Figure 2a}
\end{figure*}

\begin{figure*}
    \vspace*{-2cm}
    \centering
     \resizebox{\textwidth}{!}{
    \includegraphics{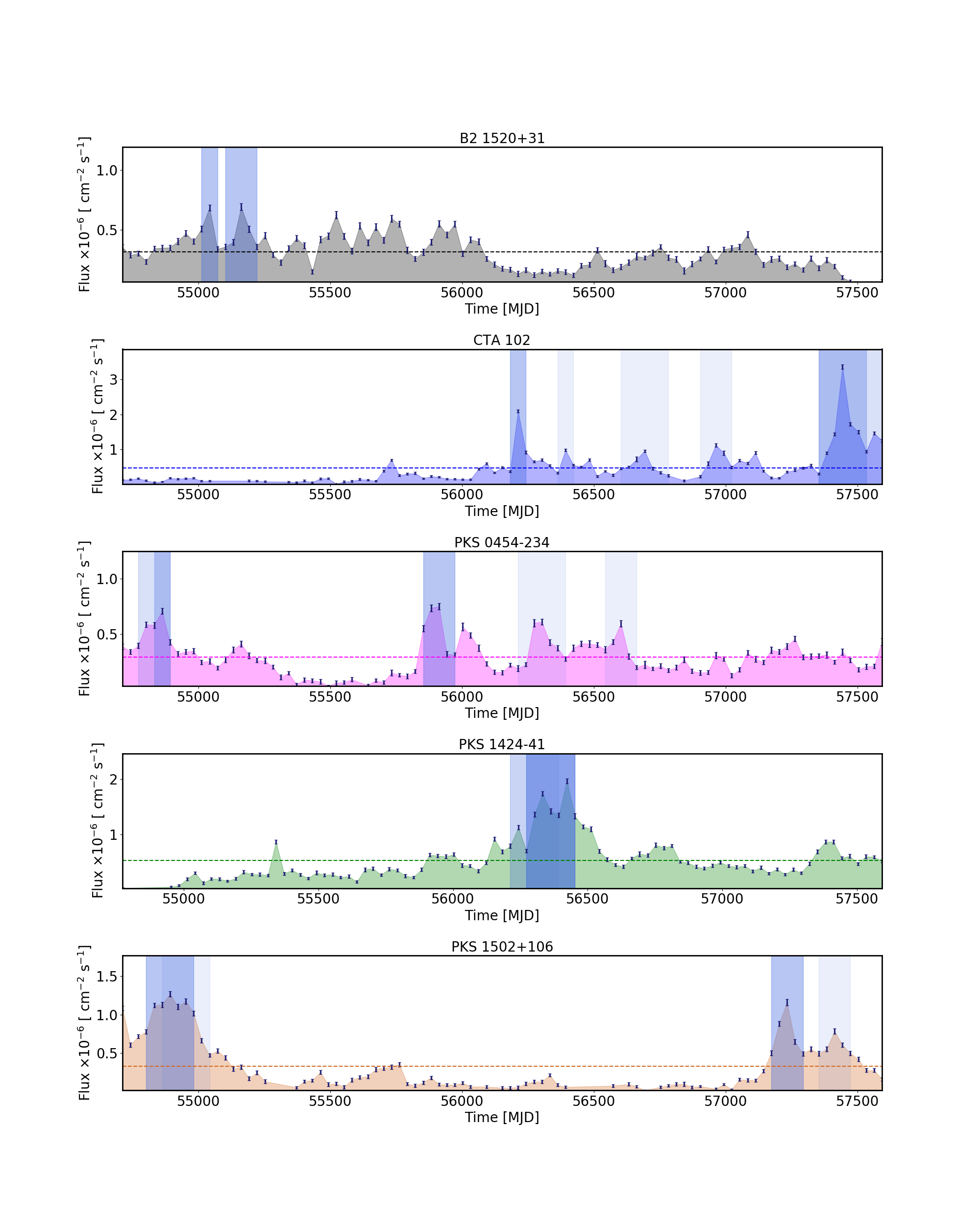}}
     \caption*{\textbf{b.} The eight year gamma-ray lightcurves for B2 1520+31, CTA 102, PKS 0454-234, PKS 1424-41 and PKS 1502+106 between August 4, 2008 (MJD 54682.66) and August 4, 2016 (MJD 57604.66) binned in monthly periods. The errors are purely statistical, and only data points with TS $\geq$ 10 are shown. The horizontal lines indicate the average flux of each source during the entire period. The blue shaded regions indicate periods of flaring activity, with the dark blue shaded regions being the time intervals studied in this investigation.}
\end{figure*}

To study the temporal behaviour of the gamma-ray flux, the eight year \textit{Fermi}-LAT data were initially binned monthly with a likelihood routine applied to each bin separately  \footnote[6]{This was implemented using the \textit{gta.lightcurve()} method in \textit{FERMIPY}.}.
The spectral parameters of all sources within 5$^{\circ}$ of the RoI centre were left free for each bin as were the normalisation factors of the background emission models.
The resulting lightcurves are shown in Figure \ref{fig: Figure 2a} along with the corresponding uncertainties. Only time intervals having  TS $\geq$ 10 were considered, which roughly equates to a significance of $3\sigma$. 

In order to pursue an analysis of the location of the emission region, we need to identify periods of flaring in our lightcurves.
There is no general consensus on how to define a flaring period (e.g. \cite{RN8}, \cite{RN50}). A study by \cite{Nalewajko_2013} defines flares as a contiguous period of time associated with a flux peak having a flux higher than half the peak value of the entire observation. \cite{RN8} proposes a simple two-step procedure of identifying blocks of data points having a flux higher than both the preceding and subsequent blocks and proceeding downwards in both directions as long as the blocks have successively lower fluxes. 

Our definition of flares is primarily designed to identify the periods of highest flux during the eight year dataset, and we define a flare by combining these two approaches. Our method identifies local peaks in flux defined as bins having a flux more than both the preceding and succeeding bin. We then keep going in both directions as long as the corresponding bins are successively lower in flux. 
We then impose the following conditions:
(1) The peak of the flare must have a flux greater than twice the average flux during the entire observation period;
(2) Each bin in the flare must also have a flux greater than  the average flux during the observation period. 
Once this is no longer satisfied, we extend the final ranges by one time bin on each side to mark the onset and end of the flares. 

Also shown in Figure \ref{fig: Figure 2a} are the time periods satisfying our definition of a flare. Although some objects show several flares based on our definition, this study focuses on the two brightest flares for each object shown in darker shaded regions. These are likely to provide sufficient statistics to search for the shortest variability timescales and investigate the presence of a spectral cut-off or energy dependence in the cooling timescales during these periods.

For the identified flare periods, we search for variability on shorter timescales by re-analysing the data with finer binning including daily, 6 hour and 3 hour bins which still satisfy the TS $\geq$ 10 criterion.
The choice of 3 hour bins as a minimum is motivated by the fact that this is roughly the time taken for the \textit{Fermi}-LAT to complete a full scan of the sky (2 orbits).  
The resulting 3 hour binned lightcurves for each flare period considered in this study are shown in Appendix \ref{sec:A1}. We will now use these flare observations to localise the origin of the gamma-ray emission in each source. 

\newpage

\section{Constraining the Size and location of the Emission region}
\label{sec:4}

\subsection{Variability timescales}
\label{subsec:4.1}
The observed flux variability can be characterised by calculating the time taken for the flux to increase or decrease by a factor of 2. Known as the doubling or halving timescale, $\tau$, this is defined by:

\begin{equation}
    \hspace*{3cm}
    \centering
    F(t)=F(t_{0}) 2^{\tau^{-1}(t-t_{0})}
    \label{eq:4}
\end{equation}

where F(t) and F($t_{0}$) are the fluxes at times t and $t_{0}$ respectively. A least squares routine was performed to provide the best fit to equation \ref{eq:4} for three consecutive flux measurements in the 3 hour binned lightcurves of each flare period. 
From these, we can calculate the intrinsic timescales, $\tau_{\text{int}}=\tau/(1+z)$, where $z$ is the redshift of each source.

The choice of three consecutive observations in the fitting procedure is motivated by it being the minimum number of points required to estimate the variability timescale given the number of free parameters in Equation \ref{eq:4}. As we are going through every single point in the lightcurve sequentially, this method should be able to give a good estimate for variability.
However, an important caveat of using Equation \ref{eq:4} is that it considers flux ratios rather than the flux values, which raises the possibility of timescales anti-correlating with the logarithm of the ratio of the fluxes for fixed time differences between observations. Nevertheless, the measurement of doubling timescales is common in the study of FSRQs (for example \cite{Foschini_2011}, \cite{Saito_2013}) and we use the same method for the purposes of comparability. As seen in Section \ref{sec:6}, the results we obtained are compatible with other studies of the same flares.

The fastest variability timescales having a statistical significance of at least $3 \sigma$ found for the flares studied are tabulated in Table \ref{tab:table2}. 
Interestingly, only 3 of the 18 fastest timescales are associated with an event corresponding to a decrease in flux.
This could be interpreted as evidence for fast-rise exponential decay (FRED) type flares, resulting from the injection of energetic particles on shorter timescales than the timescales associated with subsequent cooling processes.

\begin{table*}
\centering
\caption{Summary of the shortest intrinsic variability timescales in hours for each source during the flare periods investigated which have a significance of at least $3 \sigma$. The times listed, $\text{T}_{\text{start}}$ and  $\text{T}_{\text{stop}}$ respectively, are in MJD, with the corresponding fluxes in units of $10^{-6}$ photons $\text{cm}^{-2}\text{s}^{-1}$. The intrinsic variability timescales, $\tau_{\text{int}}$, are calculated from the observed characteristic timescales $\tau$ (see equation \ref{eq:4}) with $\tau_{\text{int}}=\tau/(1+z)$, where z is the redshift of each source. The last column indicates whether the variability event results from a rise (R) or decay (D) in the flux.}
\label{tab:table2}
\resizebox{\textwidth}{!}{
\begin{tabular}{lcccccccr} 
\hline
Source &Flare &$\text{T}_{\text{start}}$ & $\text{T}_{\text{stop}}$ & $\text{Flux}_{\text{start}}$  & $\text{Flux}_{\text{stop}}$  &   $\tau_{\text{int}}$    & Significance &Rise/Decay  \cr
 &Peak &[MJD] & [MJD] &[$10^{-6}$ photons $\text{cm}^{-2}\text{s}^{-1}$]   & [$10^{-6}$ photons $\text{cm}^{-2}\text{s}^{-1}$]  &  [hours]   & $\sigma$ &   \cr
\hline

3C 454.3 & Dec 2009 & 55191.89 & 55192.02 & 1.45 $\pm$ 0.73 & 4.48 $\pm$ 1.31 & 1.47 $\pm$ 0.32 & 4.59 & R \cr
3C 454.3 & Nov 2010 & 55516.76 & 55516.89 & 14.47 $\pm$ 1.84 & 24.68 $\pm$ 2.90 & 2.80 $\pm$ 0.39 & 7.27 & R \cr

CTA 102 & Sept 2012 & 56191.76 & 56191.89 & 1.10 $\pm$ 0.62 & 3.00 $\pm$ 0.83 & 1.45 $\pm$ 0.26 & 5.55 & R \cr
CTA 102 & Feb 2016 & 57439.89 & 57440.01 & 17.81 $\pm$ 2.09 & 4.81 $\pm$ 0.87 & 1.09 $\pm$ 0.18 & 6.02 & D \cr

B2 1520+31 & July 2009 &55046.64 & 55046.77 & 0.11  $\pm$ 0.08 &0.65  $\pm$ 0.67 &0.65  $\pm$ 0.11 &5.85 &R \cr
B2 1520+31 & Nov 2009 &55146.64 & 55146.77 &0.44  $\pm$ 0.24 &1.05  $\pm$ 0.38 &3.03  $\pm$ 0.84 &3.61 &R \cr

PKS 1510-089 & Nov 2011 & 55880.24 & 55880.37 & 0.43 $\pm$ 0.19 & 1.39 $\pm$ 0.52 & 1.79 $\pm$ 0.27 & 6.57 &R \cr
PKS 1510-089 & Feb 2012 & 55966.24 & 55966.37 & 1.33 $\pm$ 0.65 & 7.42 $\pm$ 3.66 & 1.39 $\pm$ 0.41 & 3.40 &R \cr

PKS 1502+106 & Feb 2009 & 54876.51 & 54876.64 & 2.79 $\pm$ 0.75 & 0.81 $\pm$ 0.34 & 0.86 $\pm$ 0.17 & 5.00 & D \cr
PKS 1502+106 & July 2015 & 57216.01 & 57216.14 & 0.88 $\pm$ 0.28 & 1.79 $\pm$ 0.73 & 1.23 $\pm$ 0.08 & 14.71 & R \cr

PKS 1424-41 & Jan 2013 & 56300.76 & 56300.89 & 0.46 $\pm$ 0.21 & 2.82 $\pm$ 0.91 & 0.71 $\pm$ 0.22 & 3.31 & R \cr
PKS 1424-41  & Apr 2013  & 56393.64  & 56393.77  & 0.55 $\pm$ 0.26 & 1.45 $\pm$ 0.43 & 2.56 $\pm$ 0.61  &4.21  &R \cr

3C 279 & Dec 2013 & 56646.26 & 56646.39 & 4.14 $\pm$ 1.01 & 9.09 $\pm$ 1.53 & 2.08 $\pm$ 0.17 & 12.26 & R \cr
3C 279 & June 2015 & 57196.99 & 57197.12 & 5.54 $\pm$ 1.96 & 2.06 $\pm$ 0.72 & 2.14 $\pm$ 0.65 & 3.28 & D \cr

4C 21.35 & June 2010 & 55369.64 & 55369.76 & 1.39 $\pm$ 0.50 & 3.41 $\pm$ 0.77 & 2.57 $\pm$ 0.83 & 3.10 & R \cr
4C 21.35 & Nov 2014 & 56975.14 & 56975.26 & 1.01 $\pm$ 0.48 & 1.74 $\pm$ 0.71 & 2.09 $\pm$ 0.15 & 13.54 & R \cr

PKS 0454-234 & Jan 2009 & 54840.89 & 54841.01 & 0.69 $\pm$ 0.34 & 1.70 $\pm$ 0.54 & 1.62 $\pm$ 0.28 & 5.72 & R \cr
PKS 0454-234 & Nov 2011 & 55896.01 & 55896.14 & 0.86 $\pm$ 0.44 & 2.47 $\pm$ 0.83 & 1.39 $\pm$ 0.24 & 5.73 & R \cr

\hline
\end{tabular}}
\end{table*}

\begin{table*}
\centering
\caption{Results for the size of the emission region, $\text{r}_{\text{emission}}$, obtained for both flare periods of each source. Also listed are the fastest intrinsic variability timescales, $\tau_{\text{int}}$, in hours (see Table \ref{tab:table2}) as well as the average values of the Doppler factors, $\delta$ (\protect\cite{Jorstad_2017}) (see equation \ref{eq:5}), used in the calculation. Where a Doppler factor is not available in the literature we use a value of 10, considered typical for these objects (for example \protect\cite{Foschini_2011}). For comparison, the final column shows the Schwarzschild radius, $\text{r}_{\text{s}}$, for each source, calculated from the mass of the SMBH (\protect\cite{Ghisellini_2010}).}
\label{tab:table3}
\begin{tabular}{lccccr} 
\hline
Source &Flare  &$\delta$  &$\tau_{\text{int}}$ &$\text{r}_{\text{emission}}$ & $\text{r}_{\text{s}}$ \cr 
 &Peak &  &[hours] & [$10^{13}$m] & [$10^{13}$m] \cr
\hline

3C 454.3 & Dec 2009  & 24.4 & 1.47 $\pm$ 0.32 & 3.87 $\pm$ 0.84 & 0.15 \cr
3C 454.3 & Nov 2010 & 24.4 & 2.80 $\pm$ 0.39 & 7.38 $\pm$ 1.03 & 0.15 \cr

CTA 102 & Sept 2012  & 30.5 & 1.45 $\pm$ 0.26 & 4.78 $\pm$ 0.86 & 0.15 \cr
CTA 102 & Feb 2016  & 30.5 & 1.09 $\pm$ 0.18 & 3.59 $\pm$ 0.59 & 0.15 \cr

B2 1520+31 & July 2009  & 10.0 & 0.65 $\pm$ 0.11 & 0.70 $\pm$ 0.12 & 0.37 \cr
B2 1520+31 & Nov 2009   & 10.0 & 3.03 $\pm$ 0.84 & 3.27 $\pm$ 0.91 & 0.37 \cr

PKS 1510-089 & Nov 2011 & 35.3 & 1.79 $\pm$ 0.27 & 6.82 $\pm$ 1.03 & 0.10 \cr
PKS 1510-089 & Feb 2012 & 35.3 & 1.39 $\pm$ 0.41 & 5.30 $\pm$ 1.56 & 0.10 \cr

PKS 1502+106 & Feb 2009 & 10.0 & 0.86 $\pm$ 0.17 & 0.93 $\pm$ 0.18 & 0.44 \cr
PKS 1502+106 & July 2015  & 10.0 & 1.23 $\pm$ 0.08 & 1.33 $\pm$ 0.09 & 0.44 \cr

PKS 1424-41 & Jan 2013  & 10.0 & 0.71 $\pm$ 0.22 & 0.77 $\pm$ 0.24 & 0.15 \cr
PKS 1424-41 & Apr 2013 & 10.0 & 2.56 $\pm$ 0.61 & 2.76 $\pm$ 0.66 & 0.15 \cr

3C 279 & Dec 2013  & 18.3 & 2.08 $\pm$ 0.17 & 4.11 $\pm$ 0.34 & 0.13 \cr
3C 279 & June 2015  & 18.3 & 2.14 $\pm$ 0.65 & 4.23 $\pm$ 1.28 & 0.13 \cr

4C 21.35 & June 2010  & 7.4 & 2.57 $\pm$ 0.83 & 2.05 $\pm$ 0.66 & 0.09 \cr
4C  21.35 & Nov 2014  & 7.4 & 2.09 $\pm$ 0.15 & 1.67 $\pm$ 0.12 & 0.09 \cr

PKS 0454-234 & Jan 2009  & 26.0 & 1.62 $\pm$ 0.28 & 4.55 $\pm$ 0.79 & 0.37 \cr
PKS 0454-234 & Nov 2011 & 26.0 & 1.39 $\pm$ 0.24 & 3.91 $\pm$ 0.67 & 0.37 \cr

\hline
\end{tabular}
\end{table*}

\begin{figure*}
    \centering
    \includegraphics[width=10cm]{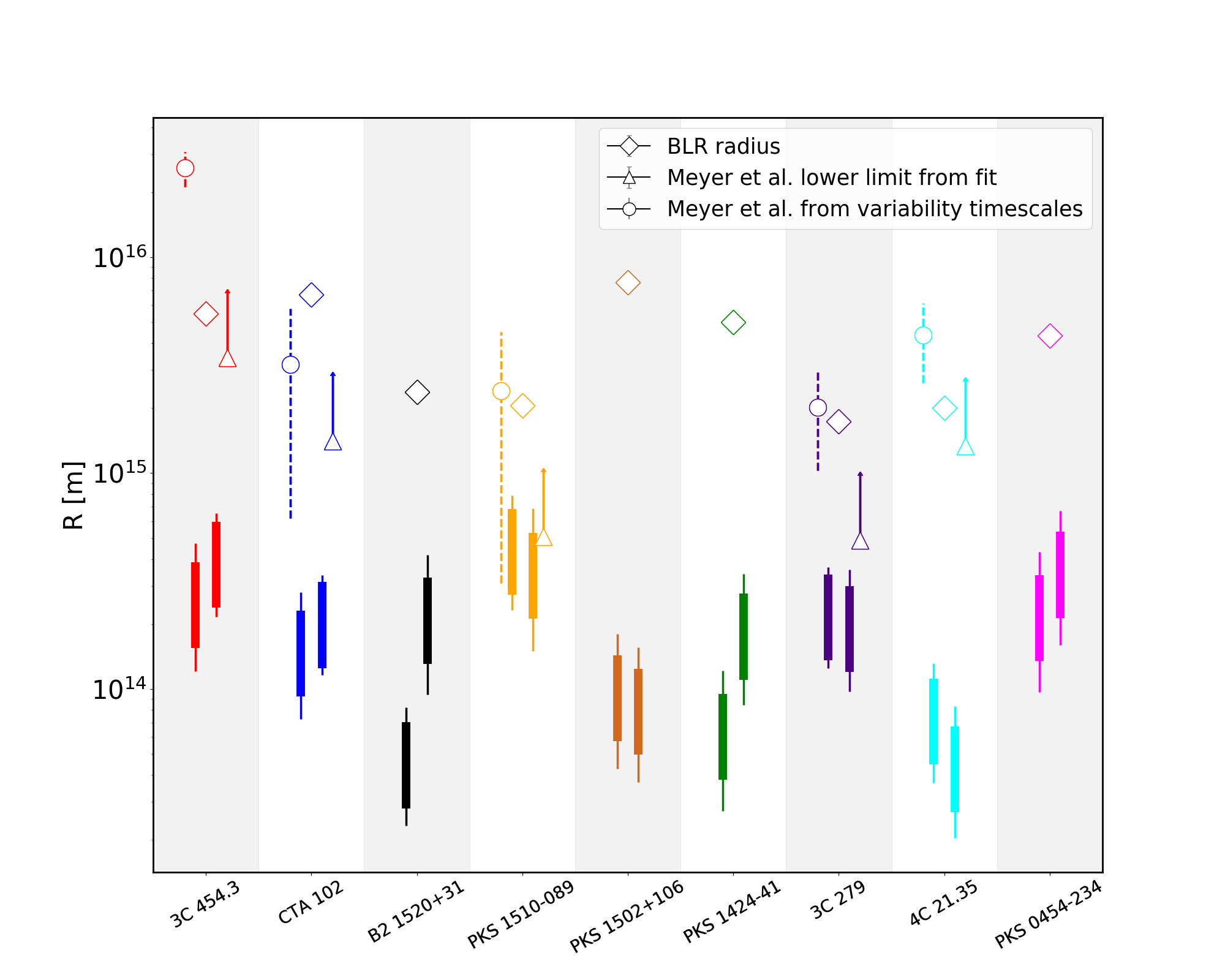}
    \caption{Limits on the distance of the gamma-ray emission regions from the central black hole obtained for both flares from each source are shown as solid shaded regions, with the earlier flare on the left.  This calculation assumes the entire width of the jet to be responsible for the emission. The circles represent the limits on the corresponding distances calculated by \protect\cite{RN8} using variability considerations under the assumption of a conical jet model. The triangles represent lower limits obtained by \protect\cite{RN8} using fits to the gamma-ray spectra (M. Meyer, private communication). The diamonds represent the radius of the BLR ($\text{R}_{\text{BLR}}$) for each source taken from \protect\cite{Ghisellini_2010}. For sources not reported in \protect\cite{Ghisellini_2010}, $\text{R}_{\text{BLR}}$ was calculated using $\text{R}_{\text{BLR}}=10^{15} \text{L}_{\text{disk,45}}^{0.5}$ m from values of $\text{L}_{\text{disk,45}}$, the disk luminosity in units of $10^{45} \text{ergs }\text{ s}^{-1}$, reported in \protect\cite{RN42}.}
    \label{fig: Fig 3.}
\end{figure*}

Using geometric arguments, the intrinsic variability timescales can be used to constrain the size of the emission region:

\begin{equation}
    \hspace*{3cm}
    \centering
    r \leq c \delta \tau_{\text{int}}
	\label{eq:5}
\end{equation}

where r is the size of the emission region, c is the speed of light and $\delta$ is the Doppler factor of the jet. Wherever possible, we use the optical measurements of $\delta$ from \cite{Jorstad_2017} for this calculation; where a measurement is not available we use a value of 10, considered typical for these objects (for example \cite{Foschini_2011}). The size of the emission region derived for each of our flares is reported in Table \ref{tab:table3}. Also given for comparison are the Schwarzschild radii for these objects calculated from the mass of the SMBH (\cite{Ghisellini_2010}).

With the size of the emission region accounted for, we then try to constrain its location. 
A small emission region does not automatically imply emission from near the central engine as over-densities of the plasma can occur throughout the jet, including within the MT.
It has been proposed that these result from magnetic reconnection events (\cite{RN18}, \cite{giannios2}) or the recollimation of the jet (\cite{RN22}).
However, a first order approximation of the distance of the emission region from the SMBH can be made by assuming a simple one-zone emission model in which the entire width of the jet is responsible for the emission. The size of the emission region, $r$, is then related to the distance of the emission region from the central engine, $R$, using:

\begin{equation}
    \hspace*{3cm}
    \centering
    r = \psi R
	\label{eq:6}
\end{equation}

where $\psi$ is the semi-aperture opening angle of the jet and has typical values between 0.1 - 0.25. (\cite{Ghisellini_2009}, \cite{2009Dermer}). 

The limits obtained are shown in Figure \ref{fig: Fig 3.} which plots the distances of the gamma-ray emission regions from the central engine obtained for the two brightest flares together with the radius of the BLR region ($\text{R}_{\text{BLR}}$) for each source (\cite{Ghisellini_2010}, \cite{RN42}). 

For comparison, we also show the corresponding distances reported in \cite{RN8} who investigated five of the sample of FSRQs studied here (M. Meyer, private communication). 
These distances were obtained using variability timescales (shown as circles) and from fits to the observed gamma-ray spectrum (shown as triangles). 
In general, we find the emission regions to be closer to the black hole than both sets of results reported in \cite{RN8} and within the BLR for all sources.

\subsection{Photon-photon pair production}
\label{subsec:4.2}

\begin{table*}
	\centering
	\caption{Summary of the mean difference in AIC values (see equations \ref{eq:8} and \ref{eq:9}) between a log parabola and power law model during the flare periods from each source. Also shown is the model the flare spectra prefer, if any; this was determined using the mean difference in AIC values, whereby a difference of greater than 2 between two models indicates that the model with the higher AIC is significantly worse than that with the lower AIC value (\protect\cite{RN21}).}
	\label{tab:table4}
	\resizebox{0.4 \textwidth}{!}{
	\begin{tabular}{lccccr} 
		\hline
		Source &Flare Peak  &Model Preferred    &$\Delta_{\text{AIC}}$  \\
		\hline
        3C 454.3 &Dec 2009  & Log parabola & -6.10 \\
        3C 454.3 &Nov 2010  & Log parabola & -39.45\\
        CTA 102 &Sept 2012  & Log parabola & -2.51\\
        CTA 102 &Feb 2016  & Log parabola & -2.22\\
        B2 1520+31 &July 2009  & Neither & -0.57\\
        B2 1520+31 &Nov 2009 & Neither & 1.09\\
        PKS 1510-089 &Nov 2011 & Neither & -0.05 \\
        PKS 1510-089 &Feb 2012  & Log parabola & -2.28\\
        PKS 1502+106 &Feb 2009 & Neither & -1.45\\
        PKS 1502+106 &July 2015 & Neither & -0.40\\
        PKS 1424-41 &Jan 2013  & Neither & -0.65\\
        PKS 1424-41 &Apr 2013 & Neither & -0.47\\
        3C 279 &Dec 2013  & Neither & -1.64\\
        3C 279 &June 2015  & Log parabola & -6.01\\
        4C 21.35 &June 2010  & Log parabola & -2.03\\
        4C 21.35 &Nov 2014  & Neither & 0.51\\
        PKS 0454-234 &Jan 2009 & Neither & -1.97\\
        PKS 0454-234 &Nov 2011  & Neither & -0.92\\
		\hline
	\end{tabular}}
\end{table*}

\begin{figure*}
    \centering
    \includegraphics[width=14cm]{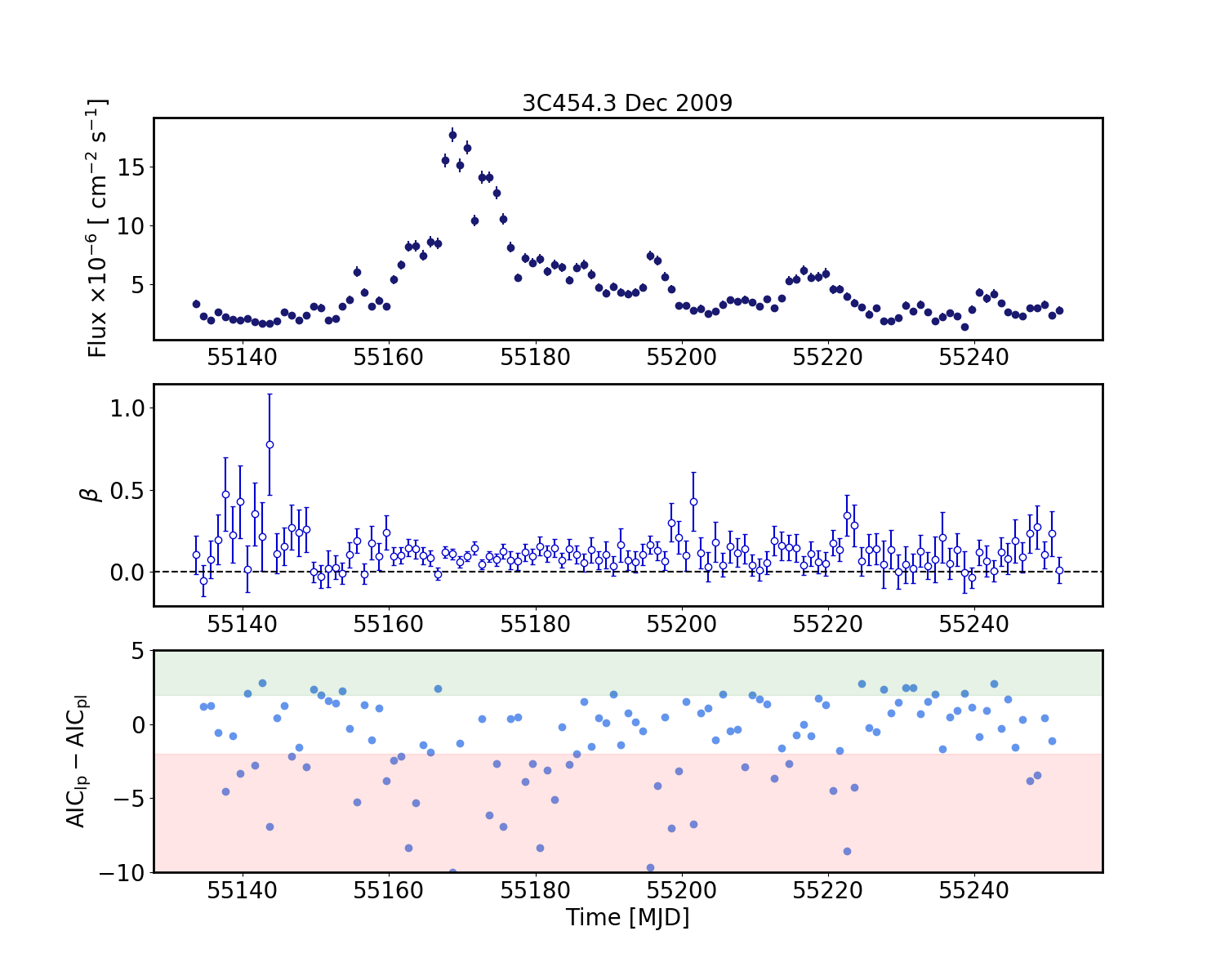}
    \includegraphics[width=14cm]{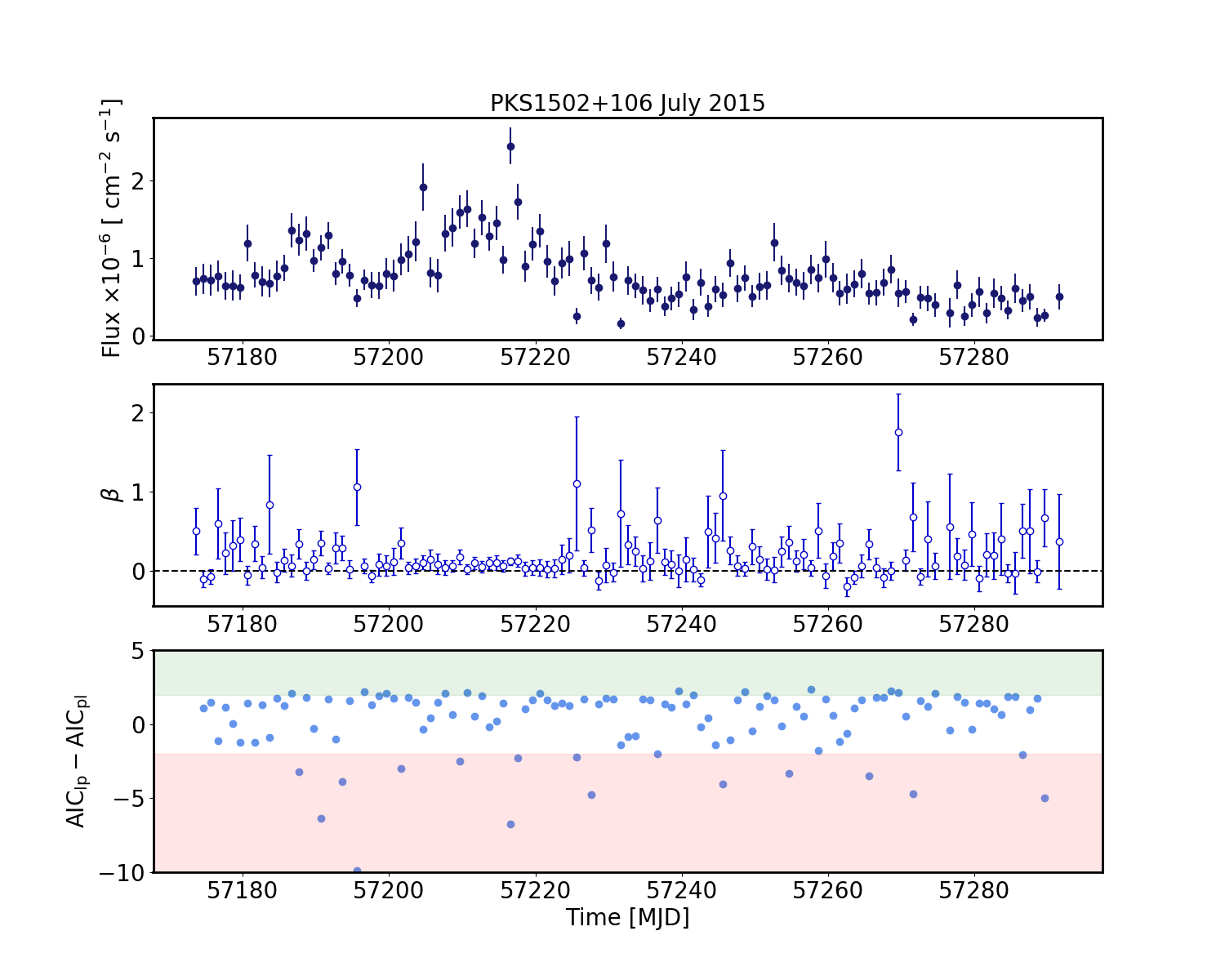}
    \caption{Upper panels: Daily evolution of flux for 3C 454.3 during its December 2009 flare (top) and  PKS 1502+106 during its July 2015 flare (bottom). Middle panels: The daily variation of the spectral parameter $\beta$ during the corresponding flare periods. The dashed horizontal line is at $\beta=0$. Lower panels : Difference in AIC values between the log parabola and power law fits to spectra observed during the flare in daily intervals. The points in the red shaded region represent daily intervals better modelled with a log parabola over a power law. The points in the green shaded region represent daily intervals favouring a power law over a log parabola. Points between the shaded regions represent daily intervals showing no significant deviation between the two models.}
    \label{fig: Fig 4.}
\end{figure*}

\begin{figure*}
    \centering
    \resizebox{\textwidth}{!}{
    \includegraphics{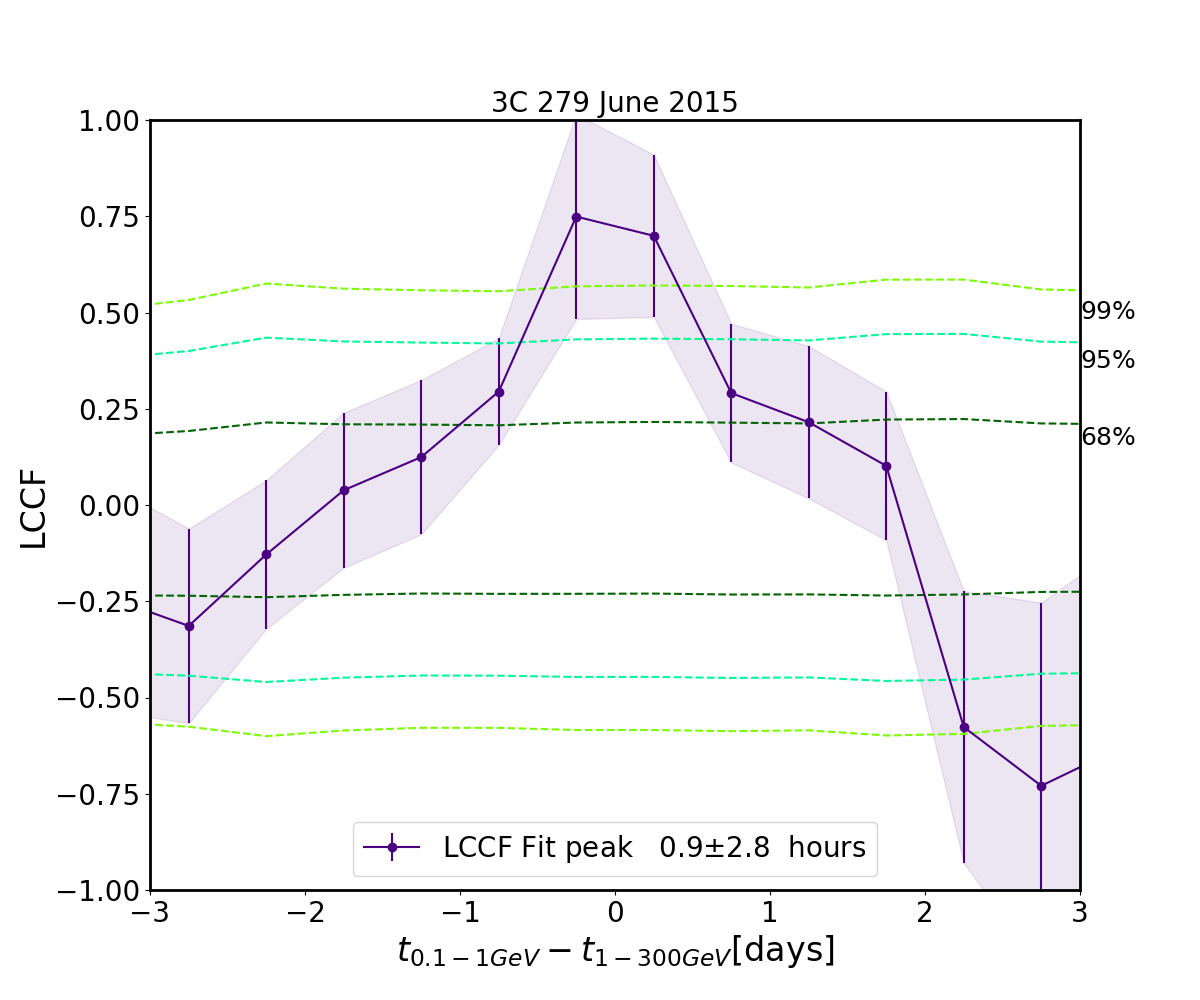}
    \includegraphics{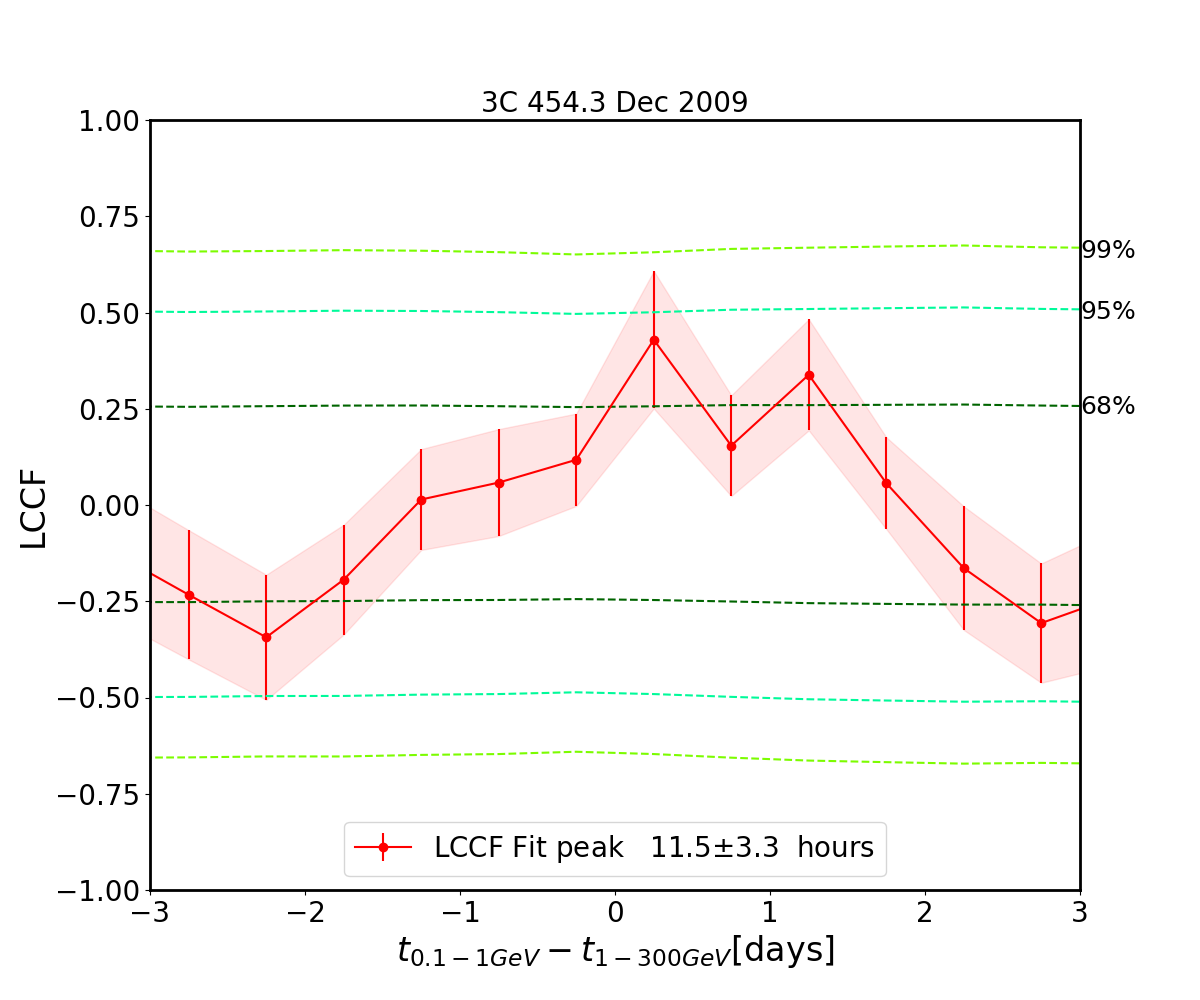}
    \includegraphics{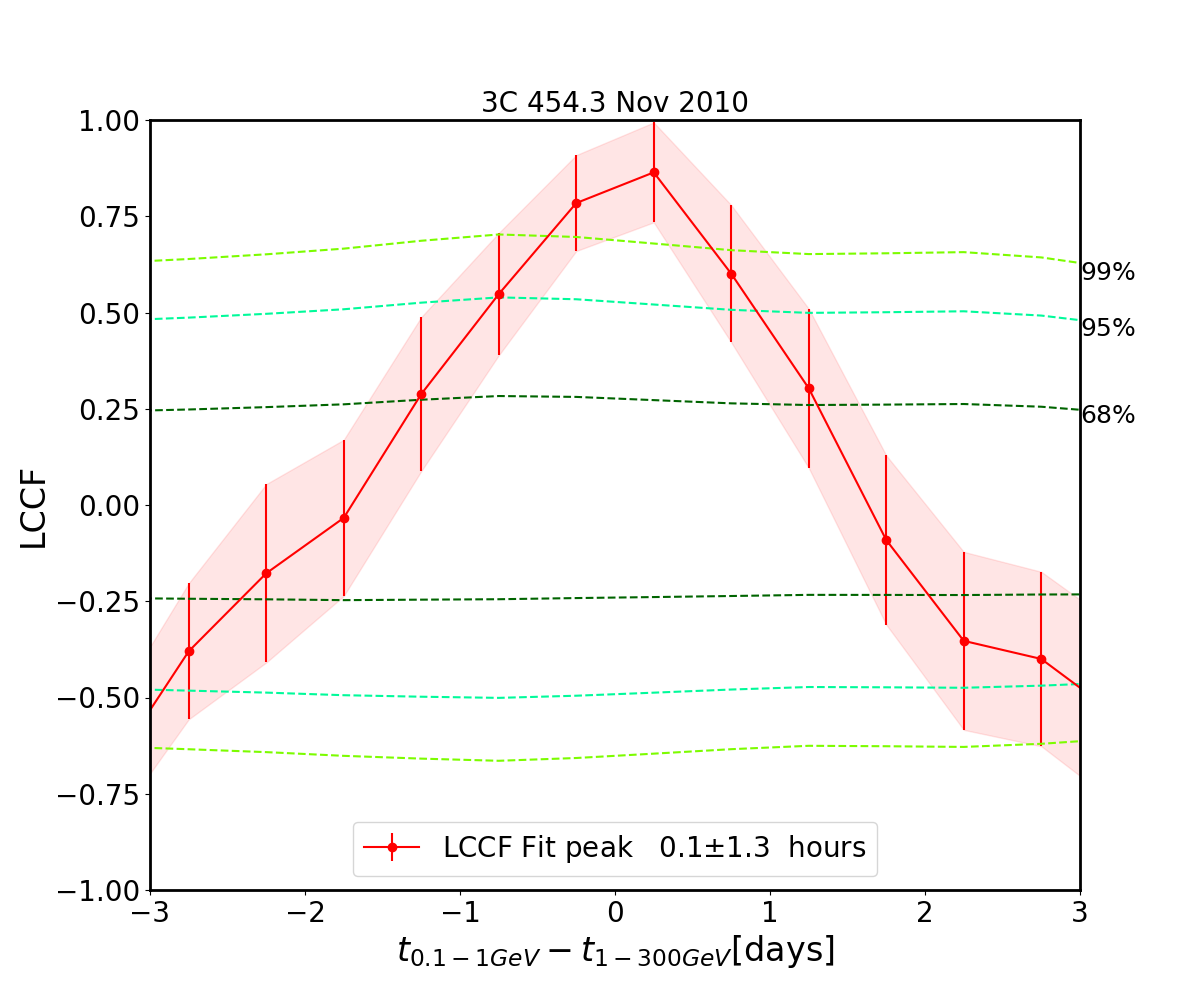}}
    \resizebox{\textwidth}{!}{
    \includegraphics{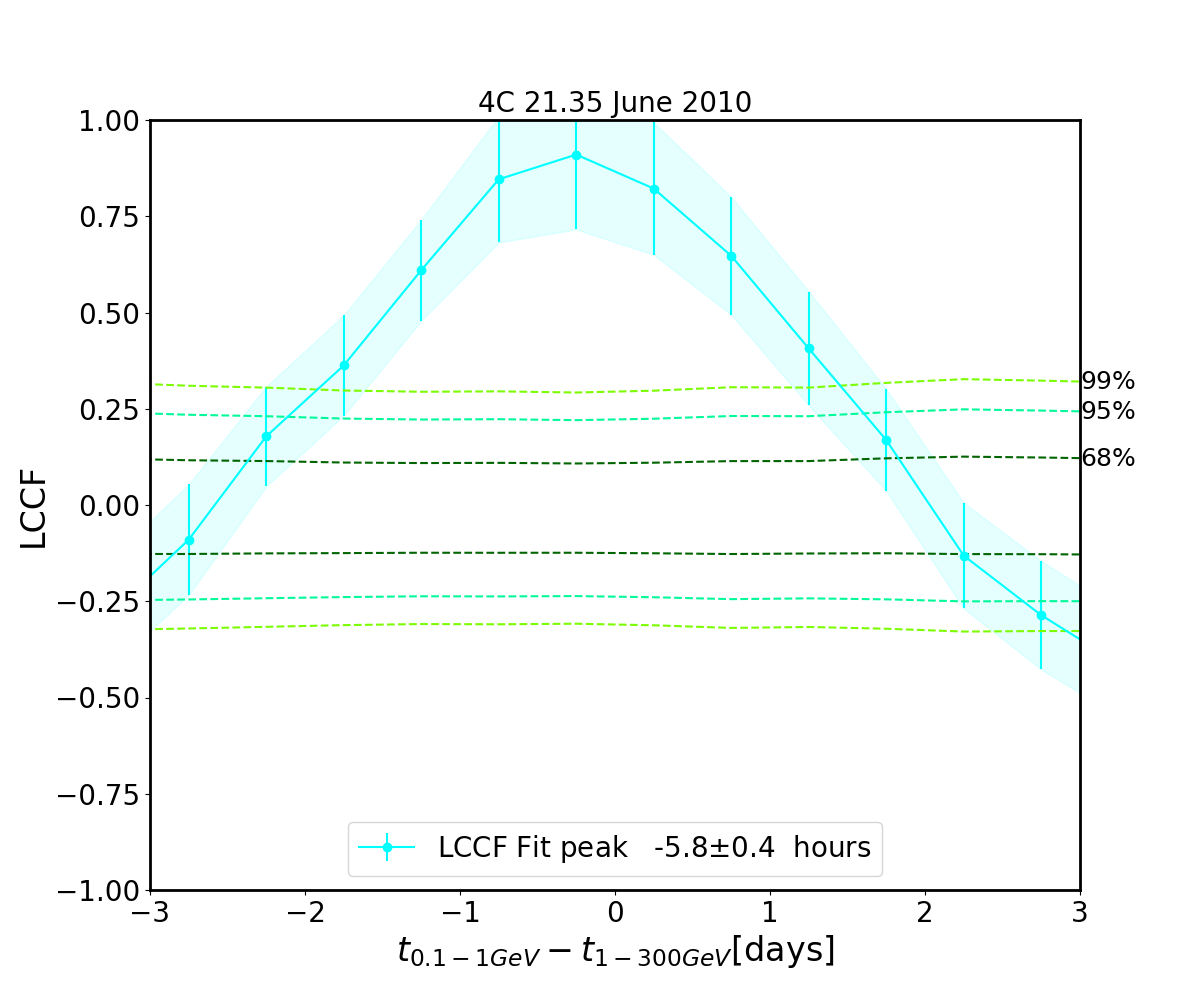}
    \includegraphics{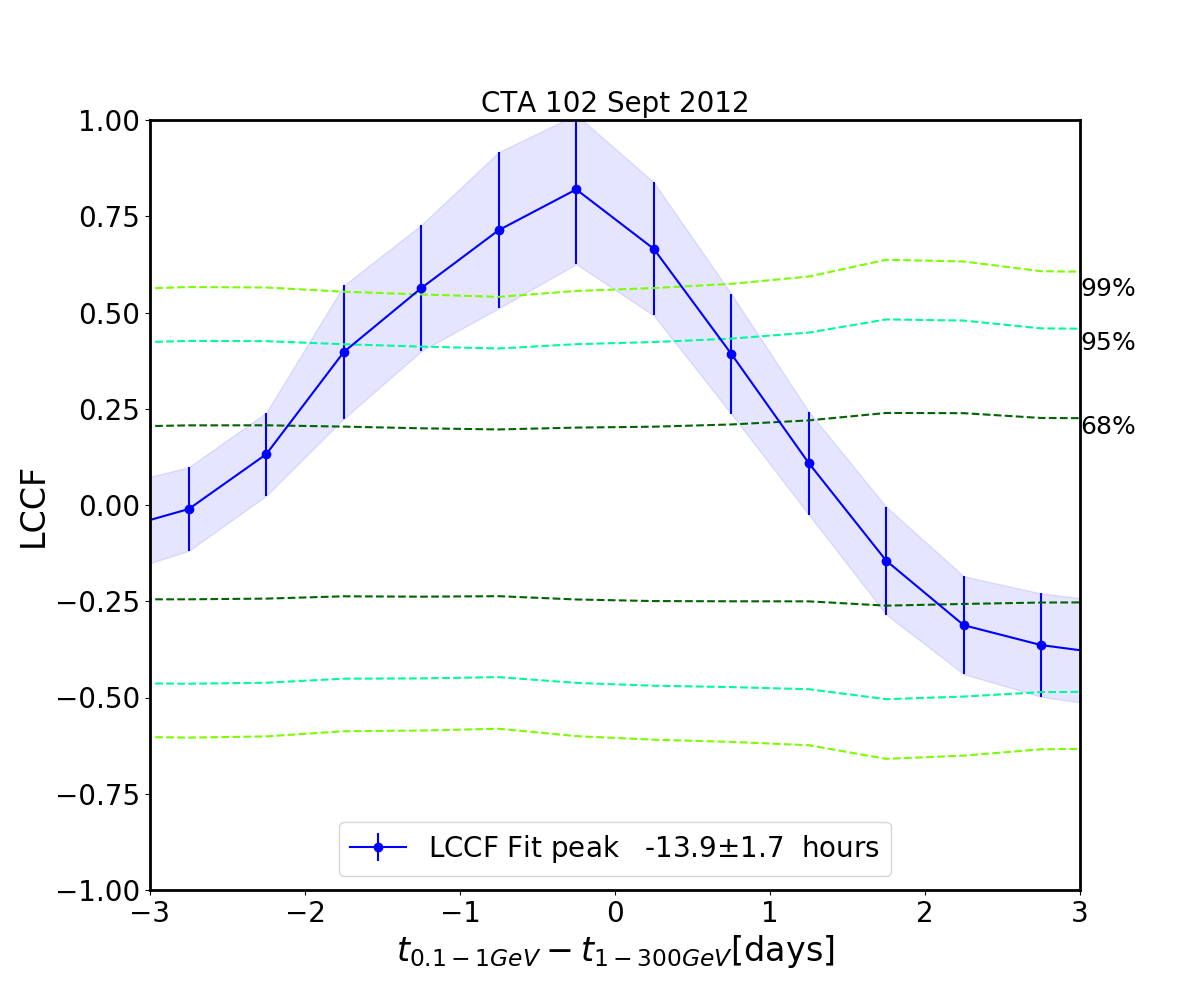}
    \includegraphics{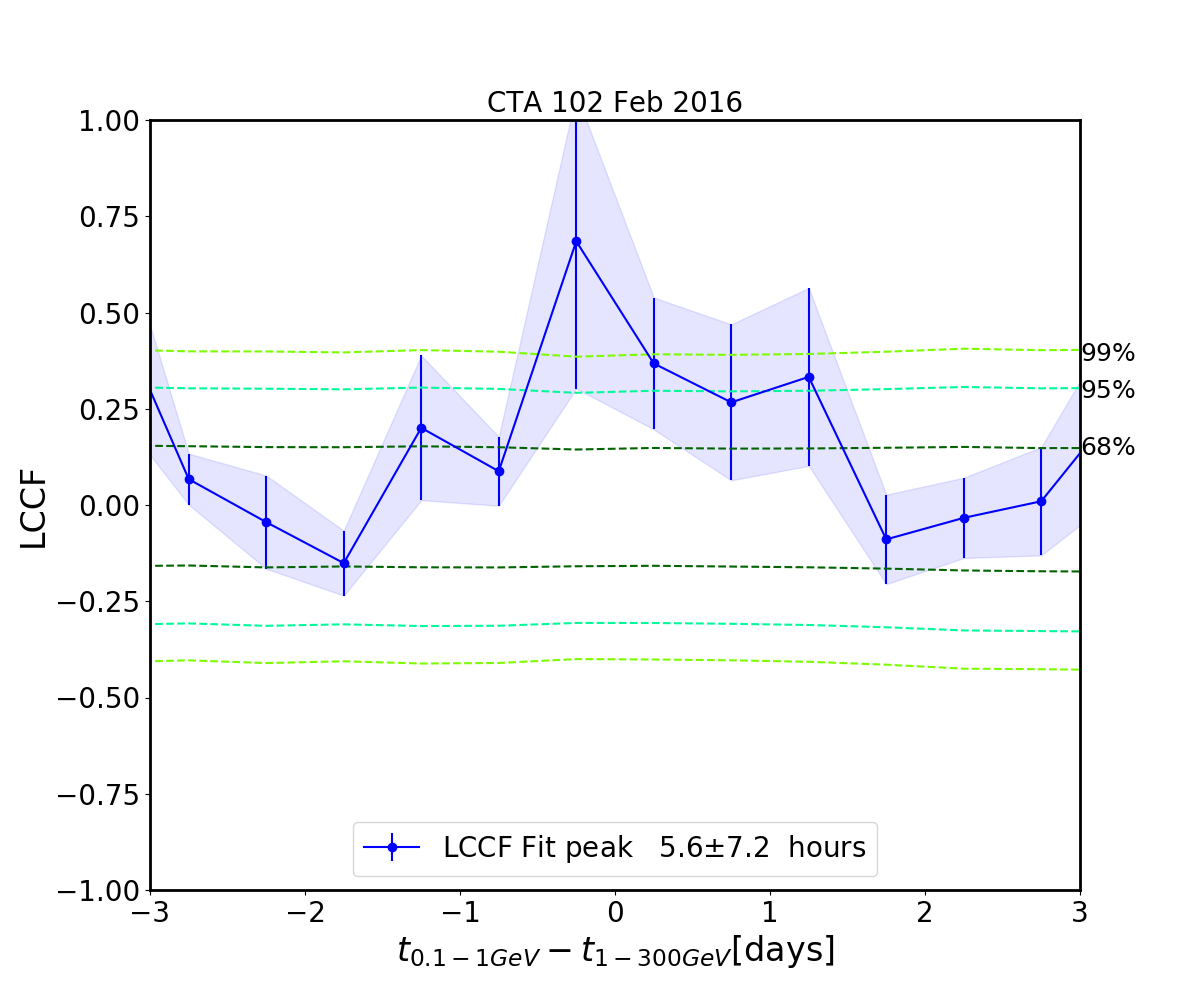}}
    \resizebox{\textwidth}{!}{
    \includegraphics{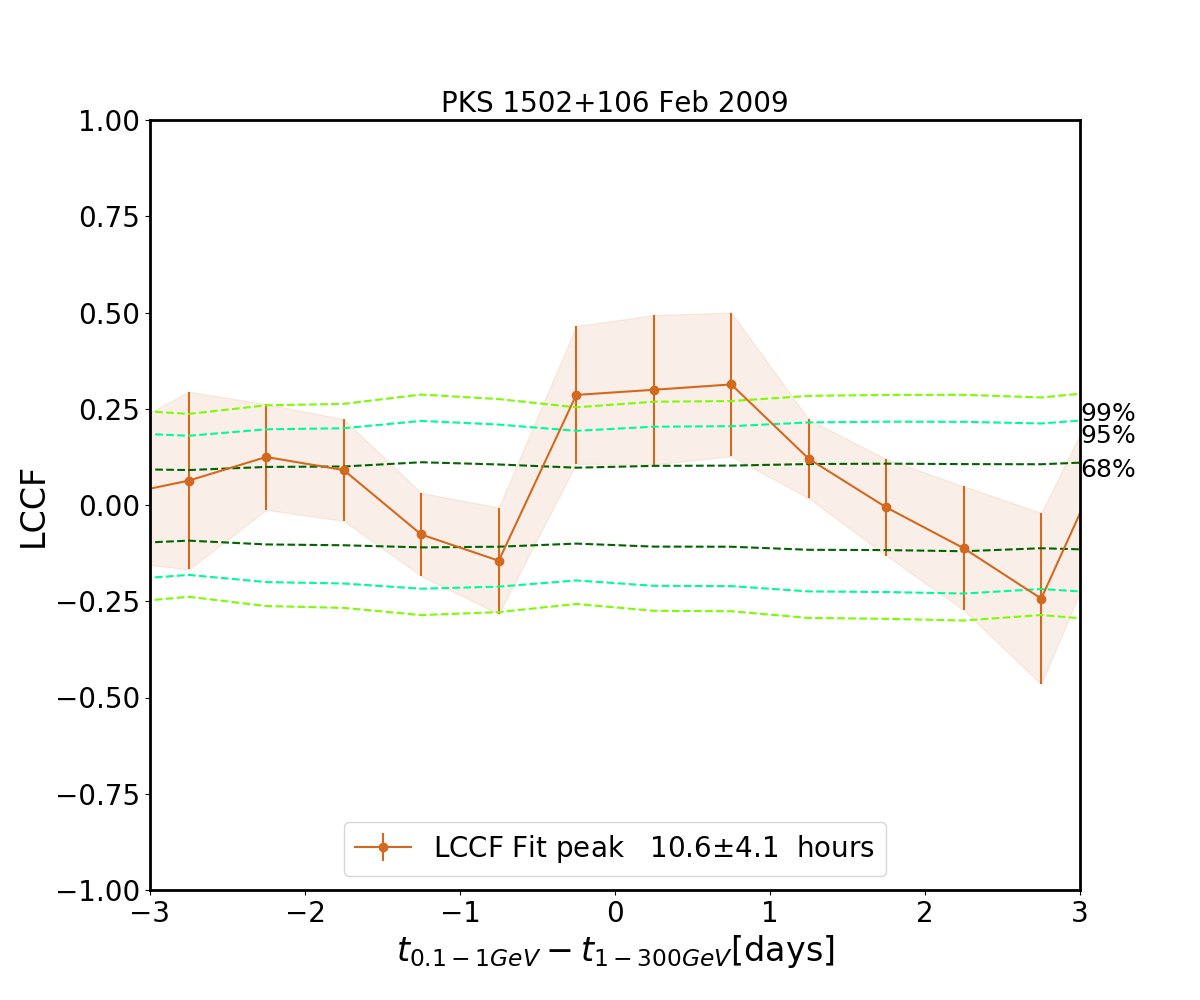}
    \includegraphics{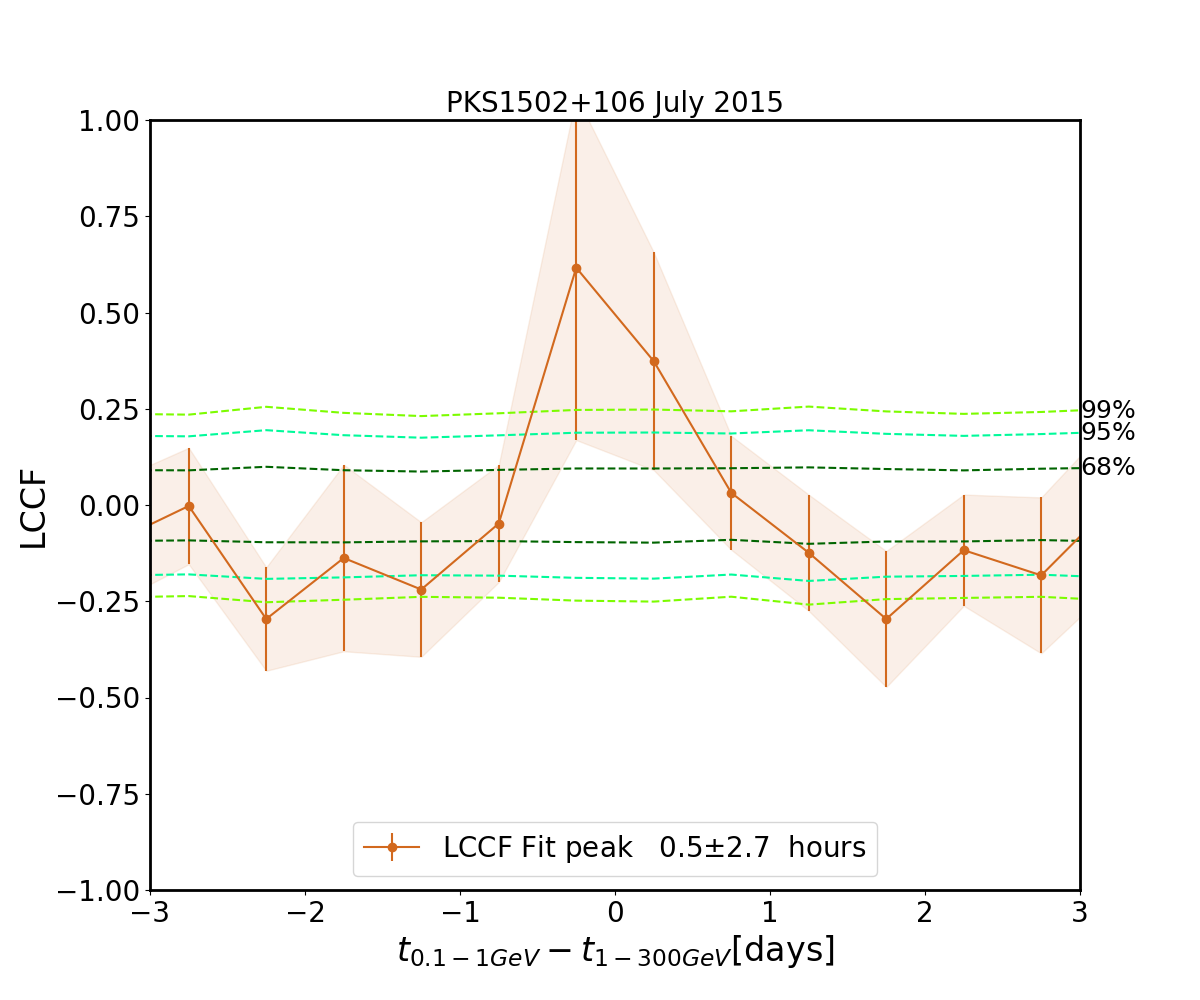}
    \includegraphics{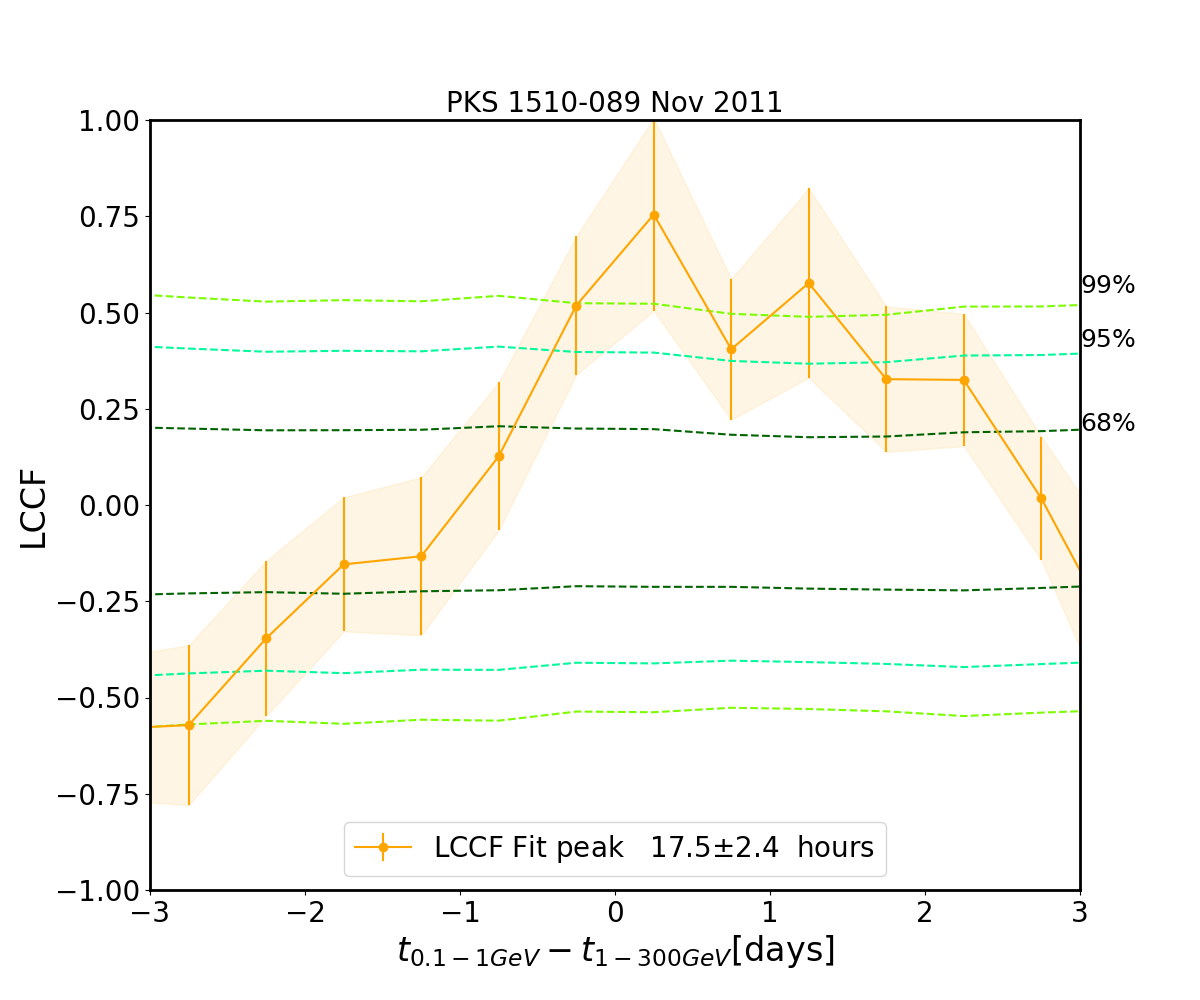}}
    \resizebox{0.33 \textwidth}{!}{
    \includegraphics[width=5.5cm]{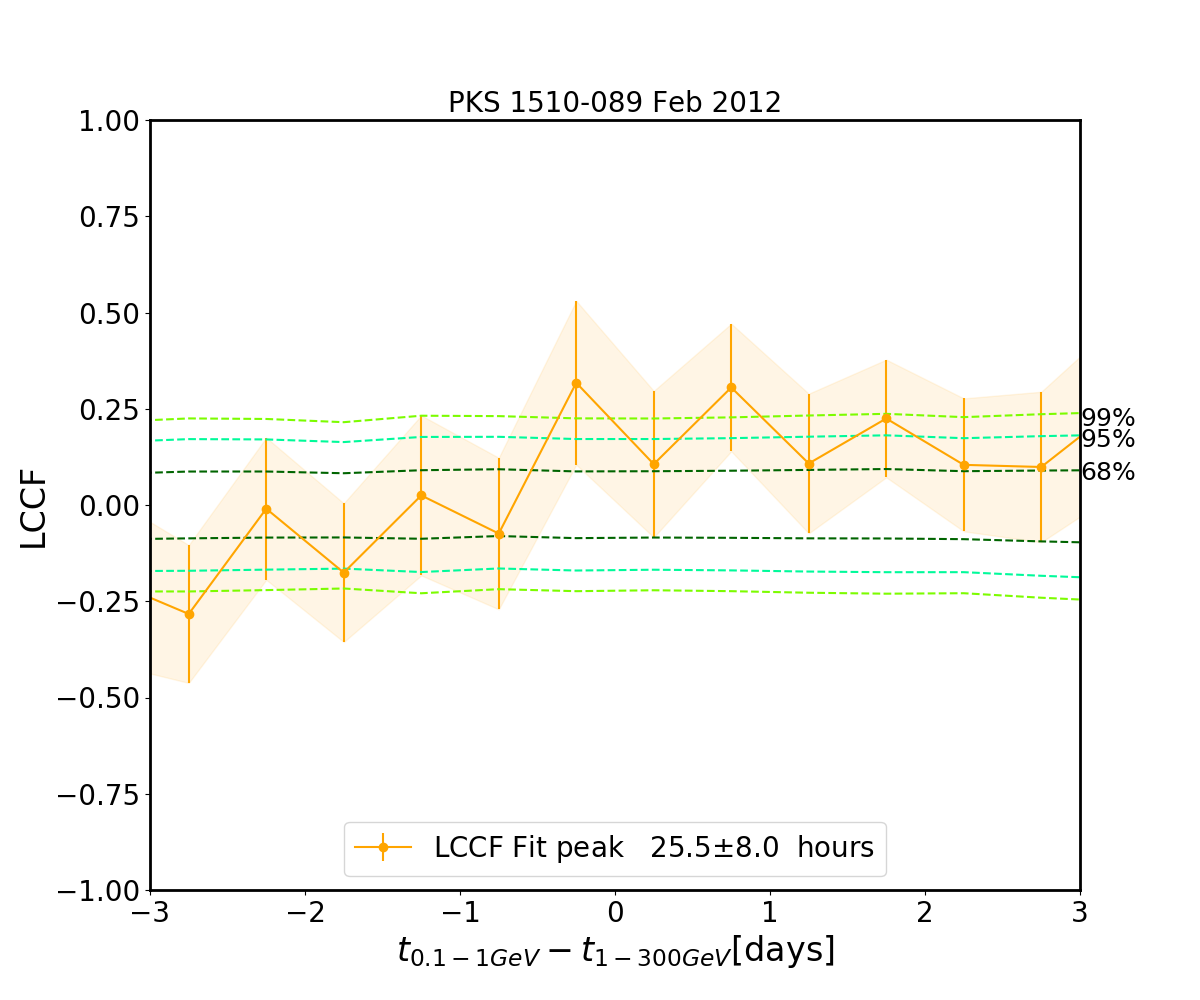}}
    \caption{Local cross-correlation functions (LCCF) calculated between the 0.1 - 1 GeV and the 1 - 300 GeV lightcurves during the flare periods. (The lightcurves, binned in six hour intervals, shown in Appendix \ref{sec:A2}.) The shaded regions indicate the error bounds of the LCCFs. The LCCFs have been fitted with a Gaussian with the time corresponding to the peak of the fit and the associated uncertainty shown in the legend. The green lines represent the \textbf{68$\%$, 95$\%$ and 99$\%$} confidence intervals (from darker to lighter shades) derived from Monte Carlo simulations.}
    \label{fig: Fig 5.}
\end{figure*}

The BLR is a photon-rich environment and the interaction between these photons and gamma-ray photons can lead to photon-photon pair production ($\gamma \gamma \rightarrow e^{+} e^{-}$). The MT has a much lower photon density than the BLR, meaning there is less likelihood of pair production in the MT compared to the BLR. Pair production manifests itself as an attenuation of the gamma-ray flux for emission coming from the inner regions of the BLR, whereas emission originating from the MT is not expected to have this spectral feature (\cite{Donea_2003}, \cite{RN28}).

Emission originating from the BLR would therefore be expected, in general, to be better described by a model with a cut-off (such as a log parabola) rather than a power law.  It should be noted that the presence of a cut-off in the spectrum does not automatically imply BLR origin of emission; it can also be the consequence of a break in the energy distribution of the emitting electrons (\cite{Dermer_2015}). 

To search for the presence of a cut-off, each flare period was re-analysed in daily bins using the routine outlined in Section  \ref{sec:2}. This helped improve statistics at the high energy (1-300 GeV) end of the spectrum. In addition to a log parabolic model (see equation \ref{eq:2}), we also fitted the spectra during the flare periods with a simple power law, defined as:

\begin{equation}
 \hspace*{3cm}
  \centering
   \frac{dN}{dE}=N_{0} \left(\frac{E}{E_0}\right)^{-\gamma}
	\label{eq:7}
\end{equation}
where $\gamma$ is the spectral index, $E_{0}$ is the pivot energy in MeV and  $N_{0}$ is the normalisation (in units of photons $\text{cm}^{-2}\text{s}^{-1} \text{MeV}^{-1}$).

To compare the fits provided by the two models we performed an Akaike Information Criterion (AIC) test  (\cite{RN36}) to determine which model fits the data better. The AIC of a model $s$ is given by:

\begin{equation}
 \hspace*{2.5cm}
  \centering
    \text{AIC}_{s} = - 2 \text{ln}  {L}_{{s}} + 2 k_{f_{s}}
	\label{eq:8}
\end{equation}

where ${L}_{{s}}$ is the likelihood of the model s given the data and ${k}_{{f}_{{s}}}$ is the number of free parameters in the model . 

In order to compare two models s and s' we use the difference in AIC values:
\begin{equation}
 \hspace*{2.5cm}
  \centering
    \Delta \text{AIC}_{s,s'}= \text{AIC}_{s}- \text{AIC}_{s'}
	\label{eq:9}
\end{equation}
which estimates how much more model s diverges from the true distribution than model s', also known as the relative Kullback-Leibler information quantities of the two models (\cite{RN52}, \cite{RN51}).
Another way of interpreting this is to consider how much data would be lost by modelling the data by model s instead of model s'. 
This method is true for both nested and non-nested models (\cite{RN53}); for example, a power law is nested in a log parabola since every parameter in a power law is also present in a log parabola.

A log parabolic model has one extra free parameter relative to a power law model and an AIC test also balances the systematic error in a model with fewer parameters with the random errors of a model having more parameters (\cite{RN13}). 
A lower AIC means a better description of the data. An AIC difference of greater than 2 between two models means that the model with the higher AIC is significantly worse than the model with the lower AIC value (\cite{RN21}). The AIC differences between the log parabolic and power law models found for each flare investigated in this work are tabulated in Table \ref{tab:table4}.

Two of the sources, 3C 454.3 and CTA 102, are found to favour a log parabola during both flares studied, suggesting emission from the BLR. 
Three further sources, namely 3C 279, 4C 21.35 and PKS 1510-089, are seen to favour a log parabola during one flare but the results are inconclusive during the other. The results for the remaining four sources are inconclusive during either flare. This broadly agrees with the results of
\cite{RN17}, who investigated the presence of a cut-off in the spectra for a sample of 106 FSRQs with the highest significance in the Third \textit{Fermi}-LAT catalog (\cite{3LAC}) including all nine sources studied in this work. Evidence was found for a cut-off in 1/3 of the sources, and it was concluded that the emission in the sample originated in regions outside the BLR.

Figure \ref{fig: Fig 4.} shows two representative sets of plots for 3C 454.3 (December 2009) and PKS 1502+106 (July 2015). The top plots show the evolution of daily flux during the course of the flares. 
3C 454.3 is seen to strongly favour a log parabolic model during this outburst, which is also in accordance with the negative mean AIC value found. PKS 1502+106 is also observed to favour a curved spectrum during some days of the flare period but this behaviour is not consistent, resulting in neither model being favoured ultimately.
We address the exact nature and implications of any cut-off on the VHE emission later on in Section  \ref{sec:5}.

\subsection{Energy dependent cooling}
\label{subsec:4.3}

Another key difference between BLR and MT emission is the energy of the seed photons in these regions. The photons in the BLR, being ultraviolet photons, are typically a factor $\sim$ 100 more energetic than the infrared photons present in the MT. 
\cite{Dotson_final} found that Inverse Compton (IC) scattering takes place in the Klein-Nishina regime when the emission region is located inside the BLR, and in the Thomson regime for emission from farther out within the MT. 

\begin{table*}
	\centering
	\caption{Results of the LCCF study between the 0.1 - 1 GeV and the 1 - 300 GeV lightcurves during the flare periods. This includes the times corresponding to the peaks of the Gaussian fit along with the associated uncertainties and their significance in percentile derived from Monte Carlo simulations. The final two columns list the spectral slopes, $\beta$ where the power spectral density PSD $\propto$ $\nu^{-\beta}$, of the original lightcurves and the mean and 95$\%$ confidence intervals of the simulated lightcurves respectively.}
	\label{tab:tableLCCF}
	\resizebox{0.75 \textwidth}{!}{
	\begin{tabular}{lcccccr} 
		\hline
		Source  &Flare  &LCCF &Timelag &Significance &$\beta_{\text{original}}$ &$\beta_{\text{simulations}}$\\
		 &Peak  &Peak &[hours] &[$\%$] & & \\
		 \hline
		 
		3C 454.3 &Dec 2009 &0.4$\pm$0.2 &11.5$\pm$3.3 &$\geq$68 &1.05 $\pm$ 0.01 &1.11 $\pm$ 0.20 \\
		3C 454.3 &Nov 2010 &0.9$\pm$0.1 &0.1$\pm$1.3 &$\geq$99 &1.22 $\pm$ 0.01 &1.27 $\pm$ 0.28 \\
		
		CTA 102 &Sept 2012 &0.8$\pm$0.2 &-13.9$\pm$1.7  &$\geq$99 &0.67 $\pm$ 0.03 &0.73 $\pm$ 0.36\\
        CTA 102 &Feb 2016 &0.7$\pm$0.4 &5.6$\pm$7.2 &$\geq$95 &1.28 $\pm$ 0.02 &1.32 $\pm$ 0.31\\
        
        PKS 1510-089 &Nov 2011 &0.8$\pm$0.2 &17.5$\pm$2.4 &$\geq$99 &0.71 $\pm$ 0.01 &0.74 $\pm$ 0.22 \\
        PKS 1510-089 &Feb 2012 &0.3$\pm$0.2 &25.5$\pm$8.0 &$\geq$68 &1.05 $\pm$ 0.01 &1.01 $\pm$ 0.17\\
        
        PKS 1502+106 &Feb 2009 &0.3$\pm$0.2 &10.6$\pm$4.1 &$\geq$68  &1.13 $\pm$ 0.02 &1.20 $\pm$ 0.28 \\
        PKS 1502+106 &July 2015 &0.6$\pm$0.4 &0.5$\pm$2.7 &$\geq$95 &0.86 $\pm$ 0.01 &0.75 $\pm$ 0.19 \\
         
        3C 279 &June 2015 &0.7$\pm$0.3 &0.9$\pm$2.8 &$\geq$99 &0.76 $\pm$ 0.01 &0.77 $\pm$ 0.16 \\
        
        4C 21.35 &June 2010 &0.9$\pm$0.2 &-5.8$\pm$0.4 &$\geq$99 &0.97 $\pm$ 0.02 &1.08 $\pm$ 0.25 \\

    	\hline
	\end{tabular}}
\end{table*}

This difference results in energy-independent electron cooling times for emission from the BLR as opposed to energy-dependent cooling timescales for regions within the MT. Cooling times are shorter at higher energies, such that emission from the MT would be expected to have, in general, a time-lag on timescales of a few hours between the cooling of the MeV and GeV components of the flare.

To investigate this, we re-analysed our flare periods in two distinct energy ranges: 0.1 - 1 GeV (low energy) and 1 - 300 GeV (high energy), binned in six hourly intervals using the procedure outlined in Section \ref{sec:2}. Six hour bins were chosen as a compromise to allow for sufficient events for analysis (especially at high energies) while still enabling the detection of short timescale variability. The resulting high- and low-energy lightcurves are shown in Appendix \ref{sec:A2}.

Local cross-correlation functions (LCCFs; \cite{Welsh_LCCF}) were then applied to the high- and low-energy lightcurves to search for correlations in the data. The use of LCCFs was motivated by the fact that this technique is independent of differences in sampling rates of the two lightcurves. 
Unlike Discrete Correlation Functions (DCFs; \cite{RN19}), LCCFs are intrinsically bound in the interval [-1,1] and have also been found to be more efficient than DCFs in the study of correlations (\cite{Max_Moerbeck_CCFs}).
There were not sufficient statistics to enable a LCCF analysis for all flares: this was the case for both the flare studied from B2 1520+31, PKS 0454-234 and PKS 1424-41, and the November 2014 flare from 4C 21.35. 
Furthermore, the LCCF for the  December 2013 flare from 3C 279 did not exhibit a clear peak making it difficult to draw any conclusions.

The LCCFs obtained from flares of the remaining sources are shown in Figure \ref{fig: Fig 5.}.
Also shown is the peak of the LCCFs along with the corresponding uncertainties, both derived from Gaussian fits. 
While the peaks of the Gaussian fits give a first order determination of the uncertainty, these do not account for the effects of correlated red-noise between the datasets (\cite{Uttley}).

In order to provide a better estimate of the significance of the observed peaks, Monte Carlo simulations were performed to generate 1000 artificial low energy lightcurves matching the probability distribution and power spectral density (PSD) of the observations using the method outlined in \cite{Emmanoulopoulos}\footnote[7]{The code was developed from Connolly, S. D., 2016, Astrophysics Source Code Library, record ascl:1602.012. See https://github.com/samconnolly/DELightcurveSimulation .}.
Each simulated lightcurve was cross-correlated with the corresponding observed high energy lightcurves and the 68$\%$, 95$\%$ and 99$\%$ confidence intervals obtained are shown in Figure \ref{fig: Fig 5.} with the results summarised in Table \ref{tab:tableLCCF}. With the exception of the December 2009 flare from 3C 454.3, the February 2009 flare from PKS 1502+106 and the February 2012 flare of PKS 1510-089, all correlations are found to have a significance of $\geq 95\%$.

A peak at 0 indicates an absence of time-lag implying BLR origin of the gamma-ray emission.
This is found to be compatible with observations from the June 2015 flare from 3C 279, the November 2010 flare of 3C 454.3, the February 2016 flare of CTA 102 and the July 2015 flare from PKS 1502+106.
A positive time-lag on the other hand implies that the low energy flux is delayed with respect to the high energy flux. Under the assumption that the flux increase in both energy bands occurs at the same time, this points towards MT origin of emission and is seen for both flares from PKS 1510-089, the December 2009 flare from 3C 454.3 and the February 2009 flare from PKS 1502+106.

Two flares, namely the June 2010 flare from 4C 21.35 and the September 2012 flare from CTA 102, show evidence of a negative temporal lag indicating that the changes to the low energy flux precede any changes to the high energy flux. 
While a negative temporal lag does not constrain the location of the emission to either the BLR or MT, it can be interpreted as evidence indicating the MeV and GeV components of the flare have different origin.
Evidence of a negative time lag between the high- and low-energy lightcurves in FSRQs has also been reported in other studies (for example \cite{RN11}, \cite{Cohen_DCF}).

\section{VHE Emission}
\label{sec:5}

We now investigate the very high energy (VHE) photon emission from the sample of FSRQs and discuss its implications on the location of the emission region. 
For this study, VHE photons are defined as photons having an energy $E_{\gamma} \geq$ 20 GeV in the rest frame of the source. The observation of VHE photons is generally difficult to explain if the emission is assumed to be coming from the inner regions of the BLR as photon-photon pair production would make the escape of the high energy photons less probable (\cite{Donea_2003}, \cite{Liu_BLR_abs}, \cite{3c279_bot}). 

As a first step we performed a binned maximum likelihood analysis on the entire eight year data set in the 20 GeV-300 GeV, 50 GeV-300 GeV and 100 GeV-300 GeV energy ranges, using the point source and diffuse emission models outlined in Section \ref{sec:2}. The positions and spectral definitions of all sources in the RoI were once again taken from the 4FGL catalog (\cite{2019arXiv190210045T}).
The resulting flux values and test statistic of sources having a TS $\geq$  10 (which roughly equates to a detection significance of 3$\sigma$) for the different energy ranges are listed in Table \ref{tab:table5}. As expected, both flux and detection significance decrease with increasing threshold energy. Two of the three sources found to have a TS $\geq$  10 above 100 GeV are among the FSRQs detected by ground-based instruments: PKS 1510-089 (\cite{HESSPKS1510_discovery}) and 3C 279 (\cite{Errando_final}). PKS 0454-234, while not yet detected by ground-based instruments, is an interesting candidate for such observations.

\begin{figure*}
    \vspace*{-2cm}
    \centering
    \resizebox{\textwidth}{!}{
    \includegraphics{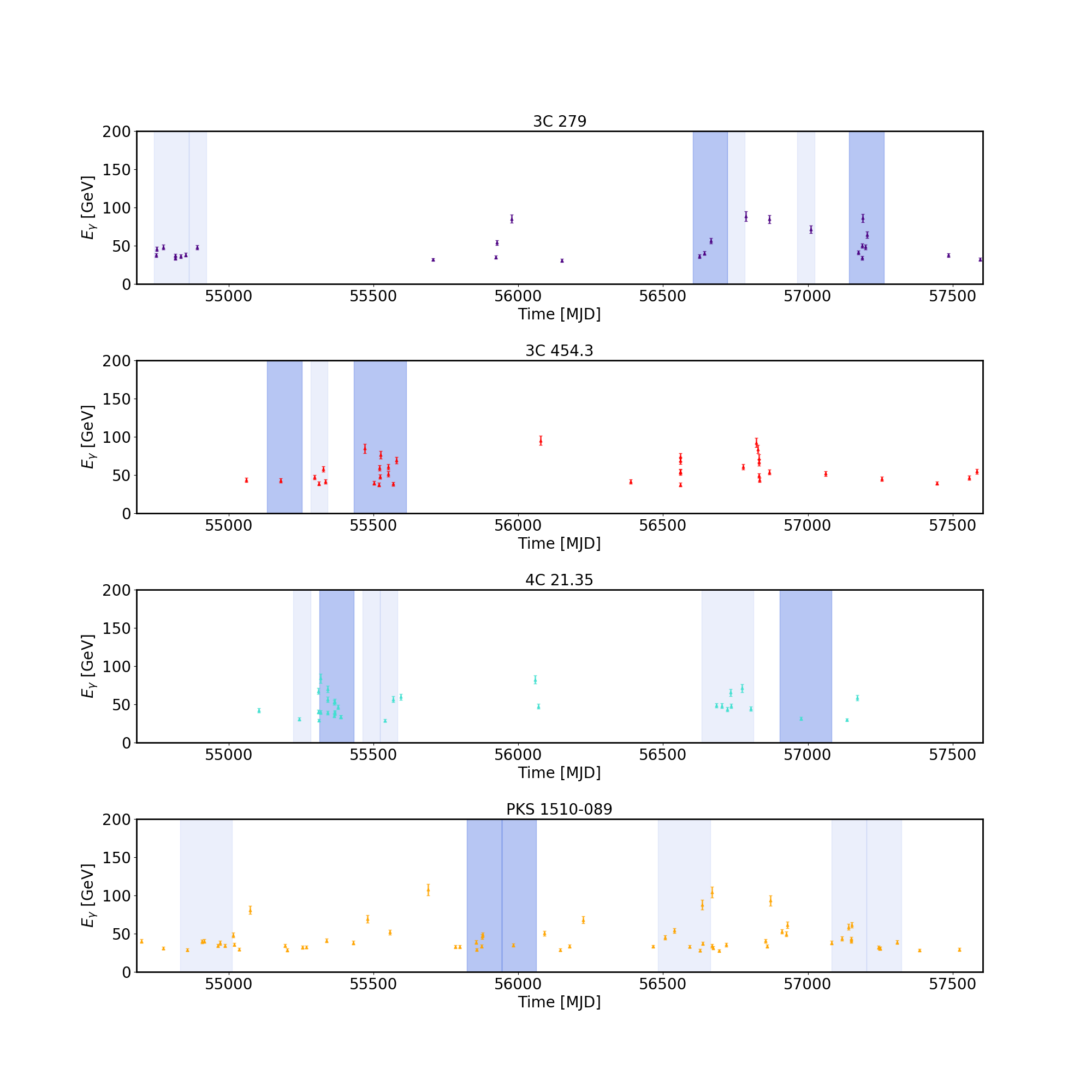}}
    \caption{\textbf{a.} The detected energy, $\text{E}_{\gamma}$, of the individual high energy photons detected with the \textit{Fermi}-LAT over the entire eight year observation period as a function of time for 3C 279, 3C 454.3, 4C 21.35 and PKS 1510-089. All energies are in the rest frame of the galaxy. Only photons with energy $\text{E}_{\gamma} \geq$ 20 GeV and a probability of  $\geq$ 99$\%$ for originating from each source are shown. Also shown as blue shaded regions are the time intervals which satisfy our definition of a flare period (see Section \ref{sec:3}), with the darker shaded regions being the time intervals studied in this investigation.}
    \label{fig:Figure 6.}
\end{figure*}

To check that the VHE emission is associated with the source, we used the \textit{Fermi} tool  \textit{gtsrcprob}, which calculates the probability of each photon being associated with a source in the RoI. Before this step, it was necessary to account for the diffuse components using another \textit{Fermi} tool \textit{gtdiffrsp} and adding the response to the input data. We restrict ourselves to a radius of 0.1$^{\circ}$ around each source and consider only photons having a $\geq$ 99 $\%$ probability of originating from the sources. 
Figure \ref{fig:Figure 6.} shows the lightcurves of the VHE photons emitted by the sample over the entire eight year observation period with the time periods satisfying our definition of flares (see Section \ref{sec:3}) again shown as shaded regions. In most cases VHE photon emission is seen to occur during the flare events. There are instances (for example PKS 0454-234) in which there is VHE photon emission outside the flare periods. As discussed later, this could indicate that the VHE photons are emitted from a different location than the lower energy emission studied previously, but at the very least shows that GeV flares are not necessarily a predictor of VHE emission, and vice versa. This reinforces the requirement for comprehensive sky surveys in the VHE regime (\cite{hassan2017extragalactic}).

As discussed in Section \ref{subsec:4.2}, emission coming from the BLR is expected to have an intrinsic cut-off due to photon-photon pair production. We now attempt to quantify the nature of this cut-off and study its implications for the location of the emission region. 
The \textit{Fermi} tool \textit{gtobssim} is used to simulate observations for the sample of FSRQs taking into account IRFs and the spacecraft pointing history.
These simulations assume intrinsic absorption due to BLR photons and the energies of the simulated photons, when compared to the energies of the observed photons, should reveal whether this assumption is correct and if the observed VHE photons are indeed compatible with BLR origin.

We specify the energy distribution for our simulations by starting with the eight year averaged spectra obtained in Section \ref{sec:2} and concentrate on the energy range 20 GeV-300 GeV. Attenuation due to extragalactic background light (EBL) is also accounted for; we use the EBL opacities, $\tau$, stated in the \cite{RN10} model in this study to calculate the likely attenuation. This model has been found to be compatible with the upper limits from gamma-ray astronomy (for example \cite{Mazin_EBL}, \cite{RN29}). 

The intrinsic absorption due to photons present in the BLR is accounted for by choosing a number of cut-off energies, ${E}_{\text{cut}}$, evenly spaced in the interval 10 GeV-30 GeV. The resultant differential flux used for simulations is given by:

\begin{equation}
 \hspace*{2cm}
    \centering
   \frac{dN}{dE}=N_{0} \left(\frac{E}{E_0}\right)^{-\alpha - \beta \text{ln} \left(\frac{E}{E_0}\right)} e^{-\tau} e^{-\left(\frac{E}{E_{\text{cut}}}\right)}
	\label{eq:10}
\end{equation}

where once again ${E_0}$ is the pivot energy in MeV, $N_{0}$ is the normalisation (in units of photons $\text{cm}^{-2}\text{s}^{-1} \text{MeV}^{-1}$) and $\alpha$ and $\beta$ the spectral index and curvature respectively.

\begin{figure*}
    \vspace*{-2cm}
    \centering
    \resizebox{\textwidth}{!}{
    \includegraphics{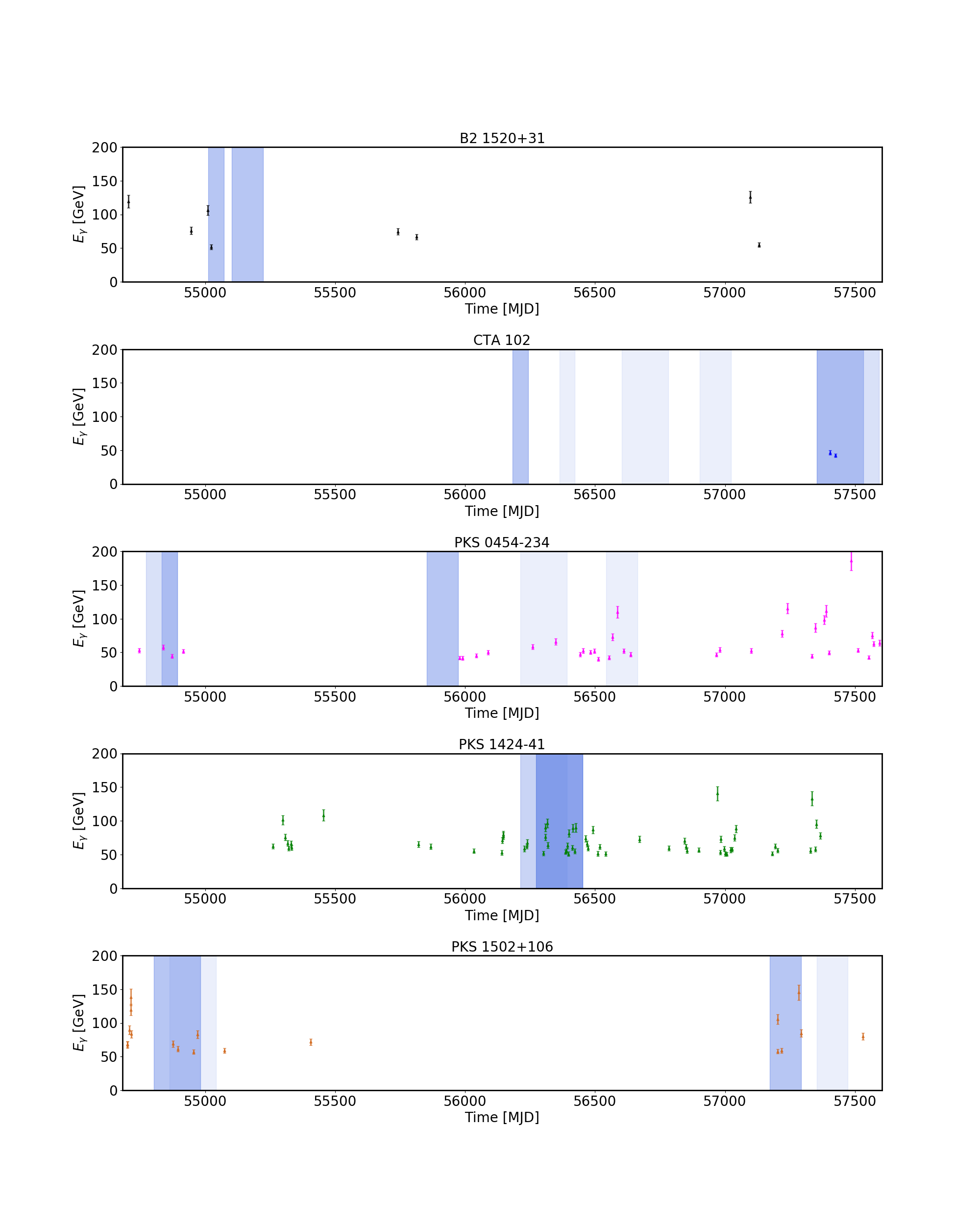}}
     \caption*{\textbf{b.} The detected energy, $\text{E}_{\gamma}$, of the individual high energy photons  detected with the \textit{Fermi}-LAT over the entire eight year observation period as a function of time for B2 1520+31, CTA 102, PKS 0454-234, PKS 1424-41 and PKS 1502+106. All energies are in the rest frame of the galaxy. Only photons with energy $\text{E}_{\gamma} \geq$ 20 GeV and a probability of  $\geq$ 99$\%$ for originating from each source are shown. Also shown as blue shaded regions are the time intervals which satisfy our definition of a flare period (see Section \ref{sec:3}), with the darker shaded regions being the time intervals studied in this investigation.}
\end{figure*}

\begin{figure*}
    \centering
    \resizebox{\textwidth}{!}{
    \includegraphics[width=8cm]{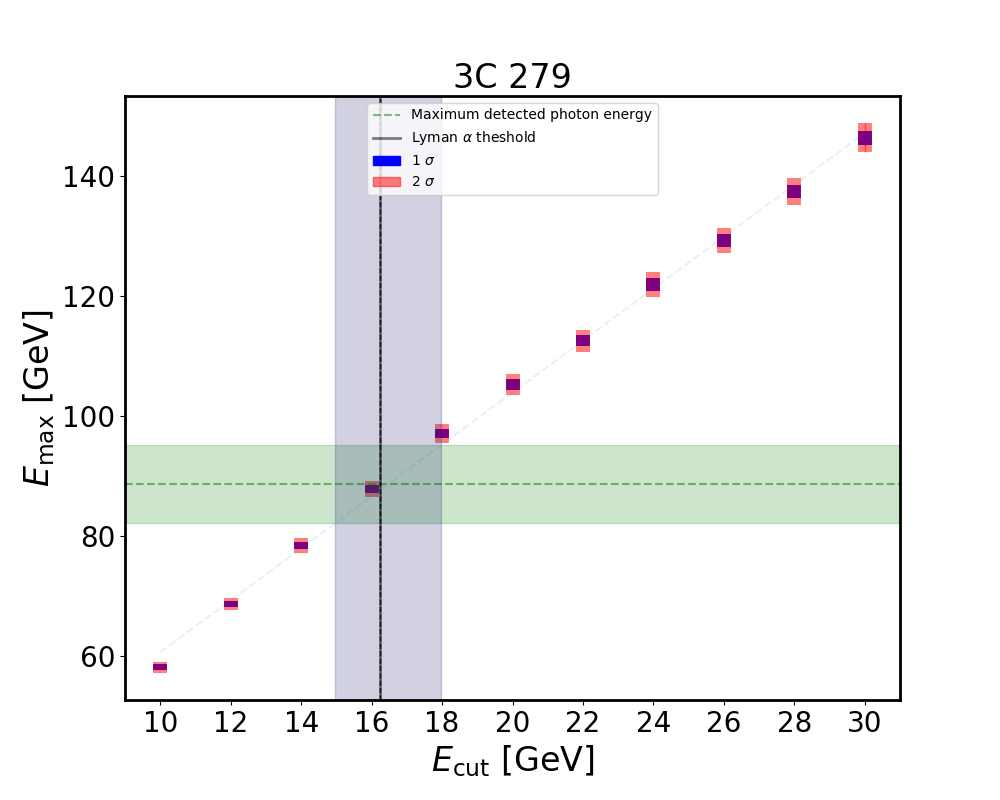}
    \includegraphics[width=8cm]{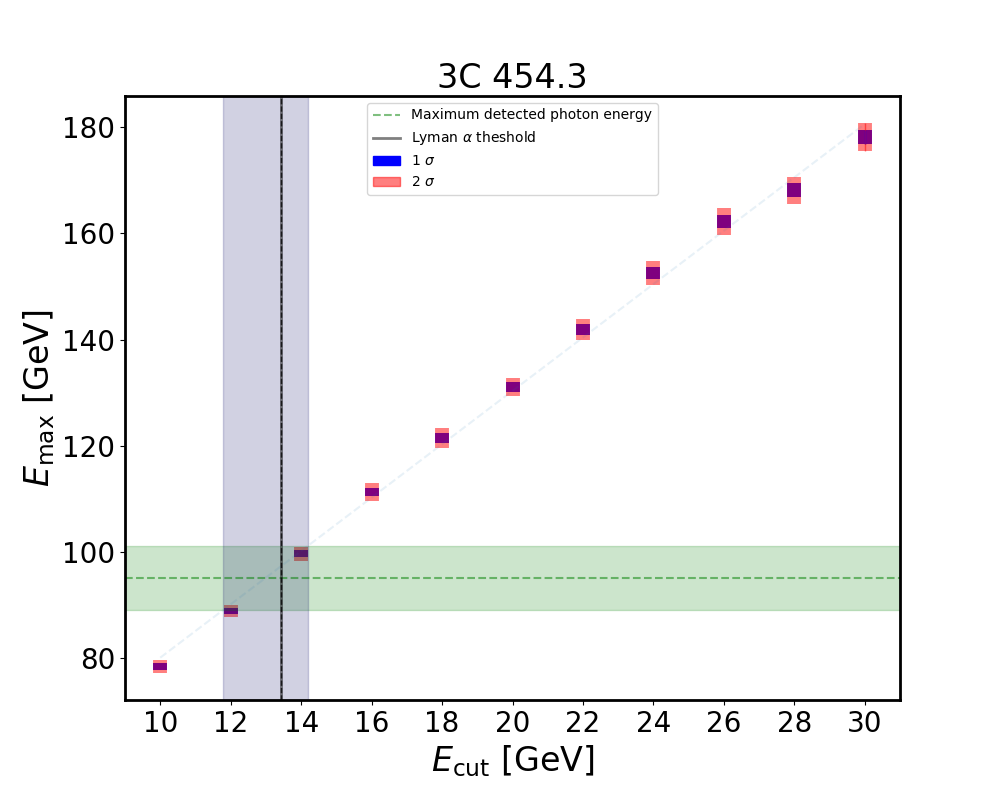}
    \includegraphics[width=8cm]{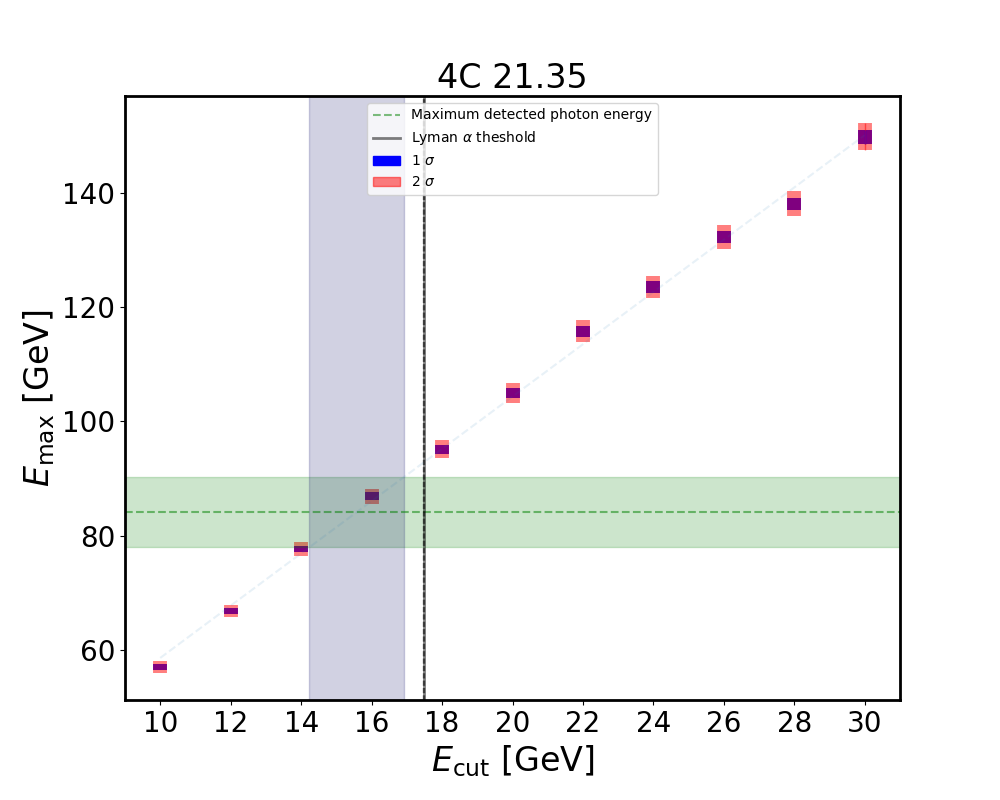}}
    \resizebox{\textwidth}{!}{
    \includegraphics[width=8cm]{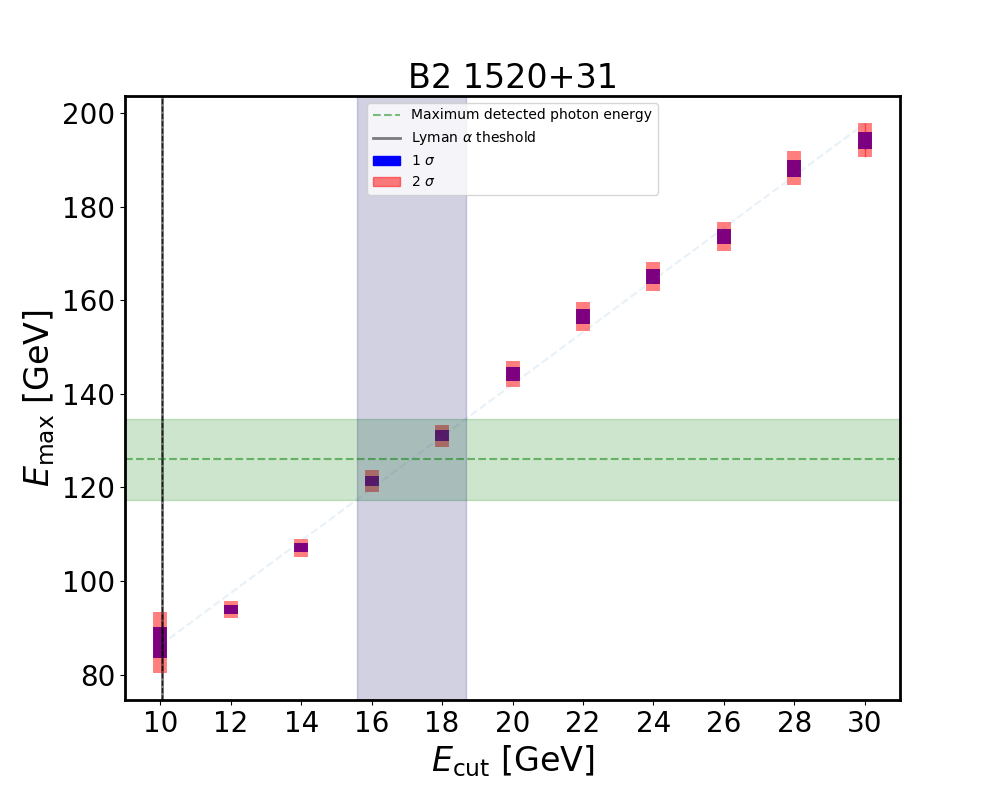}
    \includegraphics[width=8cm]{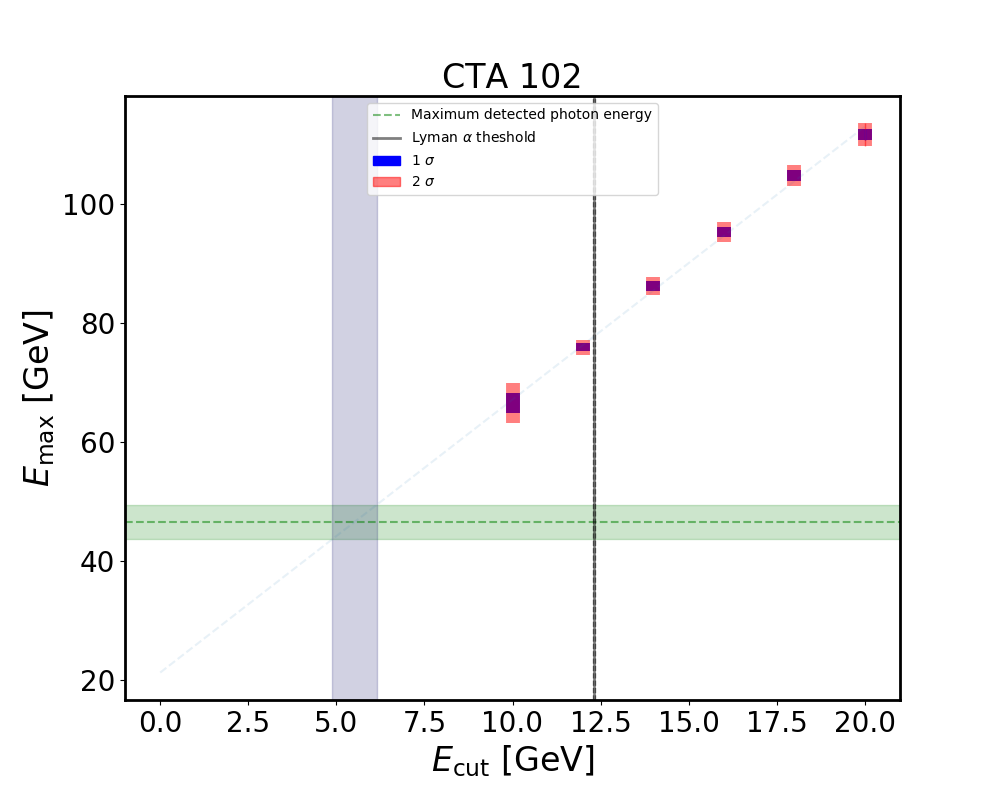}
    \includegraphics[width=8cm]{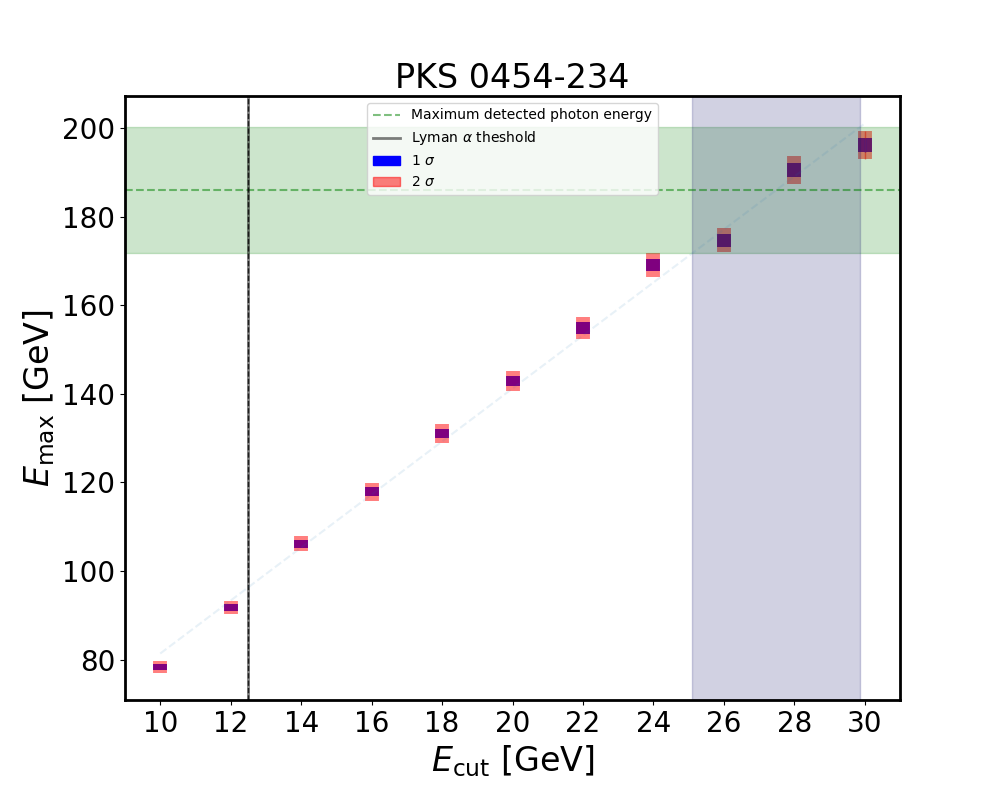}}
    \resizebox{\textwidth}{!}{
    \includegraphics[width=8cm]{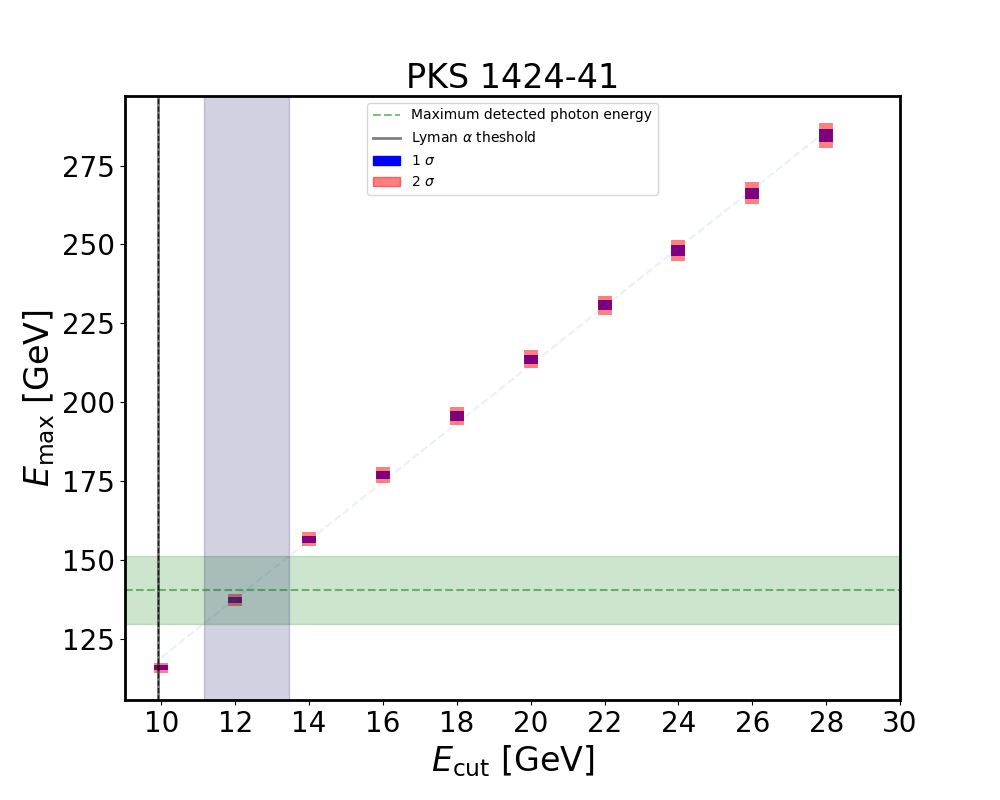}
    \includegraphics[width=8cm]{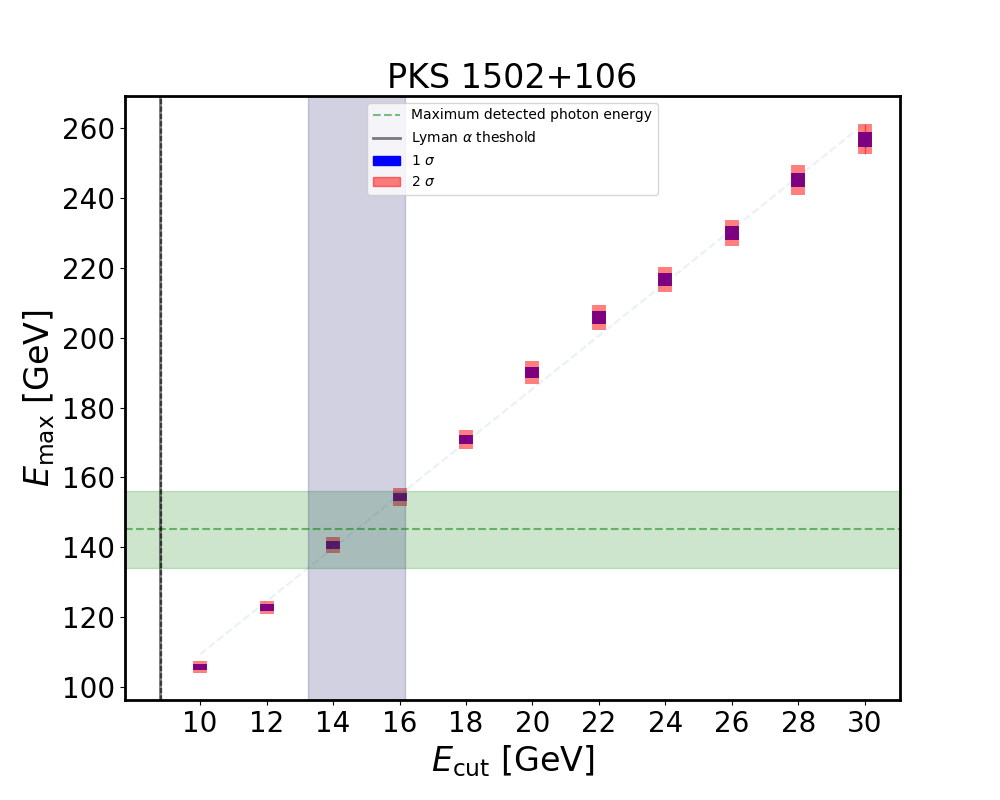}
     \includegraphics[width=8cm]{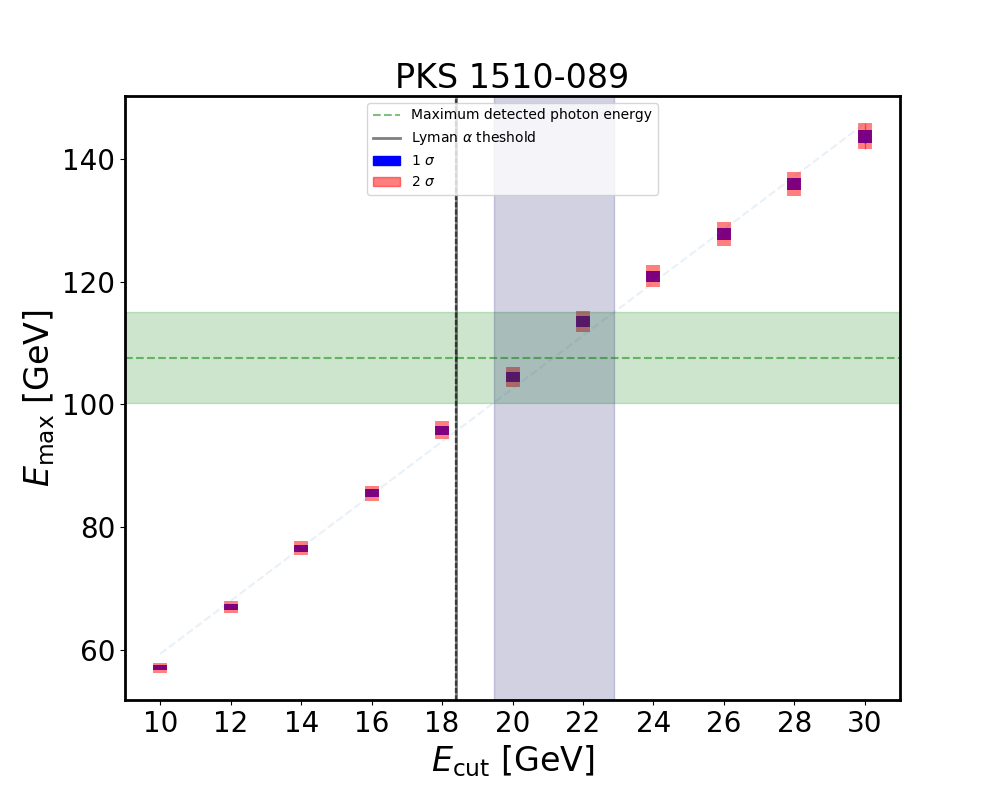}}
    \caption{The plots show the energy distributions of the most energetic photons, $E_{\text{max}}$, obtained from Monte Carlo simulations as a function of the cut-off energy, $E_{\text{cut}}$, used in the simulations (see equation \ref{eq:10}) for all sources investigated in this study. The blue and red shaded regions represent the 1$\sigma$ and 2$\sigma$ confidence intervals respectively. The green dashed line is the energy of highest energy photon observed with the \textit{Fermi}-LAT during the eight year observation period for each source along with the corresponding uncertainty. The cut-off energy range which best agrees with the observations is shown as the vertical blue shaded region. For comparison, the black vertical line is the expected intrinsic cut-off energy due to interaction with Lyman alpha photons calculated as 25/(1+z) GeV where z is the redshift of the source. }
    \label{fig:Fig7}
\end{figure*}

Using 1000 simulations for each source with different seed variables, the energy distribution of the most energetic photons simulated was determined. These are shown in Figure \ref{fig:Fig7}, where we plot the 1$\sigma$ and 2$\sigma$ confidence intervals as a function of the different cut-off energies used in the simulations. Also shown is the energy of highest energy photon observed with the \textit{Fermi}-LAT during the eight year period for each source with its corresponding uncertainty. 

The cut-off energy range which best agrees with the observation was then determined and compared to the expected onset of intrinsic cut-off due to interaction with Lyman alpha photons in the BLR ($E_{\text{Ly}\alpha}=\frac{25}{1+z}$ GeV for a source at redshift z (\cite{RN3})). 
In the case where the observed and expected photon energies are compatible, the VHE photon emission observed with the \textit{Fermi}-LAT is compatible with BLR origin; this is the case for for 3C 279, 3C 454.3 and 4C 21.35.
However, for the other 6 sources, the emission is constrained to parsec scale distances from the central engine, i.e. within the MT.

\section{Discussion}
\label{sec:6}

\begin{table*}
	\centering
	\caption{The average flux and test statistic (TS) values (see equation \ref{eq:1}) obtained from a likelihood analysis of the eight year (MJD 54682.66 - MJD 57604.66) \textit{Fermi}-LAT observations of each source above an energy threshold of $E_{\gamma} \geq 20$ GeV, $E_{\gamma} \geq 50$ GeV and $E_{\gamma} \geq 100$ GeV respectively. Only sources having a TS $\geq$ 10 in each energy range are shown.}
	\label{tab:table5}
	\resizebox{0.4 \textwidth}{!}{
	\begin{tabular}{lcr} 
		\hline
		Source  &Flux  &TS \cr
		 &[ $10^{-6}$ photons $\text{cm}^{-2}$ $\text{s}^{-1}$]  & \cr
		\hline
		 &20 GeV-300 GeV\\\cline{1-3}
         3C 454.3 &5.71 $\pm$ 0.86 &536  \\
		 CTA 102 &0.76 $\pm$ 0.32 &89 \\
		 B2 1520+31 &1.42 $\pm$ 0.41 &85 \\
		 PKS 1510-089 &11.47 $\pm$ 1.51 &820 \\
	     PKS 1502+106 &2.77 $\pm$ 1.38 &213\\
		 PKS 1424-41 &7.85 $\pm$ 0.89 &837 \\
		 3C 279 &5.97 $\pm$ 1.18 &386\\
		 4C 21.35  &6.89 $\pm$ 1.13 &501 \\
		 PKS 0454-234 &6.26 $\pm$ 1.05 &504  \\
		 \cline{1-3}
		 &50 GeV-300 GeV\\\cline{1-3}
		 3C 454.3 &0.52 $\pm$ 0.35 &25   \\
		 PKS 1510-089 &4.42 $\pm$ 1.35 &119 \\
		 PKS 1424-41 &0.42 $\pm$ 0.20 &39  \\
		 3C 279 &2.31 $\pm$ 0.95 &65 \\
		 4C 21.35  &1.19 $\pm$ 0.54 &69 \\
		 PKS 0454-234 &2.14 $\pm$ 0.79 &84  \\

        \cline{1-3}
		 &100 GeV-300 GeV\\\cline{1-3}
		 PKS 1510-089 &2.42 $\pm$ 1.18 &39  \\
		 3C 279  &1.28 $\pm$ 1.02 &16 \\
		 PKS 0454-234 &0.39 $\pm$ 0.29  &22  \\
    	\hline
	\end{tabular}}
\end{table*}

\begin{table*}
	\centering
	\caption{Summary of the results from the different methods used to constrain the location of the emission region for both flare periods from each source. These methods are the measurement of the shortest variability timescales for the flare periods, the search for evidence of a cut-off in the flare spectra and an investigation into energy dependence in cooling timescales. The final column lists whether the VHE ($\text{E}_{\gamma} \geq 20$ GeV) photon emission observed with the \textit{Fermi}-LAT is compatible with BLR origin. Inconclusive results are due to the flare spectra favouring neither a power law nor a log parabolic model or due to the lack of statistics at high energies preventing obtainment of LCCFs. Multi-zone indicates evidence of multiple emission regions but no physical constraints on the location. }
	\label{tab:table6}
	\resizebox{\textwidth}{!}{
	\begin{tabular}{lcccr} 
		\hline
		Source &Sizes of emission region  from variability timescales[$10^{13}$m]* &Spectral cut-off &Energy dependent cooling &VHE photons  \cr 
		 &Flare 1,  Flare 2 &Flare 1,  Flare 2 &Flare 1,  Flare 2 &from BLR \cr
		\hline
         3C 454.3 &3.87 $\pm$ 0.84,    7.38 $\pm$ 1.03  &BLR, BLR &MT, BLR &Compatible  \\
		 CTA 102 &4.78 $\pm$ 0.86,    3.59 $\pm$ 0.59 &BLR, BLR &Multi-zone, BLR &Incompatible \\
		 B2 1520+31 &0.70 $\pm$ 0.12,   3.27 $\pm$ 0.91  &Inconclusive, Inconclusive &Inconclusive, Inconclusive &Incompatible\\
		 PKS 1510-089 &6.82 $\pm$ 1.03,    5.30 $\pm$ 1.56 &Inconclusive, BLR &MT, MT &Incompatible  \\
	     PKS 1502+106 &0.93 $\pm$ 0.18,    1.33 $\pm$ 0.09 &Inconclusive, Inconclusive &MT, BLR &Incompatible \\
		 PKS 1424-41 &0.77 $\pm$ 0.24,    2.76 $\pm$ 0.66 &Inconclusive, Inconclusive &Inconclusive, Inconclusive &Incompatible   \\
		 3C 279 &4.11 $\pm$ 0.34,    4.23 $\pm$ 1.28 &Inconclusive, BLR &Inconclusive, BLR &Compatible\\
		 4C 21.35  &2.05 $\pm$ 0.66,    1.67 $\pm$ 0.12 &BLR, Inconclusive &Multi-zone, Inconclusive &Compatible\\
		 PKS 0454-234 &4.55 $\pm$ 0.79,    3.91 $\pm$ 0.67 &Inconclusive, Inconclusive &Inconclusive, Inconclusive &Incompatible \\
    	\hline
	\end{tabular}}
	\begin{tablenotes}
      \small
      \item * The variability timescales imply extremely compact emission regions. Assuming the entire width of the jet to be responsible for the emission, all  timescales are compatible with BLR origin of emission.
    \end{tablenotes}
\end{table*}

\subsection{Individual Sources}
\label{subsec:6.1}
In order to draw conclusions regarding the location of the emission region, we now combine the findings from the methods discussed in the previous two sections and study their implications for each source individually. The results are summarised in Table \ref{tab:table6}.

\subsubsection{3C 454.3}

The variability timescales for 3C 454.3, the brightest of the sources, predict a compact emission region of size (3.87 $\pm$ 0.84) $\times$ $10^{13}$ m for the December 2009 flare and (7.38 $\pm$ 1.03) $\times$ $10^{13}$ m for the November 2010 flare. Both of these are compatible with emission from the inner regions of the BLR in a simple one-zone model. 
This conclusion is reinforced by evidence that the spectra of both flares favour a log parabolic model over a power law, particularly in the case of the November 2010 flare. 

This bright flare also provides sufficient statistics for a LCCF to be obtained; this shows no evidence for a time-lag between the high- and low-energy emission, again supporting a BLR origin of the emission. 
However, the LCCF for the December 2009 flare indicates the low energy flux is delayed with respect to the high energy flux with evidence of a time-lag of 11.5 $\pm$ 3.3 hours. Assuming the flux increase in both bands takes place at the same time, this favours emission from the MT.

As seen in Figure \ref{fig:Figure 6.}, both flare periods studied in this work are observed to be accompanied by the emission of very high energy ($\text{E}_{\gamma} \geq 20$ GeV) photons. The most energetic photon from this source with energy 95.1 $\pm$ 6.1 GeV was emitted on MJD 56076.89 and is outside of the flare periods studied here. Monte Carlo simulations when compared to the energy of this photon show that the cut-off energy that best agrees with the observations is 13.0 $\pm$ 1.2 GeV, which is compatible with the expected cut-off of 13.441 $\pm$ 0.001 GeV due to interaction with Lyman alpha photons in the BLR. We note that 3C 454.3 has not been detected in the energy range $\text{E}_{\gamma} \geq 100$ GeV with ground-based gamma-ray telescopes and an analysis of the \textit{Fermi}-LAT data over the entire observation period also found no significant emission in this energy range.

The November 2010 flare was studied in \cite{Foschini_2011} and the $2 - 3$ hour intrinsic variability timescales reported is compatible with the 2.80 $\pm$ 0.39 hours result for the same flare observed in this investigation.
It should also be noted that the \cite{Foschini_2011} calculation requires the successive measurements to have a difference in flux significant at at least 3$\sigma$.
A separate investigation of 3C 454.3 from August 2008 to January 2010 by \cite{RN24} also found variability timescales of a few hours and constrained the size of the emission region to $\text{R} < 3.5 \times 10^{13} (\delta)/10$ m  $= 8.54$ $\times$ $10^{13}$ m for $\delta=24.4$ (\cite{Jorstad_2017}), a factor $\sim$ 1.8 times larger than the upper limits obtained in this study.
Both investigations conclude that the emission region is within the BLR. The same conclusion is also reached in a study of the November 2010 flare by \cite{vercelone_3C454_2010}.

Combining the results of our analyses, we conclude that the gamma-ray emission in 3C 454.3 predominantly comes from regions within the BLR.
However, the December 2009 flare exhibits energy dependence of the cooling timescales, suggesting the possibility of multiple simultaneously active emission regions both within the BLR and the MT.
This agrees with the findings of the multi-wavelength study of the same flare by \cite{pacciani_3C454_dec_2009} who concluded that explaining the gamma-ray observations corresponding to the peak of the flare requires models more elaborate than a simple one-zone emission model.
An investigation of the June 2014 flare from 3C 454.3 by \cite{RN33} also suggests the presence of multiple emission regions and constrains the location of the emission to be outside the BLR. 
As seen in  Table \ref{tab:table6}, the possibility of multiple simultaneously active emission regions is not a property unique to 3C 454.3 but a general feature found in our sample.

\subsubsection{CTA 102}

The September 2012 flare from CTA 102 shows a shortest variability timescale of 1.45 $\pm$ 0.26 hours while the February 2016 flare shows an even shorter timescale of 1.09 $\pm$ 0.18 hours. 
Both of these imply an extremely compact emission region which, assuming the entire width of the jet to be responsible for the emission, are both compatible with emission from near the central engine. 
The spectra from both flares studied also favour a log parabola over a power law, which in principle reinforces the theory of BLR origin of the emission. 

The peak of the Gaussian fit to the LCCF obtained for the February 2016 flare is compatible with an absence of a time-lag and further evidence of emission from within the BLR.
The LCCF for the September 2012 flare shows evidence of a lag at -13.9 $\pm$ 1.7 hours, which indicates that the variations in low energy flux precede any changes to the high energy flux. 
This can be interpreted as evidence of multiple emission regions, with the MeV and GeV components having different origins for this particular flare.

Only two VHE photons are seen during the entire observation period, both of which coincide with the February 2016 flare. 
The most energetic of these photons, observed on MJD 57404.15, has an energy 46.5 $\pm$ 2.9 GeV. 
Monte Carlo simulations indicate this photon best agrees with a cut-off at 5.5 $\pm$ 0.6 GeV as opposed to the 12.32 $\pm$ 0.02 GeV expected from Lyman alpha photon interaction. 
A possible explanation for the unusually low cut-off observed in the spectrum is the absorption of gamma-rays due to pair production on Helium II recombination continuum photons (\cite{RN2}). 
This would indicate the emission originates deep inside the BLR, within a light-year from the SMBH.
\cite{RN47} also report evidence of a spectral cut-off in the energy range 9 - 16 GeV  from \textit{Fermi}-LAT observations of CTA 102 between January 2016 and April 2018. 
The feature is stated to be likely due to an intrinsic break in the energy distribution of the emitting particles and the observations were found incompatible with BLR origin of emission.

\cite{zach_cta_2017} explain the evolution of CTA 102 from late 2016 to early 2017 as being a result of the addition of a large amount of mass to the jet over a period of a few months, with the subsequent drop in the light curve due to a the ablation of the material.
From modelling the spectrum, a strong constraint on the maximum electron Lorentz factor is derived which also forces a cut-off of the IC component to be fixed at $\sim$ 20 GeV.
This is an upper limit of the maximum photon energy achievable without taking EBL absorption into consideration. 
For comparison, our Monte Carlo simulations are compatible with an expected spectral cut-off at 6.09  $\pm$ 0.60 GeV when EBL absorption is not taken into account, which is not compatible with the result of \cite{zach_cta_2017}.
In conclusion, we find evidence indicating the gamma-ray emission in CTA 102 is produced in multiple compact emission regions, some of which may be deep inside the BLR.

\subsubsection{B2 1520+31}

B2 1520+31 shows a fastest flux doubling time of 0.65 $\pm$ 0.11 hours from the July 2009 flare, the shortest variability timescale obtained from all the flares studied in this investigation and implying an extremely compact emission region of size (0.70 $\pm$ 0.12) $\times$ $10^{13}$ m for this particular flare.
The November 2009 flare has a variability timescale of 3.03 $\pm$ 0.84 hours corresponding to an emission region of size (3.27 $\pm$ 0.91) $\times$ $10^{13}$ m.
These two timescales were observed in flux measurements $\sim$ 100 days apart and if the two flares had their origin in a single event, this would suggest the emission region is expanding with a velocity of (2.97 $\pm$ 0.97) $\times$ $10^{6}$ $\text{ms}^{-1}$ $\approx$ 0.01 c.

The spectra of the flares studied show no strong preference for either a power law or a log parabolic model, making the search for a spectral cut-off inconclusive. 
The study of energy dependence in cooling timescales was also found to be inconclusive, due to a lack of photon statistics preventing analysis using LCCFs.
A total of 8 VHE photons were observed over the entire eight year observation of which 2 coincide with the July 2009 flare.
The most energetic of these photons, having energy 126.04 $\pm$ 8.66 GeV, was observed on MJD 57095.99, outside the time intervals corresponding to a flaring period. 

Monte Carlo simulations indicate that a cut-off energy at 17.1 $\pm$ 1.6  GeV best agrees with the energy of this photon. 
This is considerably higher than the 10.040 $\pm$ 0.001 GeV expected due to Lyman alpha absorption and indicates that the VHE photon emission is not compatible with BLR origin.
\cite{RN3} investigated a high energy flaring period of B2 1520+31 from April 2009. Interpolating the work of \cite{RN28}, the optical depth, $\tau_{\gamma\gamma}$, was calculated for the BLR region. 
The optical depth for gamma-rays emitted at the midpoint of the spherical BLR shell was found to be $\tau_{\gamma\gamma}=1.4$ at 35 GeV and $\tau_{\gamma\gamma}=2.0$ at 50 GeV. This further implies the VHE photons observed are likely produced at large distances from the SMBH. 

The findings, put together, suggests the gamma-ray flares are being produced in a very small emission region, which could be within the BLR. 
However, there is no further evidence to suggest BLR origin of emission since investigations of both a cut-off in the flare spectra and energy dependence of cooling timescales proved inconclusive.
Furthermore, the VHE photon emission observed with the \textit{Fermi}-LAT strongly disfavours BLR origin for these photons. 

\subsubsection{PKS 1510-089} 

The November 2011 flare for PKS 1510-089 was found to have a fastest variability timescale of 1.79 $\pm$ 0.27 hours while the February 2012 flare has a shortest timescale of 1.39 $\pm$ 0.41 hours. Assuming the entire width of the jet to be responsible for the emission, this would indicate emission from within the BLR. 
This possibility is supported by the spectrum of the February 2012 flare favouring a log parabolic model over a power law, although the November 2011 flare favours neither model significantly.

An investigation into the energy dependence of the cooling timescales shows evidence that both flares exhibit a positive time-lag between the high- and low-energy emission. Under the assumption that the flux increase in both energy bands occurs simultaneously, this in turn indicates emission from the MT.
Furthermore, the VHE photons observed with the \textit{Fermi}-LAT predict an expected cut-off energy of 21.2 $\pm$ 1.7 GeV, which is higher than the 18.38 $\pm$ 0.03 GeV cut-off expected for BLR origin emission. There is also substantial VHE photon emission outside the flare periods including the most energetic photon, of energy 107.6  $\pm$ 7.4 GeV, observed on MJD 55687.83. Indeed, PKS 1510-089 has been detected at $\text{E}_{\gamma} \geq 100$ GeV with the H.E.S.S. telescopes (\cite{HESSPKS1510_discovery}), which would also indicate emission farther from the black hole. 

An investigation of the first 3.75 years of \textit{Fermi}-LAT data for PKS 1510-089 by \cite{RN11} includes the November 2011 flare studied here and reports an even shorter variability timescale of 1.21 $\pm$ 0.15 hours by applying equation \ref{eq:4} directly to two consecutive flux measurements satisfying TS $\geq$ 10, rather than the three consecutive time bins we have used. From spectral and variability studies, he concluded that the jet was capable of simultaneously producing rapid variability gamma-ray emission at various points along the entire jet from the BLR to the MT.Both our study and that of \cite{RN11} agree on the lack of a trend between GeV flux and emission of VHE photons, which can be interpreted as further evidence of multiple emission zones with the VHE emission thought to be produced farther out in the MT. 

A study of the \textit{Fermi}-LAT data from September to December 2011 by \cite{Saito_2013} found similar results and conclusions to \cite{RN11} and also reported observed doubling timescales $\sim 1$ hour. 
Assuming a generic Doppler factor $\delta = 20$, the emission region was constrained to be of size $1.5 \times 10^{13}$ m which is smaller than the upper limit of $(6.82 \pm 1.03) \times 10^{13}$ m obtained in this work using $\delta = 35.3$ (\cite{Jorstad_2017}) from optical data. 
This emission region was thought to be located within the BLR while any VHE emission, if detected, was argued to be produced further from the central engine.
A similar conclusion was reached by \cite{Barnacka_2014} who use a two-zone model to reproduce the VHE emission observed by H.E.S.S. telescopes in March 2009. In their model the bulk of the GeV emission is found to be coming from within the BLR, while the VHE emission results from Comptonization of IR photons from the MT.
Our results support this hypothesis.

\subsubsection{PKS 1502+106}

The shortest variability timescale from the February 2009 flare of PKS 1502+106 is 0.86 $\pm$ 0.17 hours, one of only three sub-hour timescales discovered among the flares investigated. Based on this timescale, the size of the emission region is constrained to be (0.93 $\pm$ 0.18) $\times$ $10^{13}$ m.
The July 2015 flare also shows hour scale variability, with a shortest variability timescale of 1.23 $\pm$ 0.08 hours implying an emission region of size (1.33 $\pm$ 0.09) $\times$ $10^{13}$ m. 
The spectra for both flares were found to favour neither a power law nor a log parabolic model, so there is no evidence for a cut-off in the spectrum. 
The results of our LCCF study are mixed; emission from the July 2015 flare supports the premise of BLR origin with evidence for a correlation peak at 0.5 $\pm$ 2.7 hours, but the more rapid February 2009 flare shows a correlation peak  at 10.6 $\pm$ 4.1 hours, which is instead compatible with emission from within the MT.

Both the flare periods studied coincide with VHE photon emission, including the most energetic photon which has an energy 145.11 $\pm$ 11.09 GeV and was observed on MJD 57283.92 (during the July 2015 flare).  
Monte Carlo simulations reveal that a cut-off energy of 14.7 $\pm$ 1.5 GeV best agrees with this observation, which is higher than the expected cut-off of 8.803 $\pm$ 0.005 GeV due to Lyman alpha absorption of BLR photons. 
This implies that the VHE photon emission observed with the \textit{Fermi}-LAT is not compatible with the BLR and might indicate the presence of multiple emission regions. 

The complex nature of our findings agrees with the results of \cite{RN39}, whose study of \textit{Fermi}-LAT observations from PKS 1502+106 between August - December 2008 concluded that the gamma-ray emission was produced by External Compton (EC) scattering of BLR photons.  Using the flux increase between August 5 and 6, the maximum size of the emission region was constrained to be  R $\leq$ 6.8  $\times$ $10^{13}$ m which is a factor $\sim$ 6 bigger than our findings. The level of correlations found between gamma-ray, X-ray, optical and UV data during the flare and post-outburst periods, supported the conclusion that this source is likely to be at the border between BLR dissipated FSRQs and MT dissipated FSRQs. 
\cite{RN39} also suggest the large gamma-ray dominance over other wavelengths observed during the outburst is difficult to explain with a single-zone emission model.

An investigation by \cite{Max_Moerbeck_2014}, using  the first three years of \textit{Fermi}-LAT data cross-correlated with radio data, found PKS 1502+106 as one of only three sources to show a correlation at larger than 2.25$\sigma$. The radio variations were found to lag the gamma-ray variations, indicating the gamma-ray emission originates upstream of the radio emission at a distance of 22 $\pm$ 15 pc from the central engine, which is beyond the BLR for a conical jet model.  

In conclusion, this study finds PKS 1502+106 to be another example of an FSRQ with multiple simultaneously active emission regions. There is evidence of BLR emission from the short variability timescales, while the study of energy dependent cooling timescales yields different results for the two flare periods. However, the VHE photons observed with the \textit{Fermi}-LAT are clearly not compatible with a BLR origin for the emission.

\subsubsection{PKS 1424-41}

The observed variability timescales of the January 2013 flare from PKS 1424-41 indicate a gamma-ray emission region of size (0.77 $\pm$ 0.24) $\times$ $10^{13}$ m. The April 2013 flare from this source was found to have a larger emission region of size (2.76 $\pm$ 0.66) $\times$ $10^{13}$ m. 
However, in neither flare is one spectral model favoured over another, and a lack of statistics at high energies made our study of energy dependent cooling inconclusive.
Monte Carlo simulations show that the most energetic photon observed with the \textit{Fermi}-LAT, having energy 140.5 $\pm$ 10.7 GeV and observed on MJD 56970.42 outside the flare periods we studied, is compatible with a cut-off energy of 12.3 $\pm$ 1.2  GeV. 
This is just incompatible with the energy cut-off expected due to BLR emission of 9.920 $\pm$ 0.001 GeV. 

A multi-wavelength study of the April 2013 flare by \cite{pks1424} found the emission region to be located outside the BLR. Interpreting the SED using a one-zone leptonic model, the emission region was constrained to a distance of $5 \times 10^{16}$ m from the central engine. 
Emission regions within the MT can also be reconciled with the short variability timescales observed in our study if one assumes the existence of compact emission regions throughout the jet.
It has been proposed that these result from magnetic reconnection events (\cite{RN18}, \cite{giannios2}) or the recollimation of the jet (\cite{RN22}).

The gamma-ray observations corresponding to the January 2013 flare period investigated in this work were claimed to be coincident with the petaelectronvolt (PeV; 1 PeV = $10^{6}$ GeV) neutrino \textit{cascade} event IC 35 detected by the IceCube collaboration (\cite{Kadler_2016}), interpreted as evidence for hadronic emission from this object. IceCube events are classified depending on the pattern of the light seen in the detector array. \textit{Track} events result from a high energy muon travelling a large distance, forming a visible track in the detector, and have an angular resolution of $\leq 1^{\circ}$. \textit{Cascade} events, such as IC 35, are due to particle showers resulting from neutrino interactions and can be resolved to $\sim 15^{\circ}$(\cite{aartsen2014}). The larger positional uncertainty for the cascade events, raising the possibility of chance spatial coincidences between astrophysical neutrinos and potential astrophysical sources.

The IC 35 neutrino event that \cite{Kadler_2016} claimed to associated with the gamma-ray flare studied in this work was centered on the coordinates RA = 208.4$^{\circ}$ and DEC = -55.8$^{\circ}$ with a median positional uncertainty of $R_{50}= 15.9^{\circ}$. As such, there is an angular separation of $\theta=14.8^{\circ}$ between PKS 1424-41 and the neutrino \textit{cascade} event. A Monte Carlo simulations study of IceCube \textit{track} neutrino candidates, revealed that a single neutrino event within 1$^{\circ}$ of a gamma-ray source is consistent with chance coincidence (\cite{Brown_2015}). Finally, we note that the January 2013 flare period also included the emission of 8 $\text{E}_{\gamma} \geq 20$ GeV photons but that there was no reported detection of neutrino events associated with this flare.

Nonetheless, a hadronic component to the emission might explain why the leptonic approaches used throughout this study to determine the location of the emission region from the flares have proved inconclusive.

The results of our investigations, put together, implies an extremely compact gamma-ray emission region.
There is no direct evidence to suggest BLR origin as investigations into the presence of a cut-off in the spectrum and the energy dependence of the cooling timescales proved inconclusive for both flare periods studied. 
The VHE photon emission observed with the \textit{Fermi}-LAT is incompatible with BLR origin and indicates emission from within the MT.

\subsubsection{3C 279}

3C 279 shows a shortest variability timescale of 2.08 $\pm$ 0.17 hours during the December 2013 flare and 2.14 $\pm$ 0.65 hours for the June 2015 flare.
While the December 2013 flare favours neither model, the spectrum of the June 2015 flare strongly favours a log parabola over a power law and is therefore compatible with an emission region inside the BLR. 
Further evidence towards BLR origin of emission is provided by the LCCF study of the June 2015 flare showing a `lag' of 0.9 $\pm$ 2.8 hours, indicating no energy dependence in cooling timescales. The LCCF study of the December 2013 flare was inconclusive due to a lack of photon statistics.

During the flare period studied, 19 VHE photons were observed with the \textit{Fermi}-LAT. However, the maximum observed photon energy (88.6 $\pm$ 6.5 GeV) was observed on MJD 56785.70, just outside the period of the December 2013 flare. Monte Carlo simulations suggest that this corresponds to a cut-off energy of 16.5 $\pm$ 1.5 GeV, which is compatible with the expected cut-off energy of 16.234 $\pm$ 0.004  GeV due to interaction with Lyman alpha photons and indicates the VHE emission is also compatible with BLR origin.

A study of the December 2013 flare by \cite{2015ApJ...807...79H}, based on broadband spectral modelling, found the shortest variability timescales to be $\sim$ 2 hours which agrees well with our result and also places the emission region within the radius of the BLR.
\cite{rani3c279} studied the flaring activity of 3C 279 between November 2013 - August 2014 and found six bright flares superimposed on the long term outburst. The first three of these correspond to the December 2013 flare studied in this investigation. 
This flare was accompanied by the ejection of a new VLBI component, and, the 43 GHz core beyond the BLR,  is suggested as the potential source of the gamma-ray emission.
The June 2015 flare was studied by \cite{Ackermann_2016} and a flux doubling time of less than 5 minutes on top of the long term evolution of the event has been reported. 
These extremely short timescales constrain the emission region to a size of R $\leq 10^{-4} (\delta/50)$ pc = $1.13 \times 10^{12}$ m for $\delta=18.3$ (\cite{Jorstad_2017}). 

A separate investigation of the June 2015 flare was undertaken in the $\text{E}_{\gamma} \geq 100$ GeV domain with H.E.S.S. (\cite{2019}).
Using a combined fit of the \textit{Fermi}-LAT data and H.E.S.S. data to find constraints on the absorption of gamma-rays, the emission region was found to be at a distance r $\geq 1.7 \times 10^{15}$ m from the SMBH and beyond the BLR. 
The minute scale variability was attributed to small turbulent cells (\cite{giannios2}) rather than an emission region encompassing the entire width of the jet.

The \cite{2019} study used EBL optical depths from the \cite{Franceschini_2008} model and adopted a detailed study of the BLR absorption by considering different geometries in order to extrapolate the \textit{Fermi}-LAT data beyond 10~GeV. Nonetheless, the extrapolation under-predicts the H.E.S.S. flux at the highest energies by an order of magnitude, and indeed no one-zone model was able to fully describe the multi-wavelength behaviour during the June 2015 flare.

In conclusion, while the results from the December 2013 flare are inconclusive, the results from the June 2015 flare support a BLR origin for the gamma-ray emission.
The VHE emission observed more generally from this source with the \textit{Fermi}-LAT is also found to be compatible with a BLR origin.
While the presence of multiple emission regions seen in other sources has been suggested (\cite{rani3c279}), this study finds no direct evidence for emission from beyond the BLR. 

\subsubsection{4C 21.35}

The variability timescales of 4C 21.35 (also known as PKS  1222+216) are indicative of an emission region of size (2.05 $\pm$ 0.66) $\times$ $10^{13}$ m for the June 2010 flare and (1.67 $\pm$ 0.12) $\times$ $10^{13}$ m for the November 2014 flare. Under the simple one-zone model assumption, both of these indicate emission from within the BLR.
In terms of the spectral shape, the November 2014 flare favours neither model, but the June 2010 flare is better fitted by a log parabola than a power law, which in principle is further evidence for BLR origin of the emission. 

The June 2010 flare shows evidence of a LCCF peak at -5.8 $\pm$ 0.4 hours, indicating that changes to the low energy component of the emission precedes changes to the high energy component.
This suggests that, similarly to the case of CTA~102, the MeV and GeV components of this particular flare have different origins which may be interpreted as evidence for multiple emission regions.
The November 2014 flare did not have enough photon statistics to allow the study of energy dependence in cooling timescales using LCCFs.

The most energetic VHE photon was observed at an energy of 84.1 $\pm$ 6.2 GeV on MJD 55317.89 (during the June 2010 flare). 
Monte Carlo simulations show that this is indicative of a cut-off energy of 15.6 $\pm$ 1.4 GeV, which is lower than the expected cut-off at 17.48 $\pm$ 0.01 GeV due to interaction with Lyman alpha photons within the BLR.
This implies the high energy photon emission observed with the \textit{Fermi}-LAT  is, in principle, compatible with BLR origin.

The June 2010 flare of 4C 21.35 investigated in this work was detected with MAGIC (\cite{4c2135_magic}) in the energy range $70$ GeV $\leq \text{E}_{\gamma} \leq 400$ GeV.
This spectrum was found to be well described with a hard power law and also, unlike the H.E.S.S. 3C~279 spectrum (\cite{2019}), to connect smoothly with the \textit{Fermi}-LAT spectrum (\cite{RN41}), suggesting a common origin for the emission. 
The observed flux doubling times of 10 minutes also constrained the size of the emission region to R  $ \leq 2.5 (\delta/10) \times 10^{12}$ m = $1.85 \times 10^{12}$ m for $\delta=7.4$ (\cite{Jorstad_2017}). 
Assuming a standard one-zone model, this would imply an emission region well within the BLR. 
However, the dense photon fields in the BLR would lead to high opacity for the gamma-rays detected with MAGIC due to photon-photon pair production (see Section \ref{subsec:4.2}).
This contradiction is addressed in \cite{Tavecchio_4c21} who examine whether a one-zone model is a viable solution to reproduce the observed spectral energy distribution and variability of 4C 21.35 from the MAGIC detection.
Three different models are used: i) a simple one-zone model outside the BLR, ii) a two-zone model with the emission region predominantly located outside the BLR and iii) a two-zone model with the emission regions inside the BLR. The two-zone models are found to be energetically less demanding than the single zone model and the results are compatible with a scenario in which the jet undergoes recollimation at large distances from the SMBH.

The results of our investigations for 4C 21.35, together with evidence from other work, again suggests that gamma-ray emission results from from multiple compact regions along the relativistic jet.

\newpage
\subsubsection{PKS 0454-234}

PKS 0454-234 is seen to have a shortest variability timescale of 1.62 $\pm$ 0.28 hours for the January 2009 flare and 1.39 $\pm$ 0.24 hours during the November 2011 flare. 
Both timescales indicate emission from extremely compact regions in the jet of size (4.55 $\pm$ 0.79) $\times$ $10^{13}$ m and (3.91 $\pm$ 0.67) $\times$ $10^{13}$ m respectively.
However, search for evidence of a possible cut-off in the spectrum for both flares proved inconclusive.
This was also the case for the investigation into the energy dependence in cooling timescales, due to the large uncertainties in flux.

While the flare intervals studied are both accompanied by VHE emission, the most energetic VHE photon ($E$ = 185.9 $\pm$ 14.2 GeV) was observed on MJD 57486.05, when there is no evidence for any flaring activity at lower energies. 
An energy cut-off at 27.5 $\pm$ 2.4 GeV best agrees with this observation; this is significantly higher than the 12.5 GeV expected for this source due to interaction with Lyman alpha photons and suggests emission from beyond the BLR.

Interpolating the work of \cite{RN28}, who investigated a period of VHE activity of PKS 0454-234 from November - December 2012, the optical depth, $\tau_{\gamma\gamma}$, for gamma-rays emitted at the midpoint of the spherical BLR shell is $\tau_{\gamma\gamma}=0.9$ at 35 GeV and $\tau_{\gamma\gamma}=1.3$ at 50 GeV. We find a significant detection (TS = 84) of this object at $E$ > 50~GeV combining all 8 years of \textit{Fermi}-LAT data (Table \ref{tab:table5}), suggesting that there is emission originating from beyond the BLR. This agrees with the findings of \cite{RN3} who report that the shape of the SED for PKS 0454-234, in particular the large separation between the IC peak and the synchrotron peak, suggests the VHE emission is likely to be coming from large distances from the SMBH.

Thus, while the rapid variability from this object suggests that emission could originate within the BLR, there is no supporting evidence for this contention from the spectral shape or from a LCCF analysis. The evidence from the VHE emission, which is seen both during the flares studied and outside flare events, is strongly suggestive of emission originating outside the BLR. The observed high energy photon emission makes PKS 0454-234 an interesting candidate for follow-up observations with IACTs, particularly as it is one of only three objects to show evidence (at the $\sim 4~\sigma$ level) for emission above 100 GeV in the \textit{Fermi}-LAT dataset we analysed. The other two objects (PKS 1510-089 and 3C 279) have already been detected with IACTs.

\subsection{Overview and Implications}
\label{subsec:6.9}

A detailed analysis of the two brightest flares from the sample of nine FSRQs has revealed flux variability timescales of the order of a few hours, indicating gamma-ray emission from extremely compact regions.
Within the context of a simple one-zone model, these timescales are compatible with emission regions within the BLR. However, other evidence reveals a more complex picture.

The search for presence of a spectral cut-off shows evidence that 7 of the 18 flares studied favour a log parabolic model over a power law; this can be interpreted as evidence of BLR origin of emission for these flares.
The remaining flares were found to favour neither model over the other, which could indicate emission either from within the BLR or beyond it.  

A study of energy dependence in cooling timescales shows evidence of achromatic cooling in 4 flares, indicating BLR origin of emission, while a further 6 flares revealed the presence of a time-lag between the MeV and GeV components of the emission which can be interpreted as evidence of multiple emission regions. 
Of these, 4 flares (including both those from PKS 1510-089) showed evidence of a positive time-lag between the high- and low- energy flux suggesting emission from regions within the MT, and 2 showed a negative time-lag, which may be indicative of multi-zone emission.

Finally, through Monte Carlo simulations it is shown that the $\text{E}_{\gamma} \geq 20$ GeV photon emission observed with the \textit{Fermi}-LAT from most sources (the exceptions being for 3C 279, 3C 454.3 and 4C 21.35) is incompatible with BLR origin.
This implies emission regions within the MT at parsec scale distances from the central engine, and the lack of correlation between the observed GeV flare intervals and VHE photon emission detected in some sources (for example PKS 0454-234) can be interpreted as evidence of multiple emission regions.

The results of the investigations presented in this work lead to the natural conclusion that a more complex emission model than a simple one-zone leptonic model is required. 
As seen in Table \ref{tab:table6}, there is evidence to suggest the presence of multiple simultaneously active emission regions both within the BLR and the MT, in most individual sources even during the same flaring episode.
In the context of the sources studied in this work, multi-zone emission has been suggested in previous investigations (for example PKS 1510-089 (\cite{nale_multi}, \cite{RN11}), 3C 454.3 (\cite{RN33}, \cite{Finke_3c454_multi}), 3C 279 (\cite{rani3c279}) and 4C 21.35 (\cite{foschini_4c2135_multi}). The existence of multiple extremely compact and simultaneously active emission regions is seemingly a characteristic feature found in gamma-ray observations of the brightest FSRQs.

\section{Conclusion}
\label{sec:7}

This paper undertakes a temporal and spectral analysis of the gamma-ray emission from a sample of nine bright FSRQs observed with the \textit{Fermi}-LAT over the first eight years of its operation.
We consider photons detected in the energy range 100 MeV-300 GeV in the time interval MJD 54682.66 - MJD 57604.66 which corresponds to midnight on the August 4, 2008 until midnight on August 4, 2016.
During this period each source was observed to have several intervals satisfying our definition of a flare (see Section \ref{sec:3}). 
The two brightest flares from each source were investigated in detail in order to draw conclusions regarding the size and location of the emission region.

These bright flares provided sufficient statistics to allow for re-analysis in daily, 6 hourly and 3 hourly intervals while still satisfying the TS $\geq$ 10 criterion for each bin. 
The 3 hour binned lightcurves revealed variability in timescales of a few hours, with the shortest flux doubling time obtained being 0.65 $\pm$ 0.11 hours from the July 2009 flare of B2 1520+31.
These short timescales imply an extremely compact emission region of the order of $10^{13}$ m for each source.
While it should be noted that emission regions within the MT can also be reconciled with the short variability timescales observed in our study, for instance those resulting from magnetic reconnection events (\cite{RN18}, \cite{giannios2}) or the recollimation of the jet (\cite{RN22}), if one assumes that the entire width of the jet is responsible for the emission, the timescales indicate BLR origin.

The flare periods were then studied in more detail to search for the presence of a cut-off in the spectrum which can be interpreted as a consequence of photon-photon pair production within the BLR. 
An AIC test was undertaken to determine which of a power law and a log parabolic model provided a better fit to the data. 
This study finds evidence for spectral cut-off in 7 of the 18 flares investigated, supporting a BLR origin for the emission during these events. No conclusive evidence for a cut-off was found for the other 11 flares. 

This was followed by an investigation into the energy dependence in cooling timescales by applying LCCFs to search for correlations between the high energy (1 - 300 GeV) and low energy (0.1 - 1 GeV) flux.
4 flares were found to have a LCCF compatible with a peak at 0, indicating no energy dependence and implying a BLR origin for the emission. 
A further 6 flares show evidence of a time-lag between the MeV and GeV components of the emission which can be interpreted as indicating the presence of multiple emission regions.
Among these, 4 flares have a positive time-lag between the high- and low- energy flux suggesting emission regions within the MT and 2 showed evidence for a negative time-lag.
The results of the remaining flares were found to be inconclusive, with the lack of photon statistics preventing the calculation of LCCFs.

The final investigation considered very high energy ($\text{E}_{\gamma} \geq 20$ GeV) photon emission from the sample of FSRQs. 
A likelihood analysis of all  photons in the energy range 20 - 300 GeV over the entire eight year observation period revealed significant emission from all sources at a confidence level of > 5$\sigma$.
This was followed by a closer inspection of the individual photons observed.
Monte Carlo simulations were used to compare the most energetic photon observed with the \textit{Fermi}-LAT for each source to the expected photon energy distribution assuming BLR origin of emission.
Only three of the sources, 3C 279, 3C 454.3 and 4C 21.35, are found to have VHE photon emission compatible with the expected BLR Lyman alpha photon interaction suggesting that the VHE emission in the other sources is being produced in emission regions within the MT.

The apparent contradictions regarding the origin of the gamma-ray emission found in the sample can be reconciled by invoking the presence of multiple simultaneously active emission regions both within the BLR and the MT.

Future study of the gamma-ray emission from FSRQs as well as other sources in the VHE energy range is expected to improve with the construction of the Cherenkov Telescope Array (CTA; \cite{RN43}). 
The CTA is expected to provide unprecedented insight over a wide energy range of 20 GeV-300 TeV and improve on the sensitivity of current ground-based telescopes by more than an order of magnitude. 
The CTA will comprise two observatories in order to provide a full-sky coverage. The Northern array will be located in La Palma (Spain) and the Southern array will be located at Paranal (Chile).

Preliminary simulations indicate that all of the sources presented in this work should be detectable with the CTA (\cite{hassan2017extragalactic}).
The enhanced sensitivity should provide improved statistics to make even stronger conclusions regarding the nature of the emission regions.
In particular, it will be fascinating to have an improved understanding of the dominant factors responsible for the origin of the emission as well as possible reasons for the changeable location within the context of the multi-zone emission model.
Furthermore, the lack of correlations between the VHE photon emission and the GeV flares seen in some sources (for example PKS 0454-234) underlines the importance of survey, as opposed to targeted, observations of FSRQs with IACTs. 

\section*{Data availability}
This research has made use of public data and analysis tools provided by the NASA \textit{Fermi} collaboration.
In addition, this work has also made use of the NASA/IPAC Extragalactic Database (NED), which is operated by the Jet Propulsion Laboratory, Caltech, under contact with the National Aeronautics and Space Administration.

\section*{Acknowledgements}
We thank M. Meyer for kindly sharing the data for the lower limits on the distances of the emission regions from the central black hole in the study \cite{RN8} and presented in Figure \ref{fig: Fig 3.} of this work. 
We also thank the referee for their constructive feedback and suggestions that improved the quality and clarity of this manuscript.
AMB and PMC acknowledge the financial support of the UK Science and Technology Facilities Council consolidated grant ST/P000541/1.




\bibliographystyle{mnras}
\bibliography{references}

\bsp	



\newpage


\newpage

\appendix
\onecolumn
\begin{appendices}
\section{Lightcurves during flares}
\label{sec:A1}

\begin{figure*}
    \centering
    
    \resizebox{1 \textwidth}{!}{
    \includegraphics{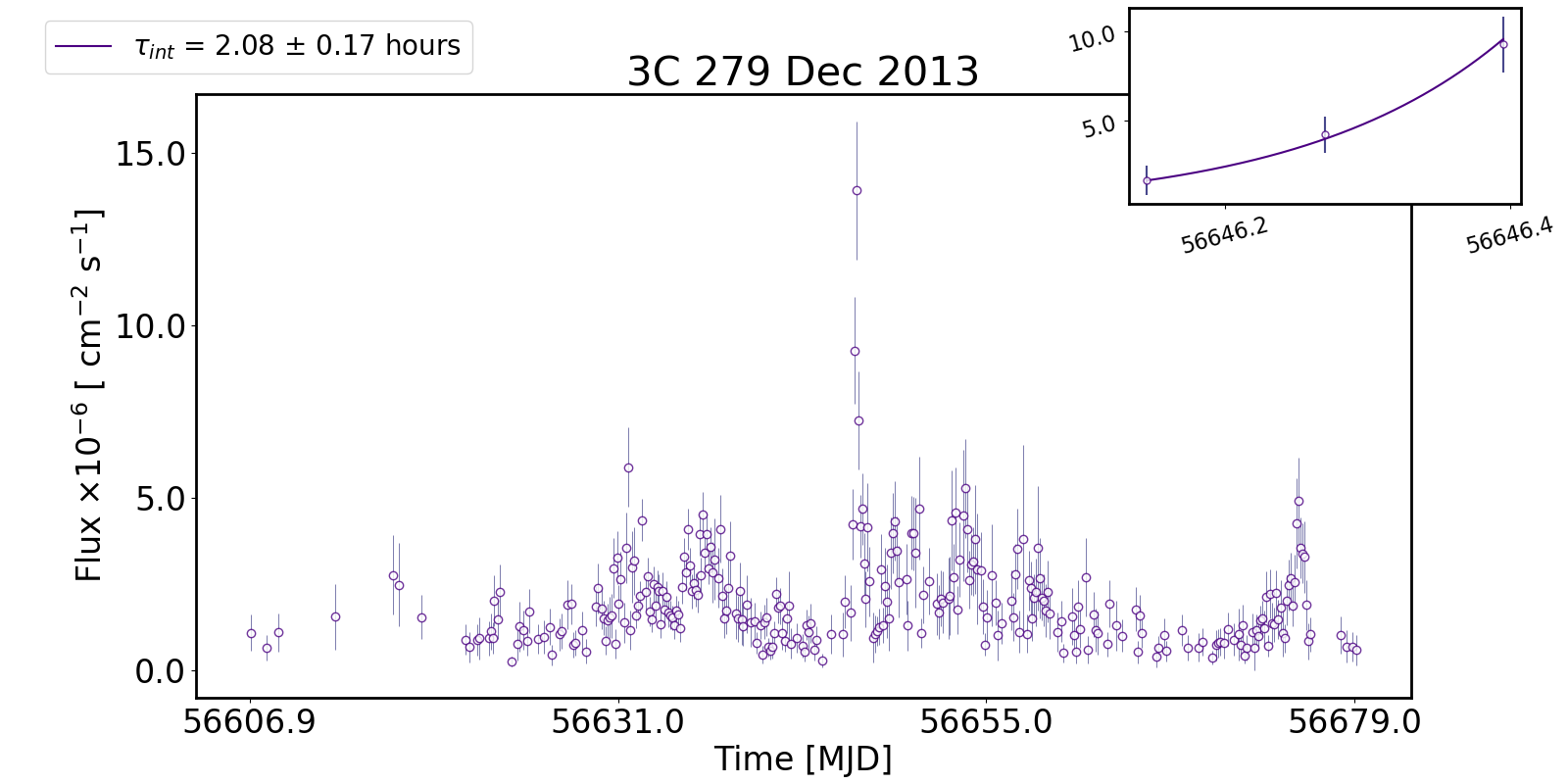}
    \includegraphics{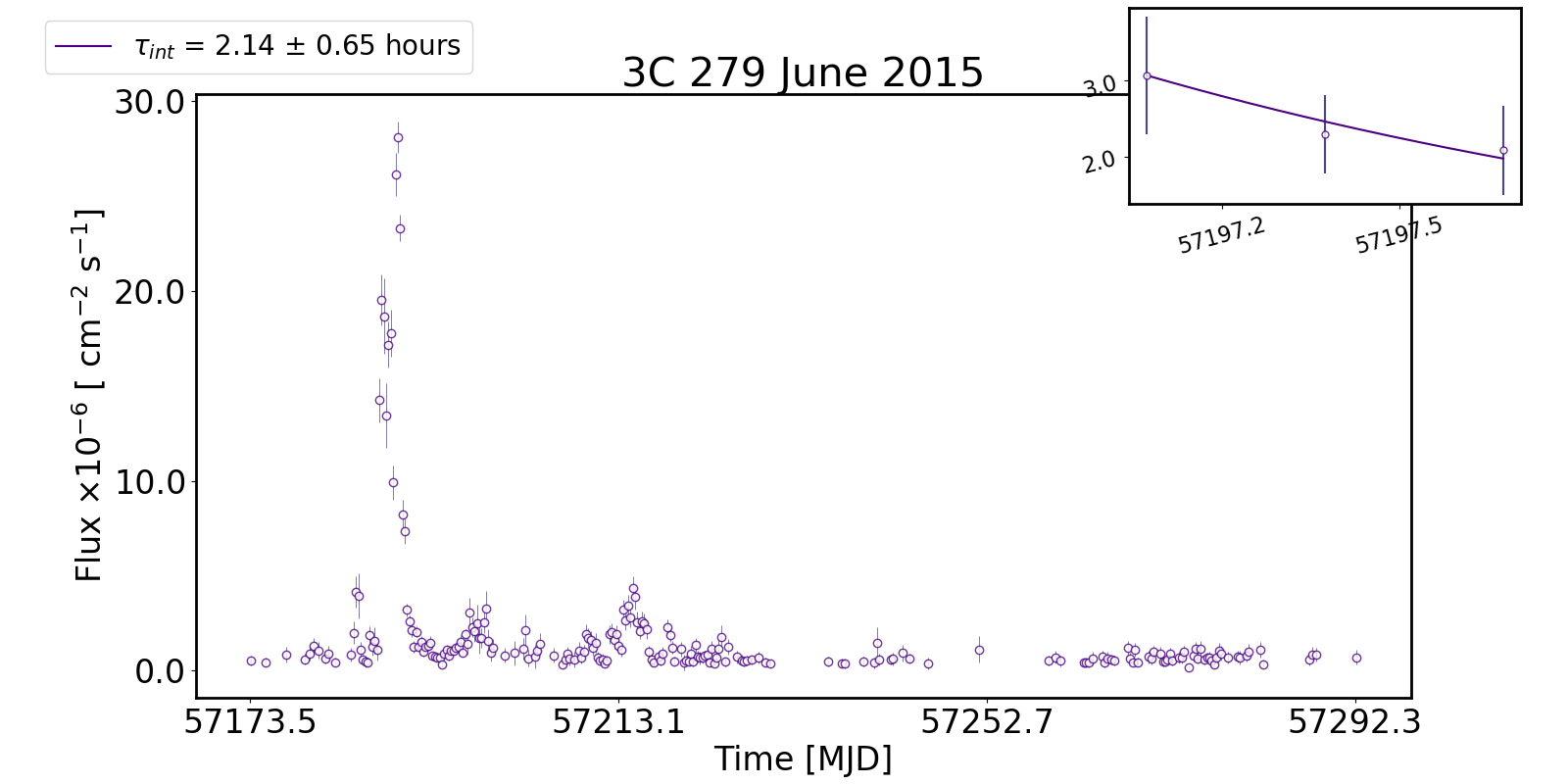}}
    
    \resizebox{1\textwidth}{!}{
    \includegraphics{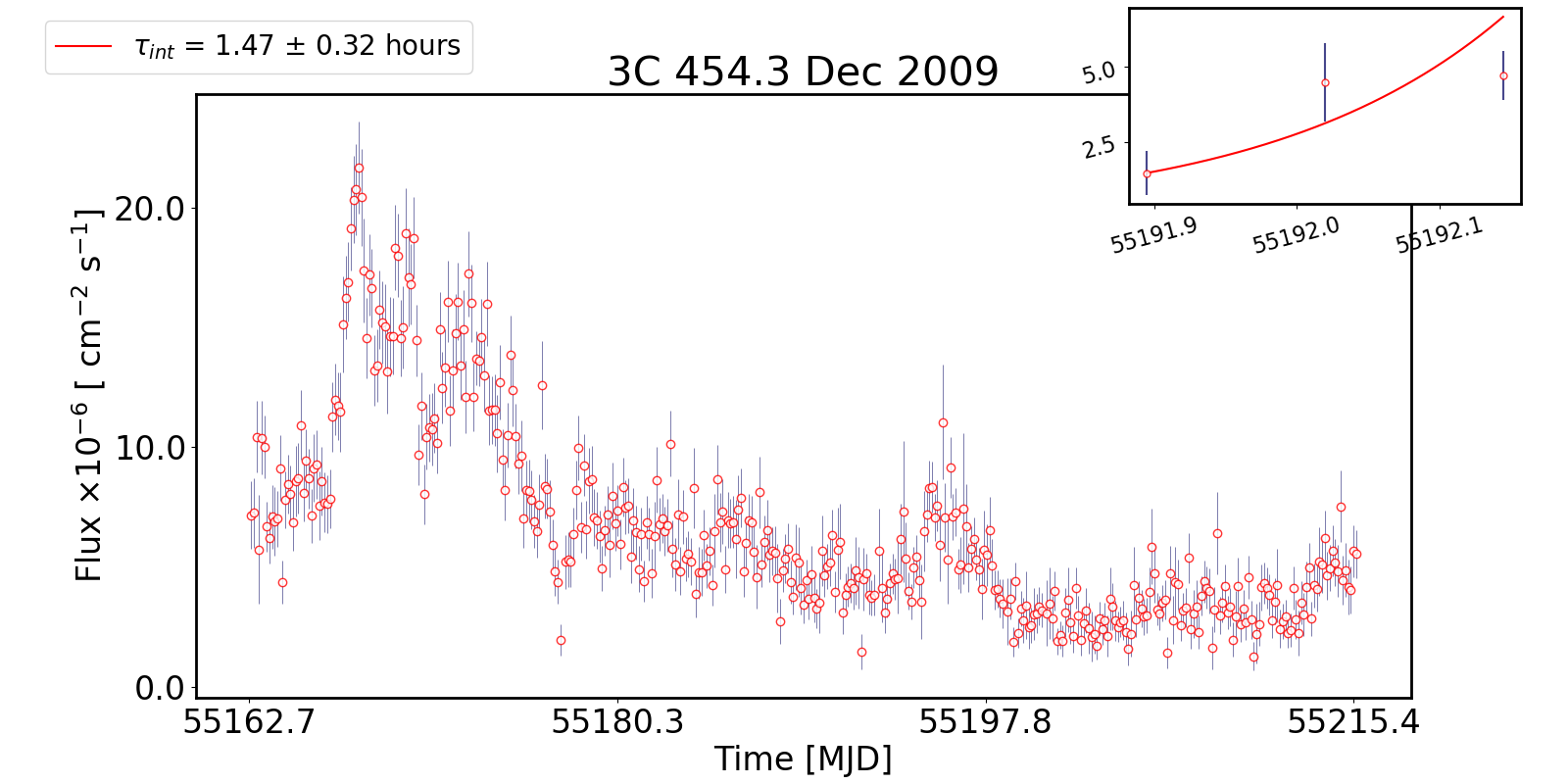}
    \includegraphics{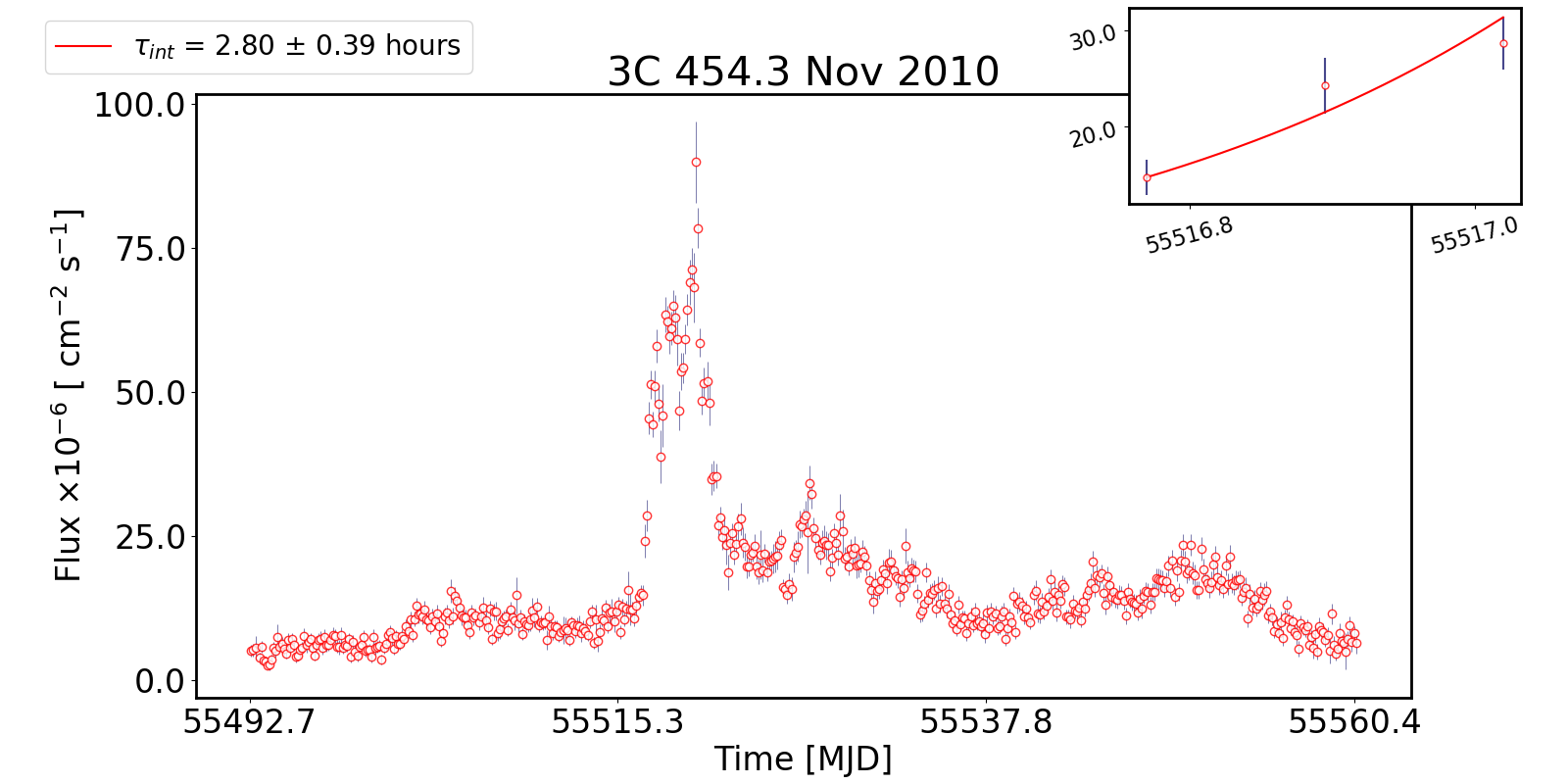}}
    
    \resizebox{1\textwidth}{!}{
    \includegraphics{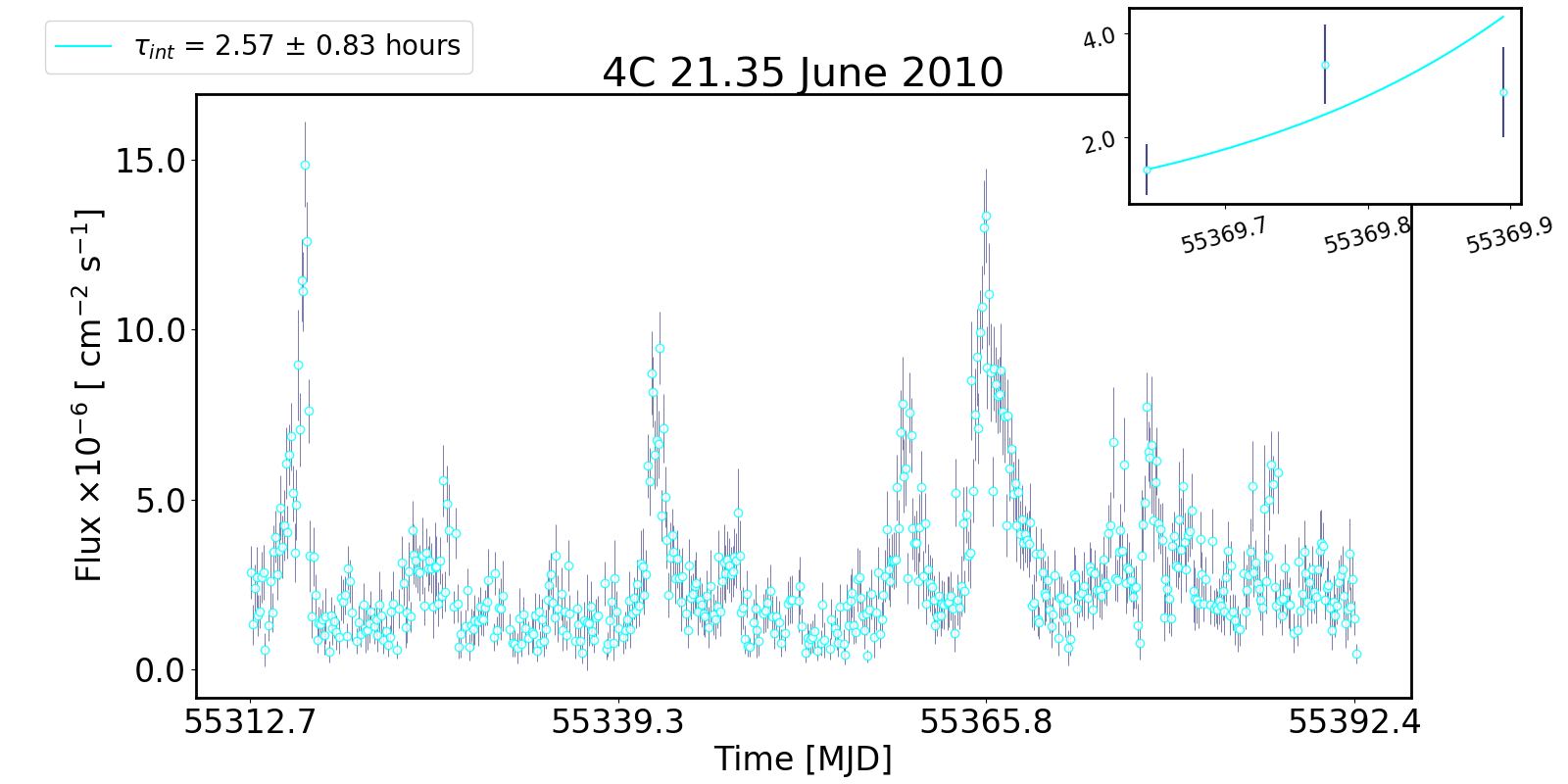}
    \includegraphics{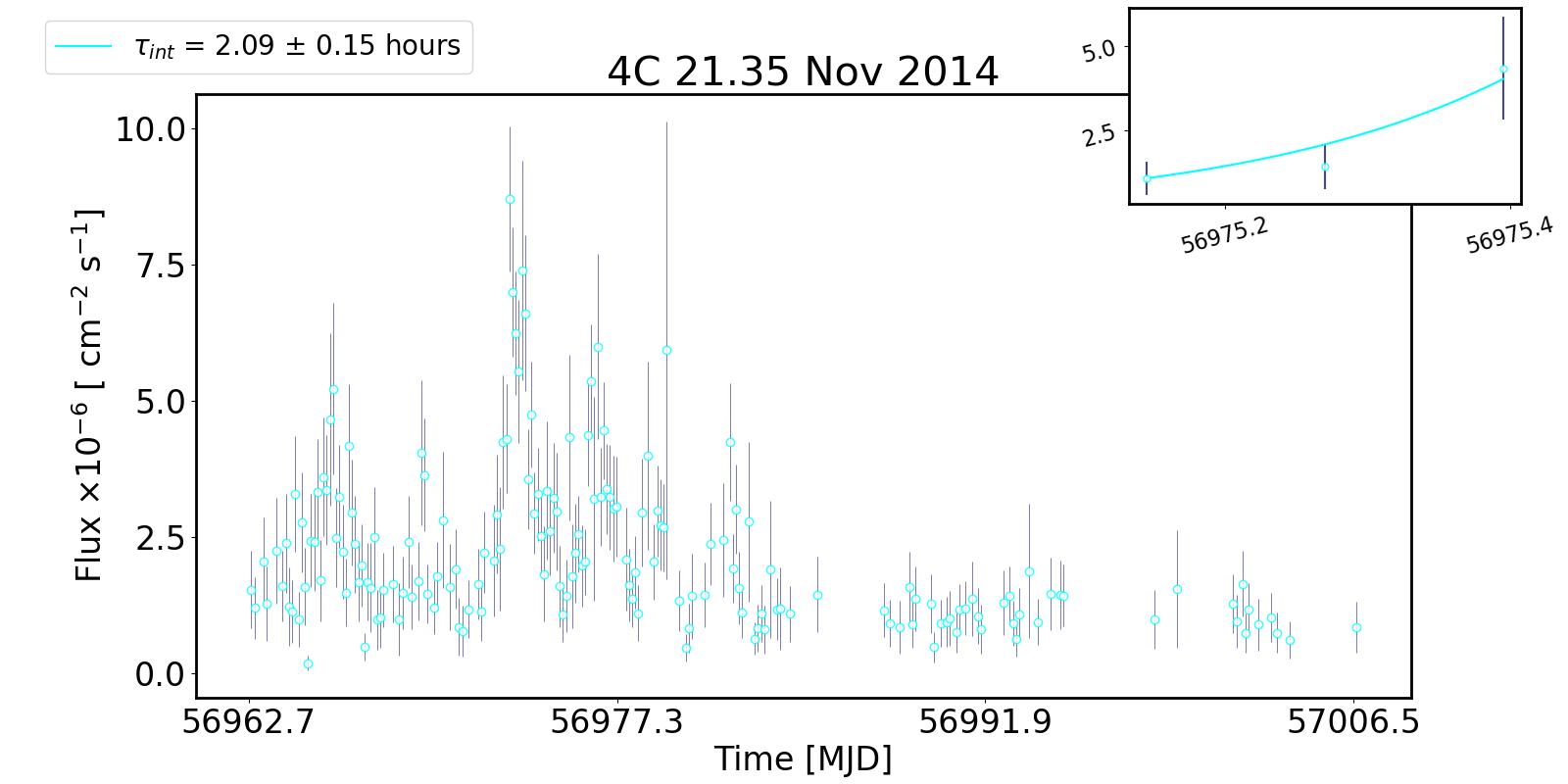}}

    \resizebox{1\textwidth}{!}{
    \includegraphics{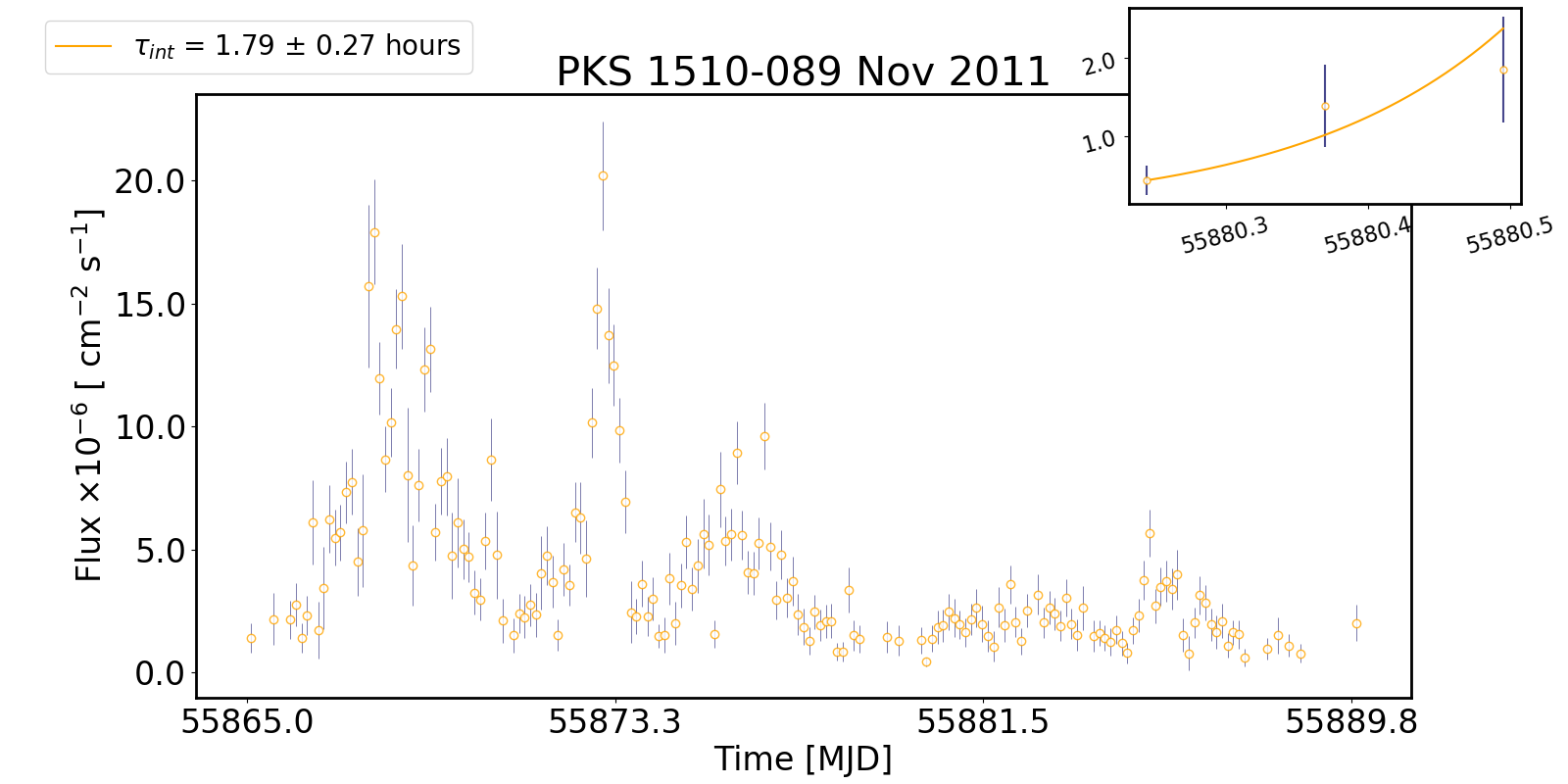}
    \includegraphics{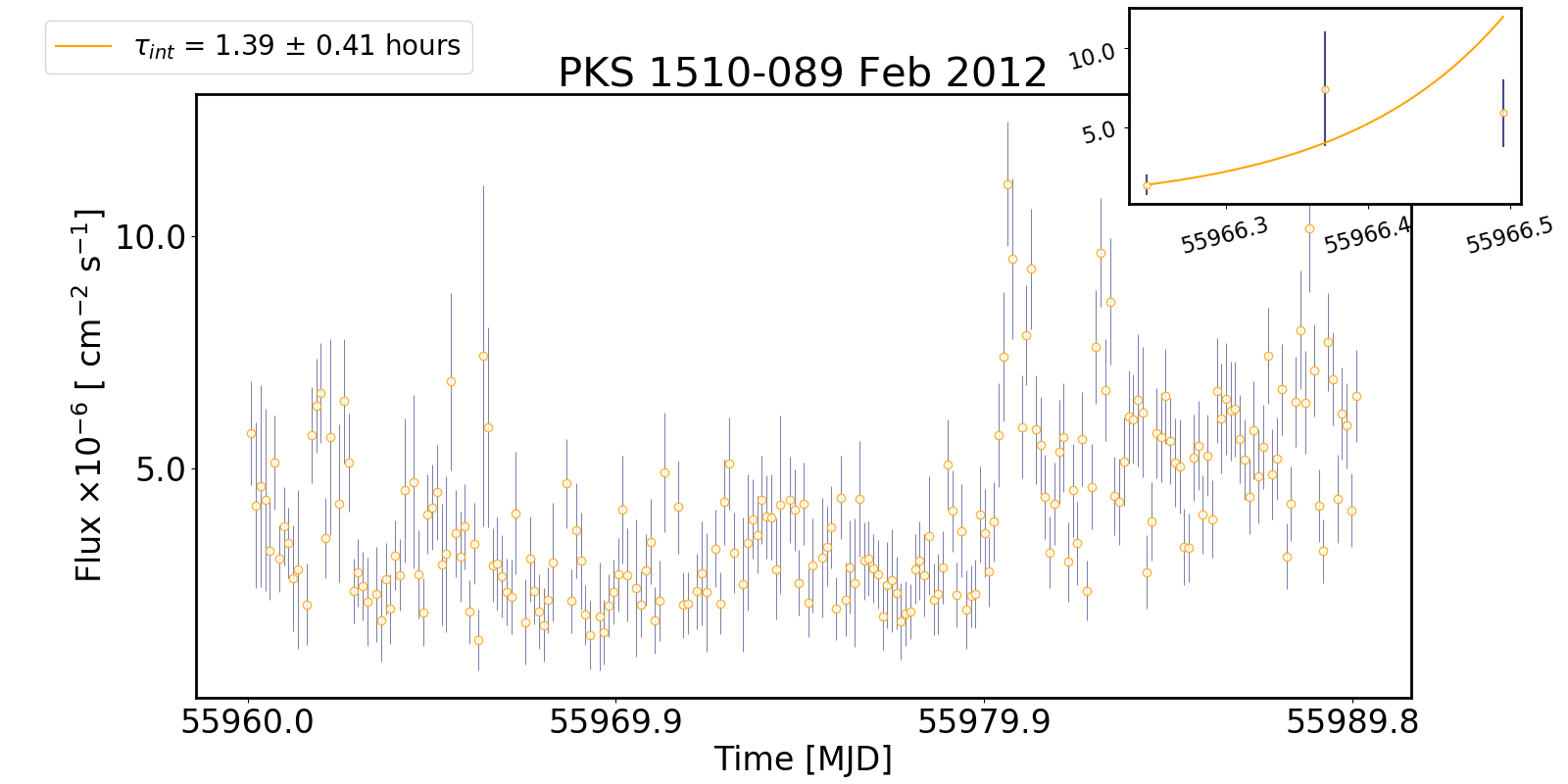}}

    \caption{Evolution of flux in 3 hr bins during each flare period considered.}
    \label{fig: A1.}
\end{figure*}

\begin{figure*}
    \centering
    
    \resizebox{\textwidth}{!}{
    \includegraphics{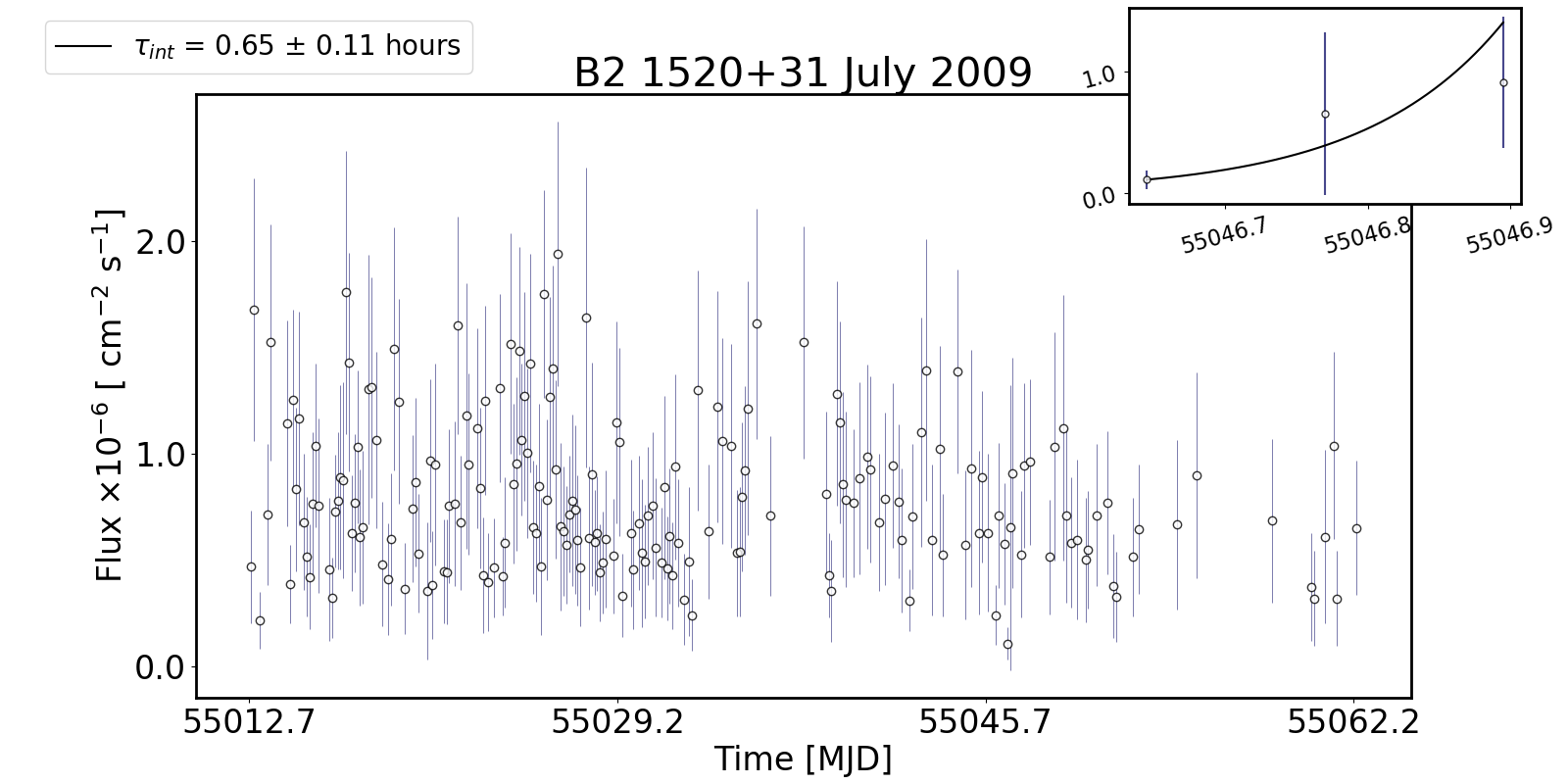}
    \includegraphics{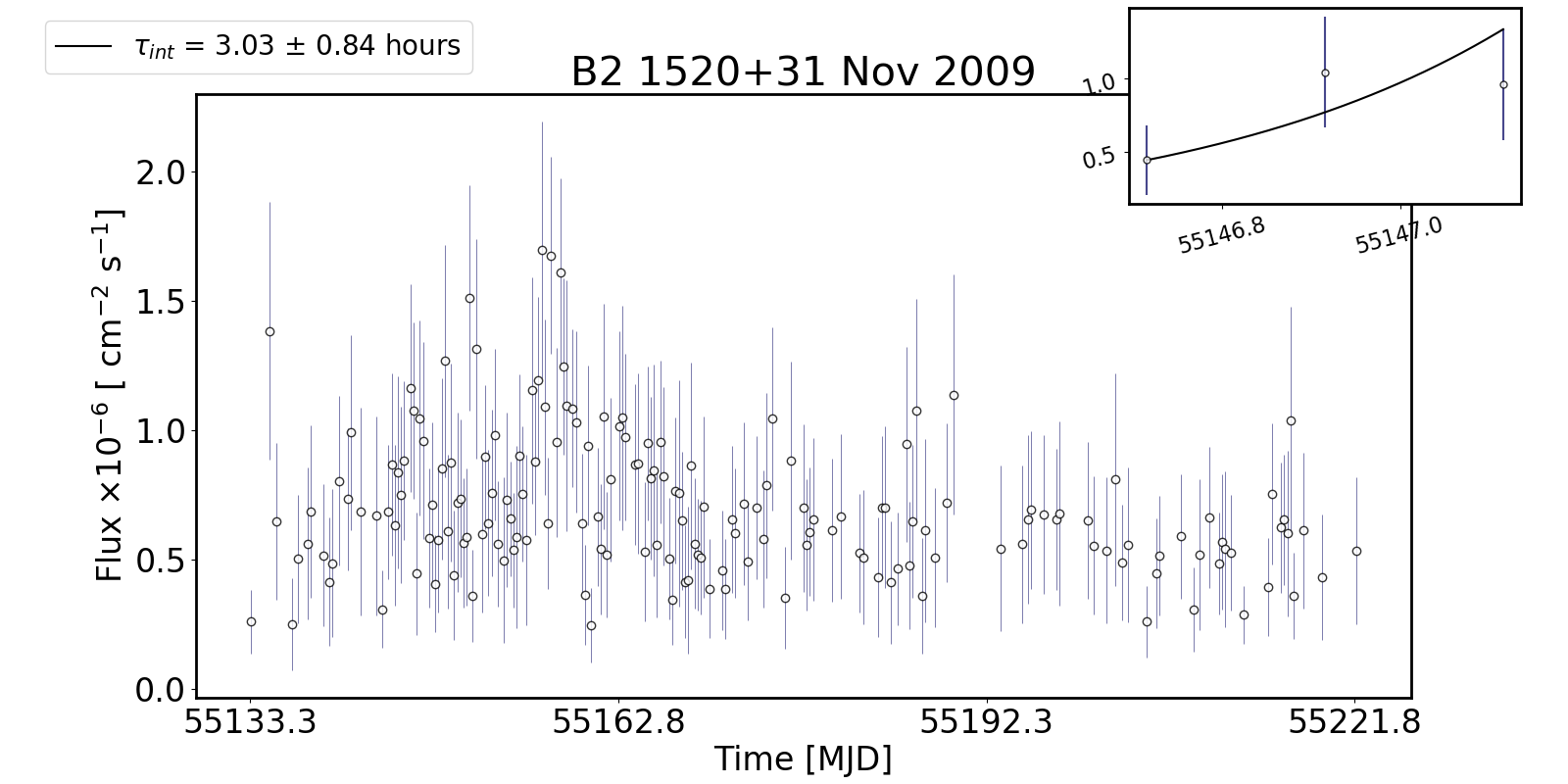}}
    
    \resizebox{\textwidth}{!}{
    \includegraphics{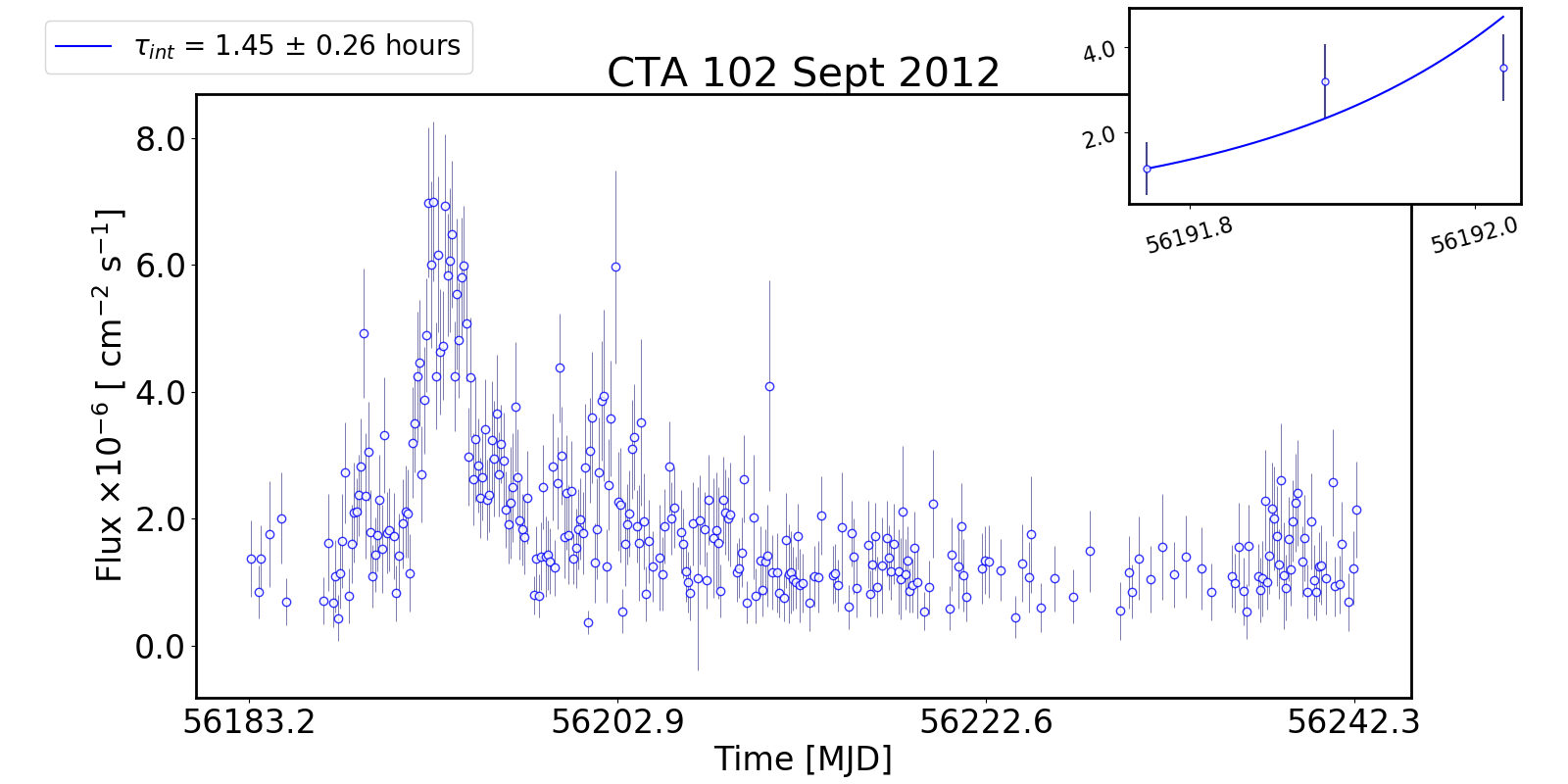}
    \includegraphics{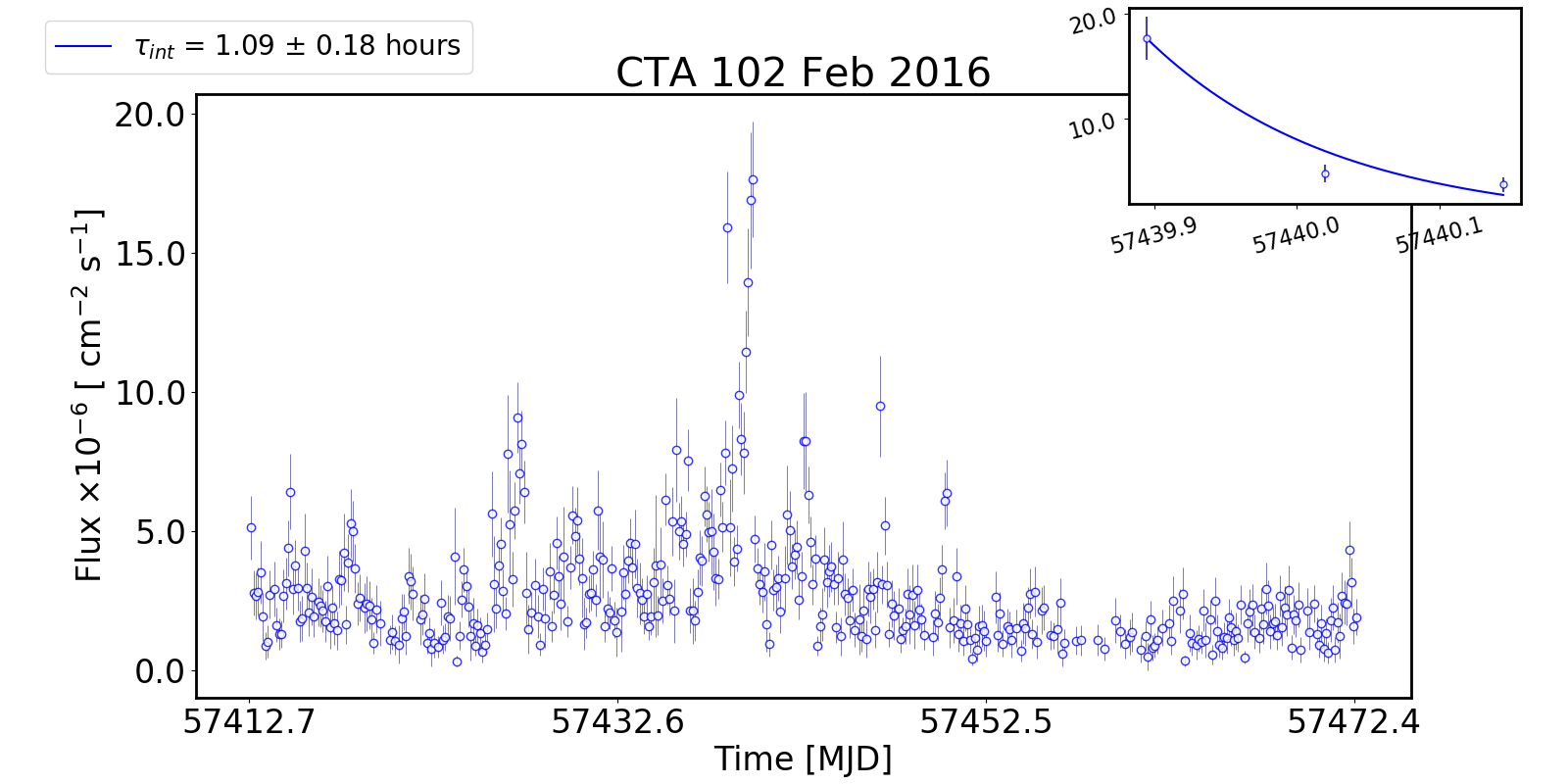}}

    \resizebox{\textwidth}{!}{
    \includegraphics{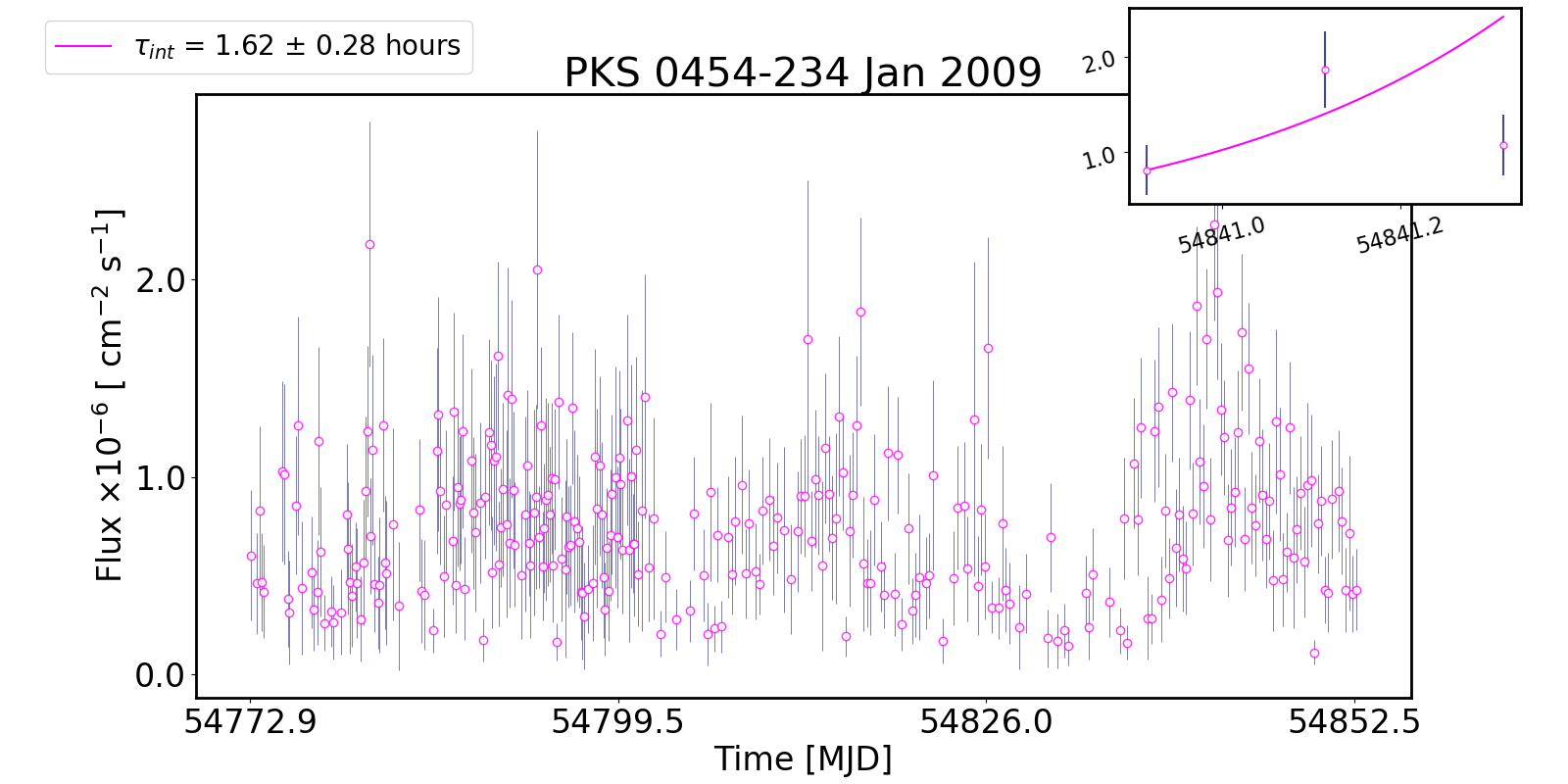}
    \includegraphics{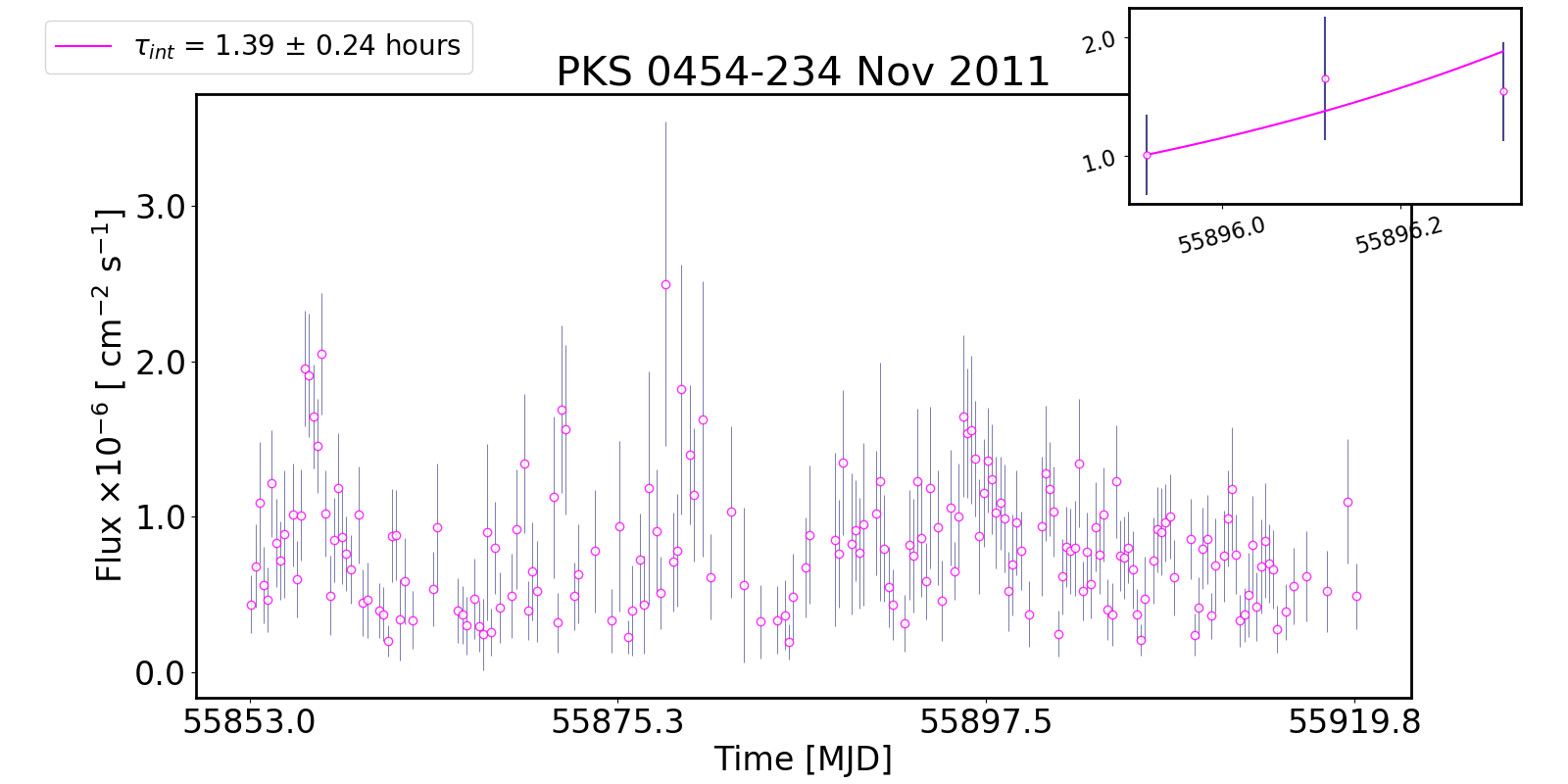}}
    
    \resizebox{\textwidth}{!}{
    \includegraphics{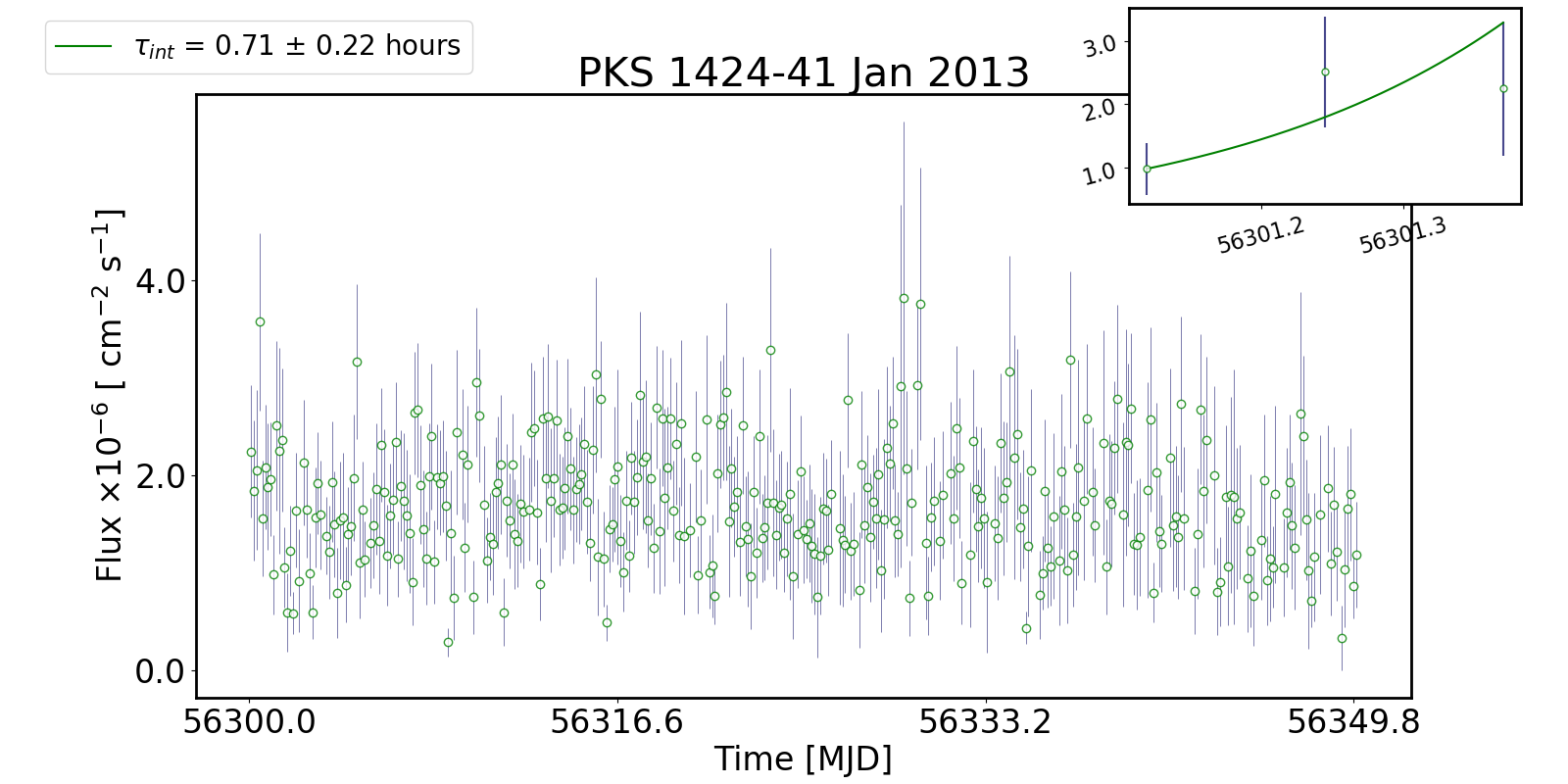}
    \includegraphics{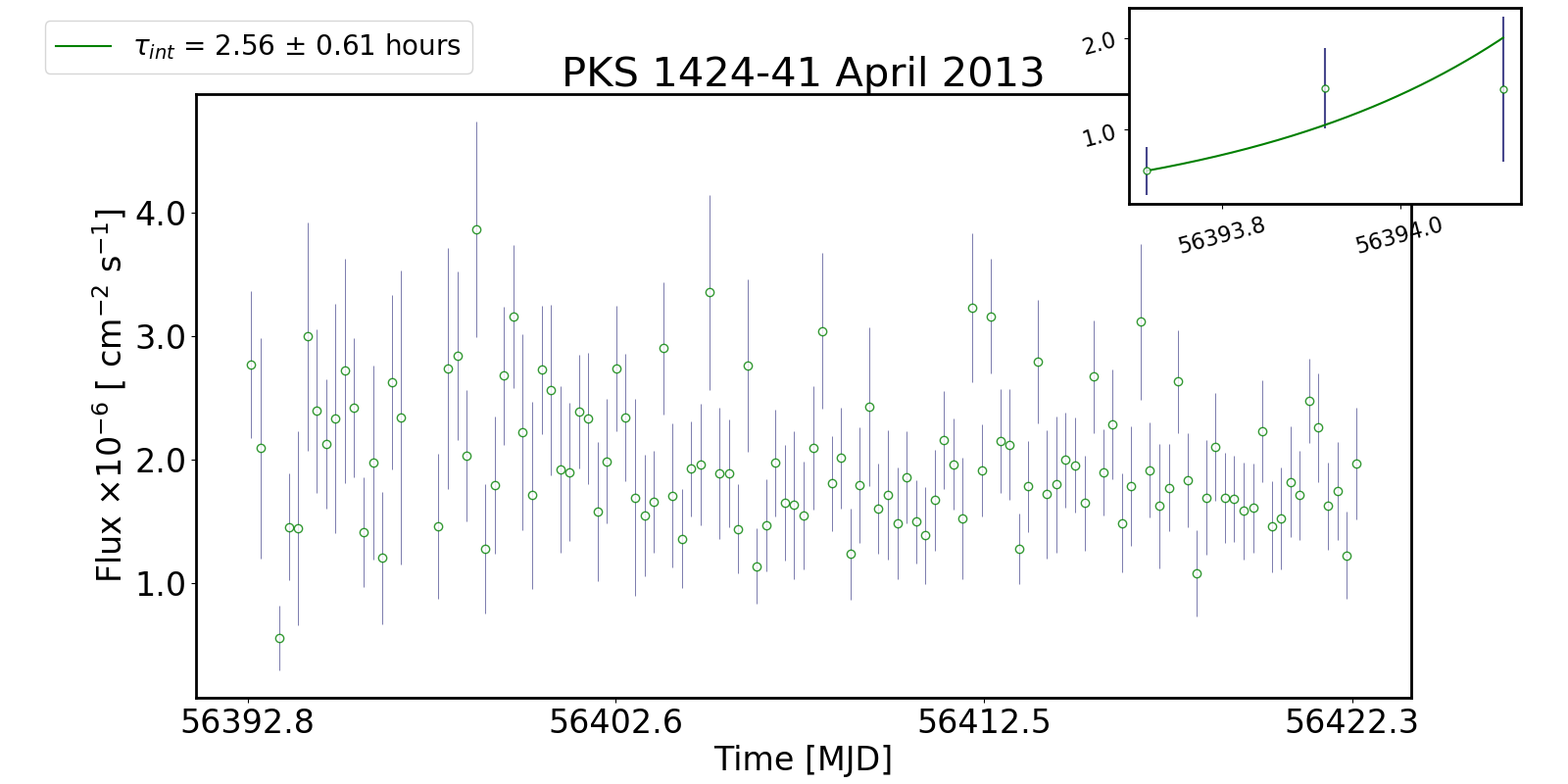}}
    
    \resizebox{\textwidth}{!}{
    \includegraphics{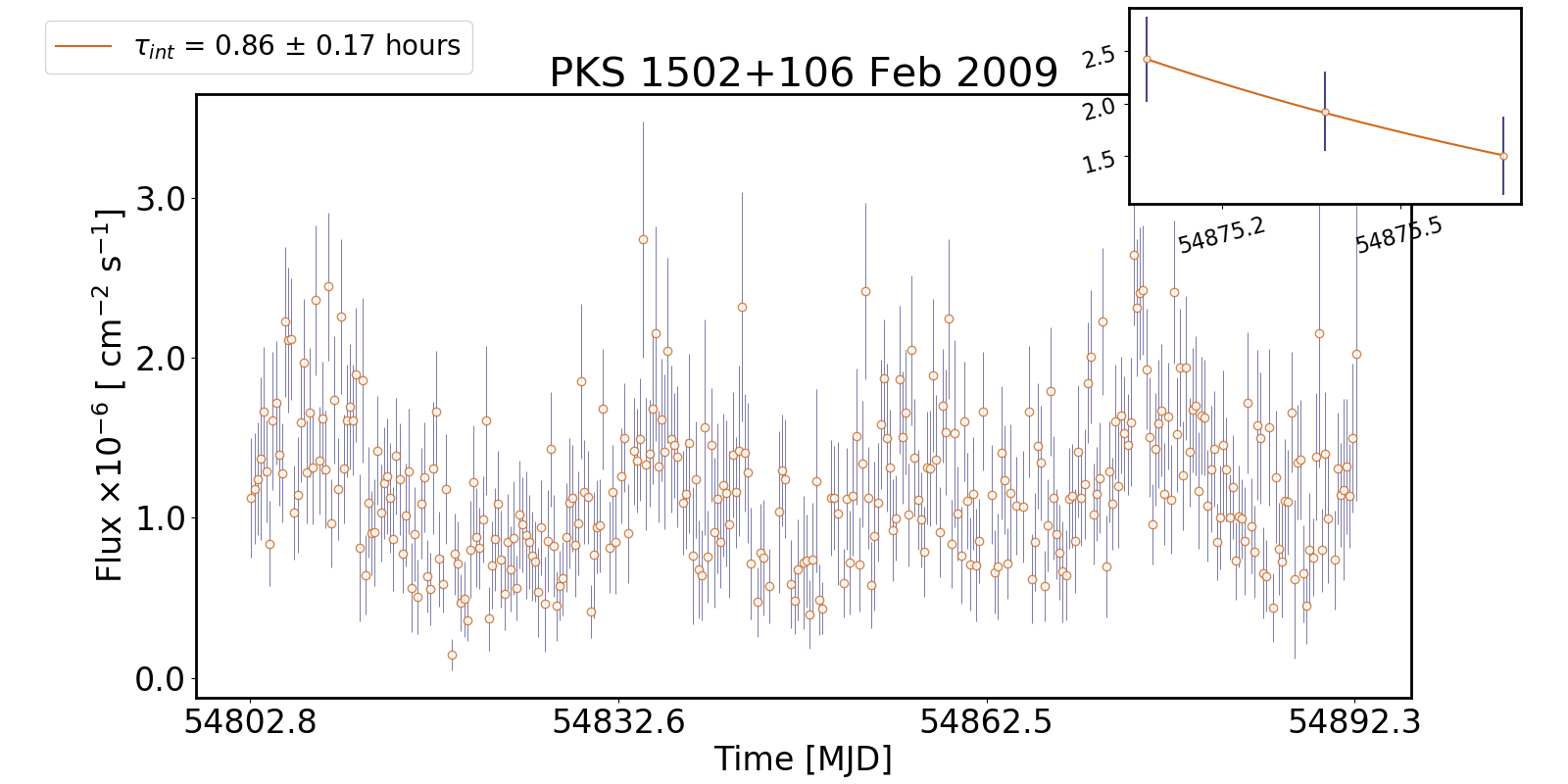}
    \includegraphics{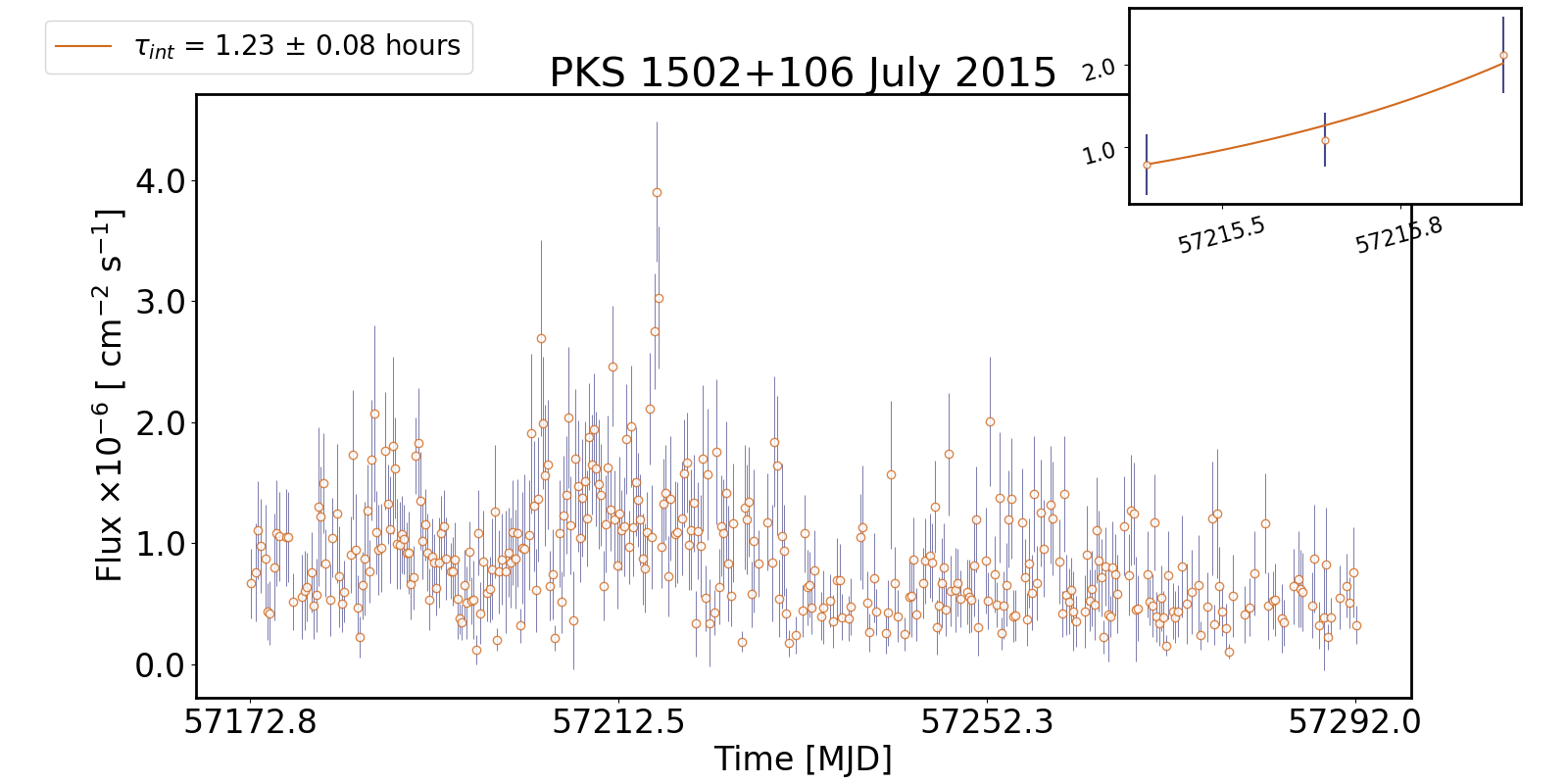}}

    \caption{Evolution of flux in 3 hr bins during each flare period considered.}
    \label{fig: A2.}
\end{figure*}

Figures \ref{fig: A1.} and \ref{fig: A2.} plots the  0.1 $\leq$ $E_{\gamma}$ $\leq$ 300 GeV lightcurves of each flare from all sources in 3 hour time bins. The error bars are purely statistical. Only data points with TS $\geq$ 10 are shown. The insets show zoomed in sections of the lightcurves containing the data points used to calculate the intrinsic timescales (shown in legend).

\section{Energy dependent Lightcurves during flares}
\label{sec:A2}

\begin{figure*}
    \centering
    \resizebox{\textwidth}{!}{
    \includegraphics{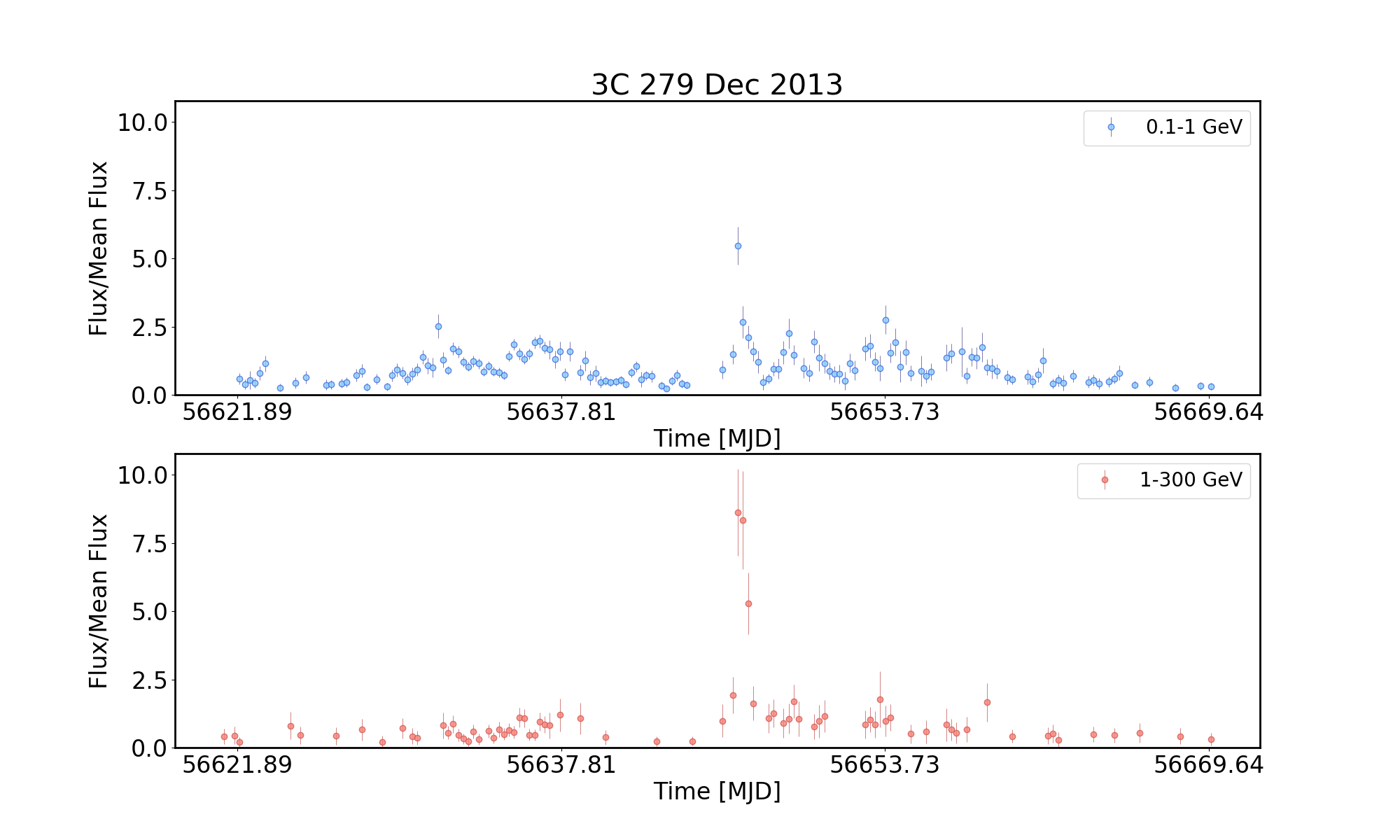}
    \includegraphics{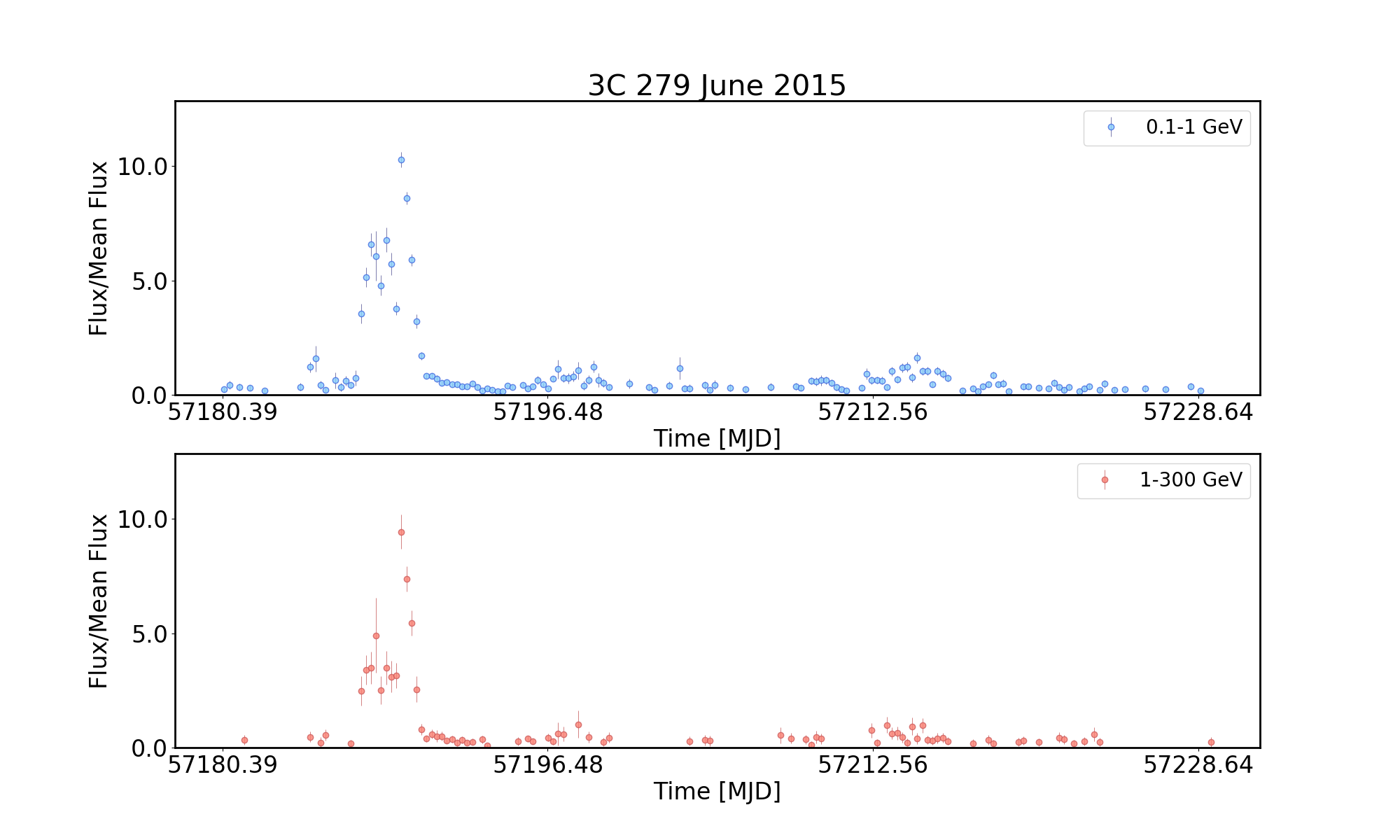}}
    
    \resizebox{\textwidth}{!}{
    \includegraphics{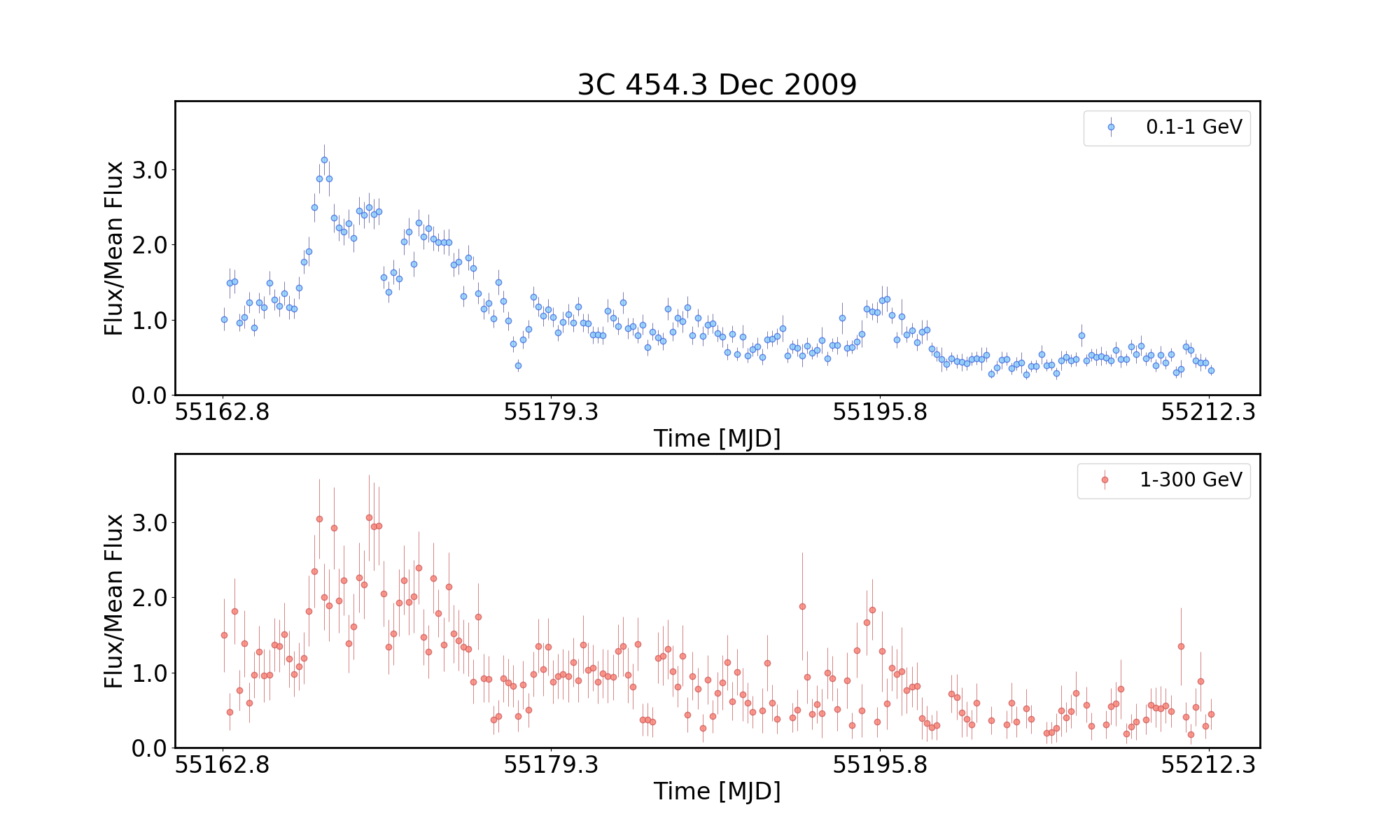}
    \includegraphics{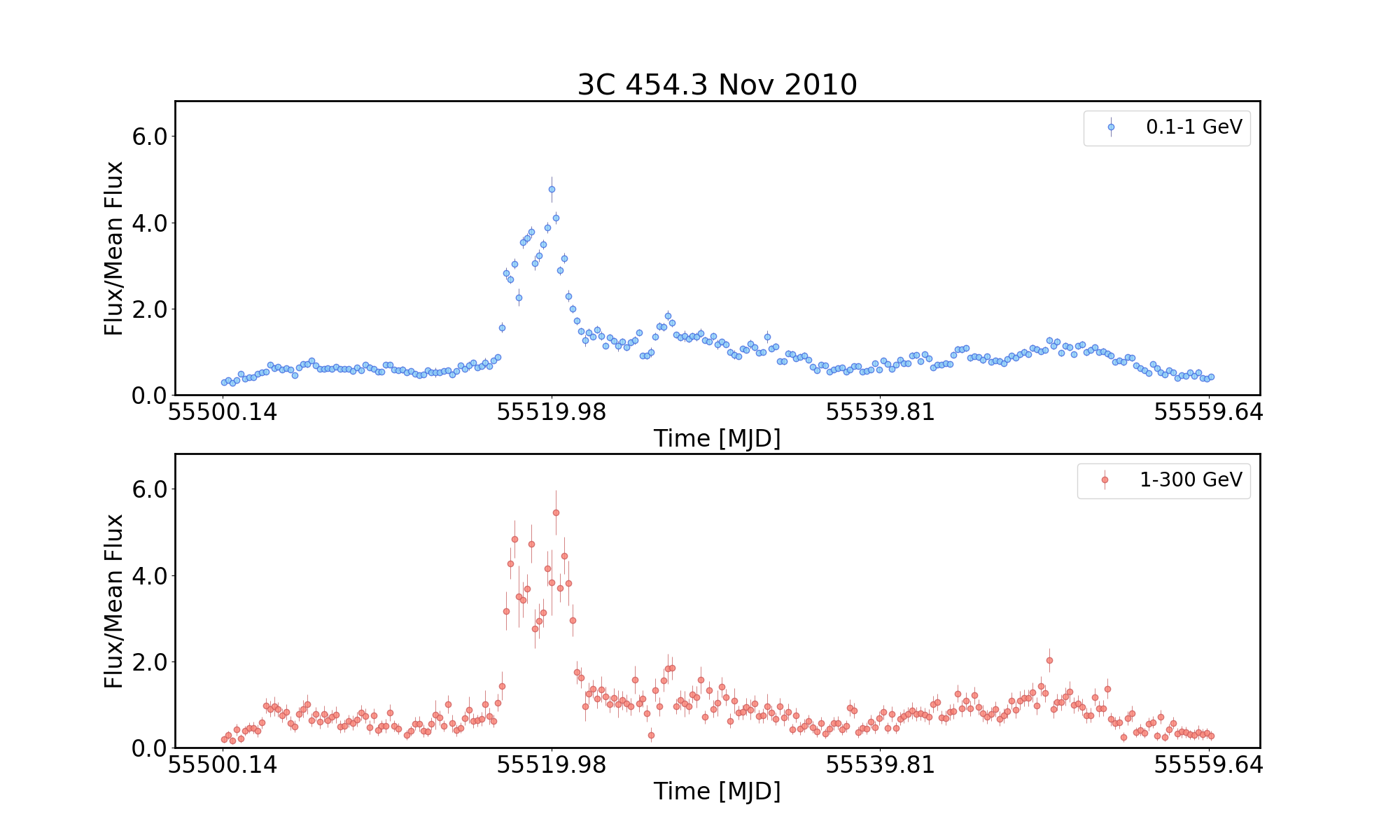}}

    \resizebox{\textwidth}{!}{
    \includegraphics{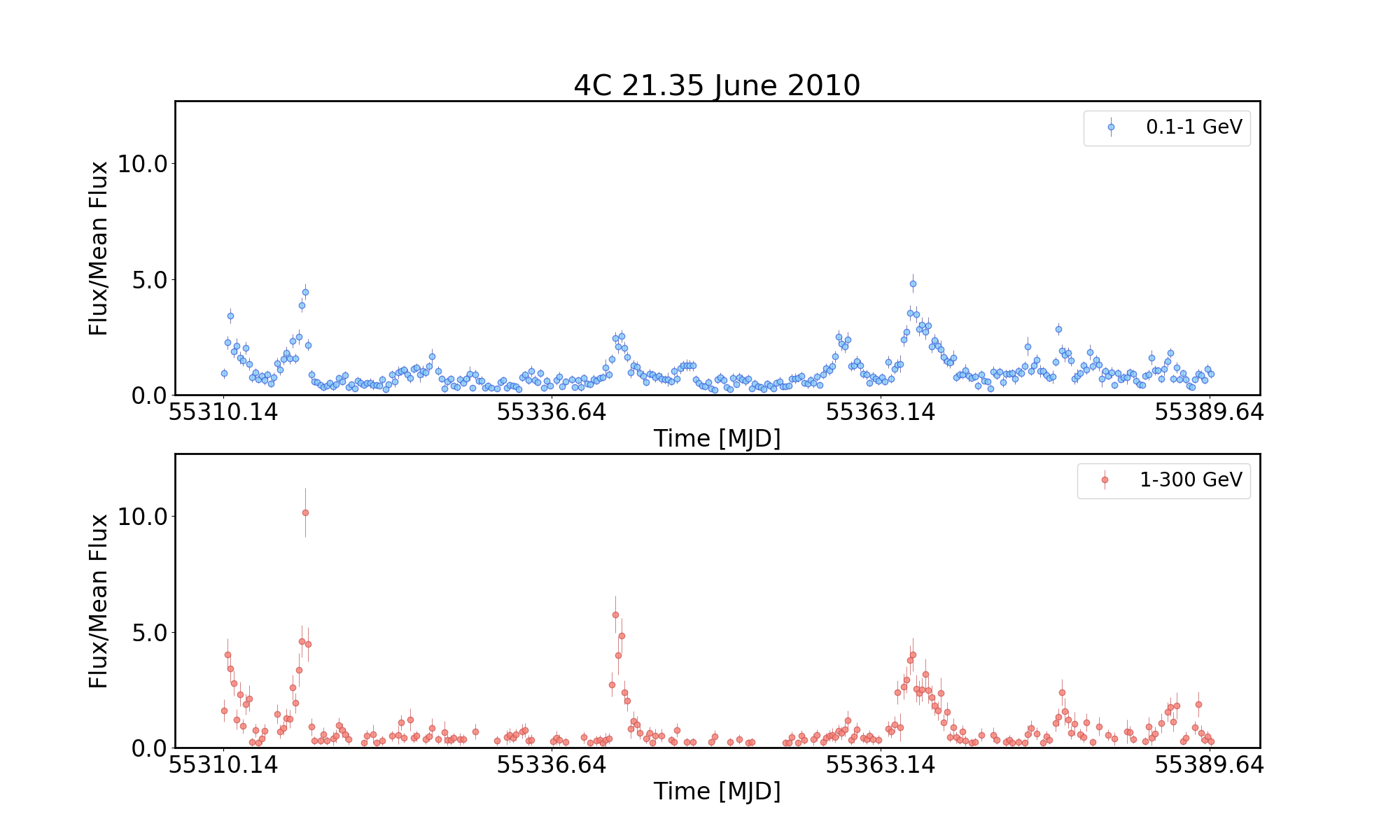}
    \includegraphics{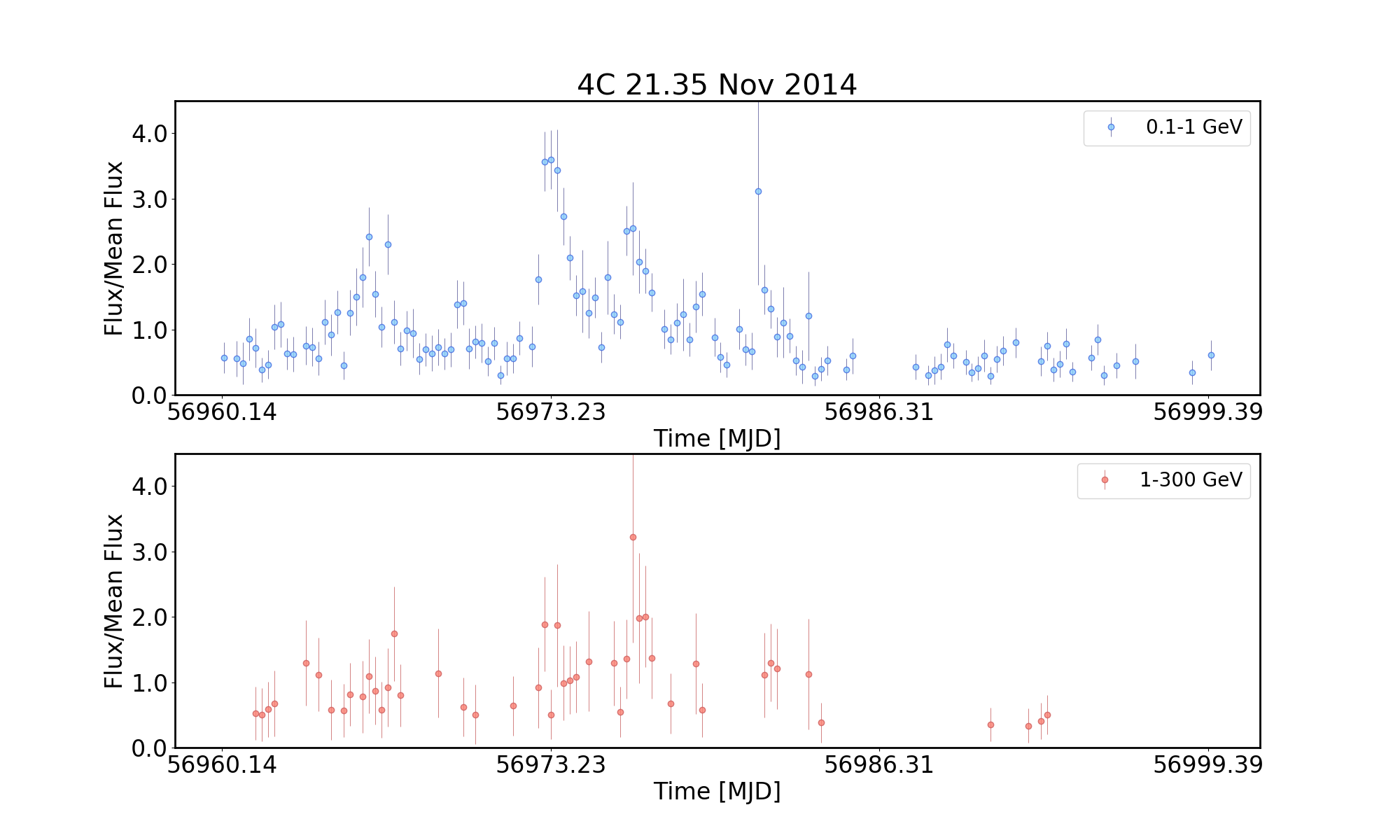}}
    
    \resizebox{\textwidth}{!}{
    \includegraphics{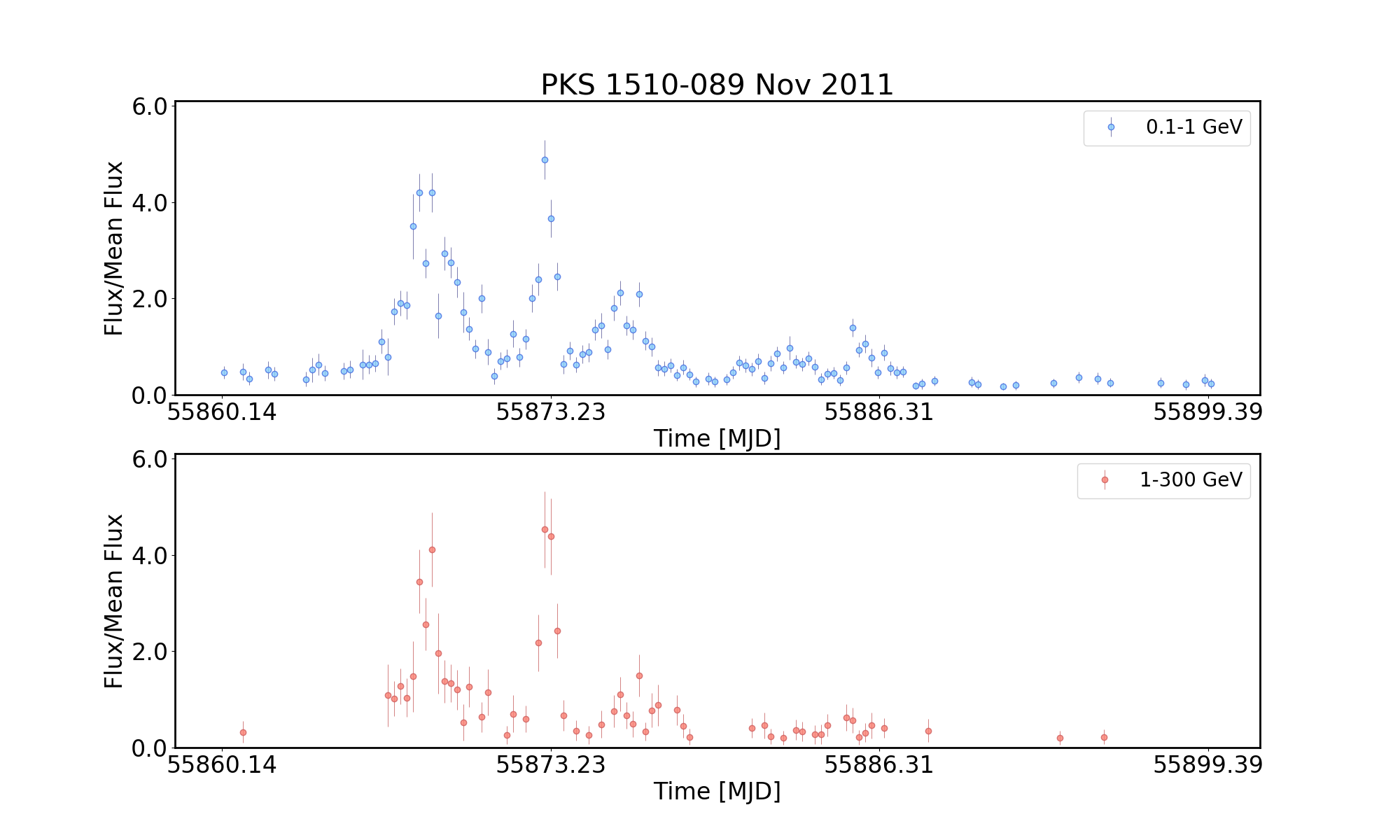}
    \includegraphics{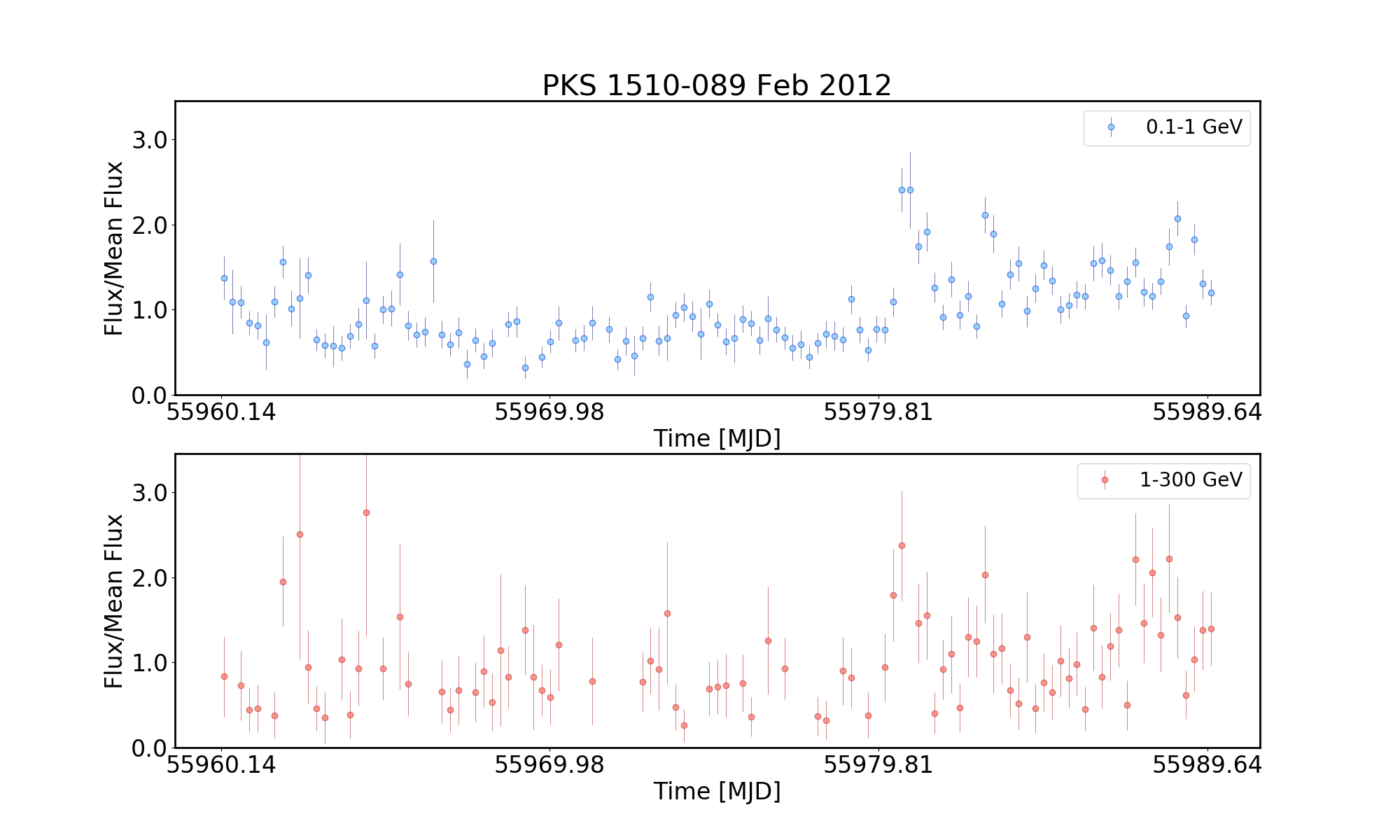}}
    
    \caption{Evolution of high- and low-energy flux in 6 hr bins during each flare period considered for 3C 279, 3C 454.3, 4C 21.35 and PKS 1510-089.}
    \label{fig: B1.}
\end{figure*}

Figure \ref{fig: B1.} and \ref{fig: B2.} plots the energy separated lightcurves of each flare from all sources in 6 hour time bins. The  low energy flux (0.1 $\leq$ $\text{E}_{\gamma}$ $\leq$ 1 GeV) is plotted as blue circles (top panel) and the high energy flux (1 $\leq$ $\text{E}_{\gamma}$ $\leq$ 300 GeV) is plotted using red circles (bottom panel). To aid visual comparison, the individual flux values have been divided by the mean flux in the corresponding energy ranges for each flare. The error bars are purely statistical. Only data points with TS $\geq$ 10 are shown.

\begin{figure*}
    \centering

    \resizebox{\textwidth}{!}{
    \includegraphics{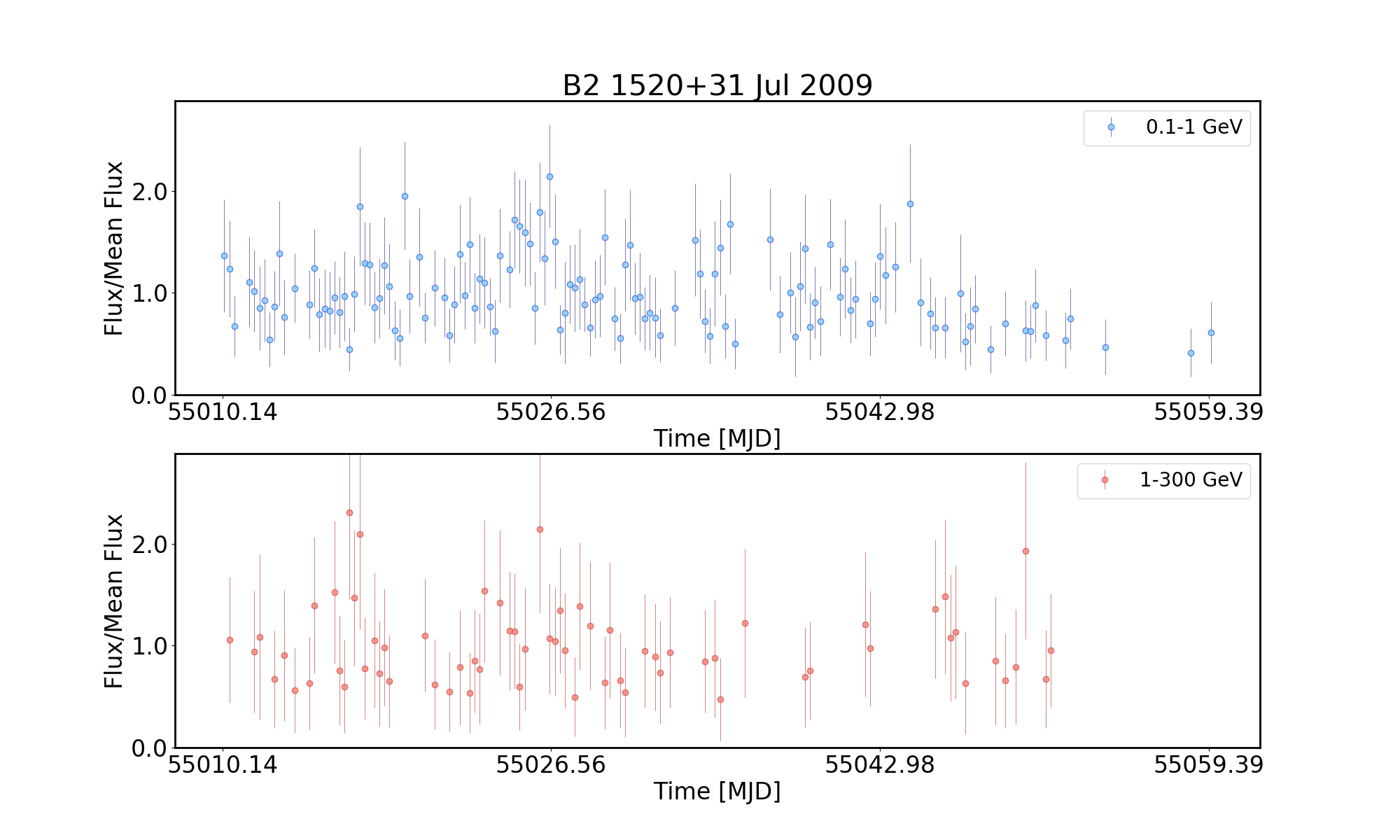}
    \includegraphics{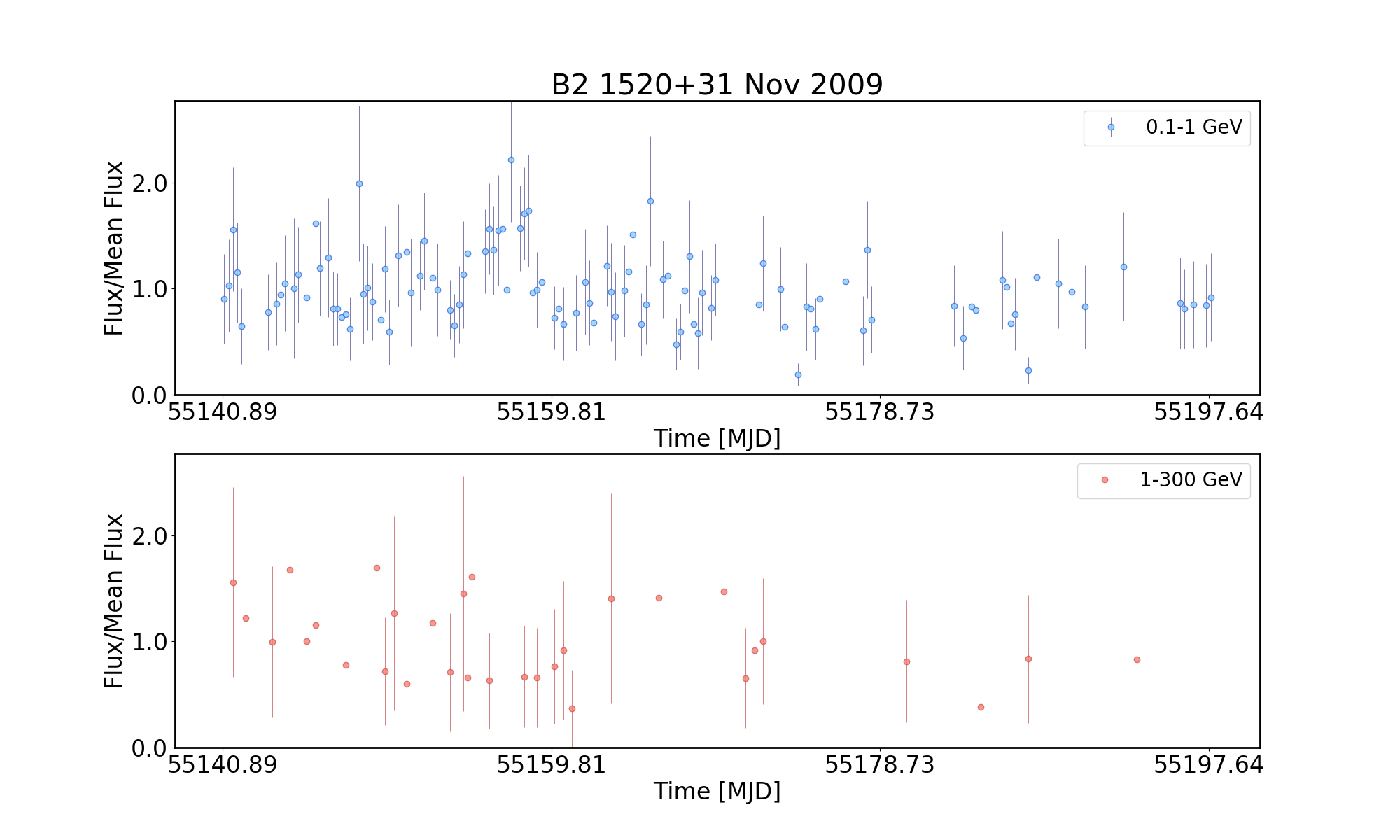}}
    
    \resizebox{\textwidth}{!}{
    \includegraphics{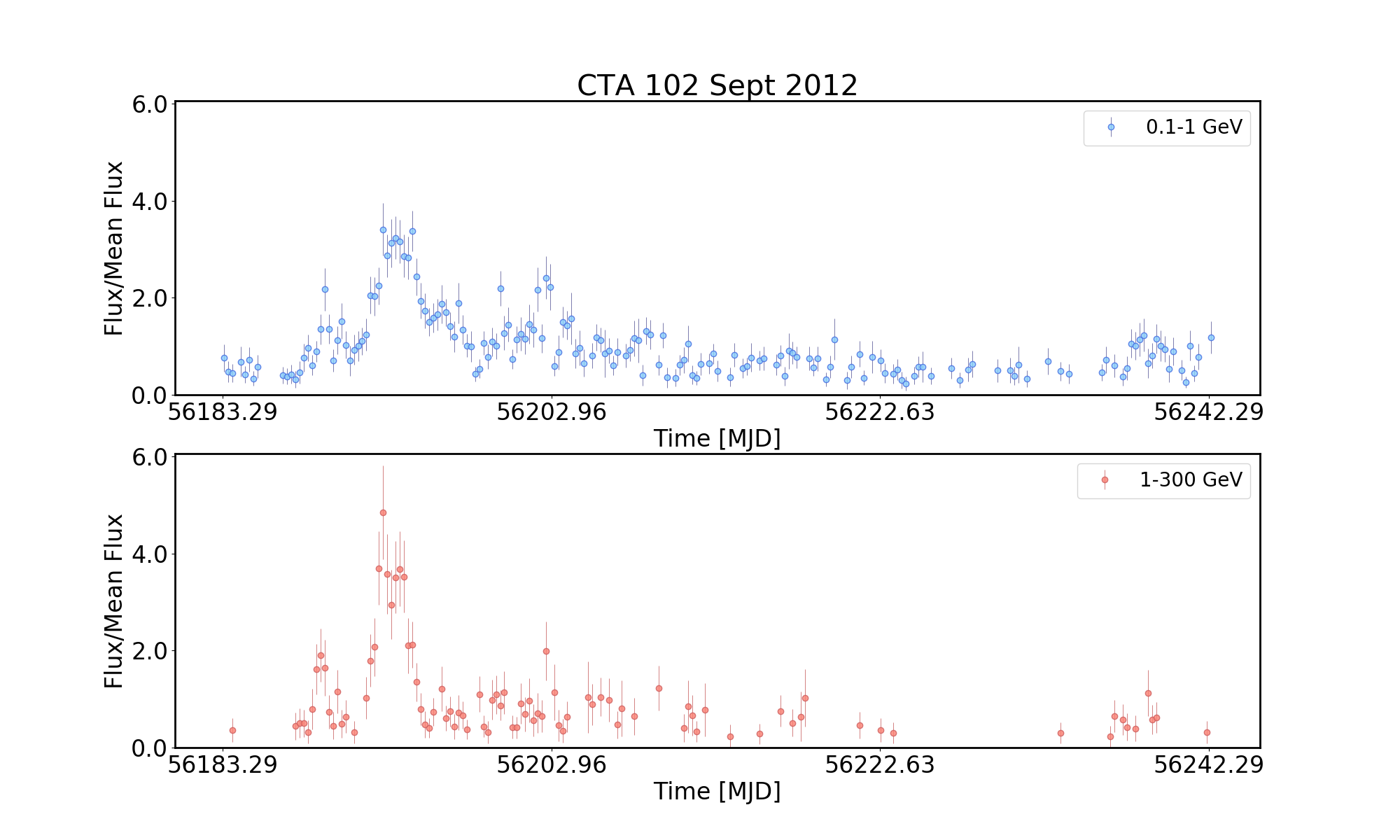}
    \includegraphics{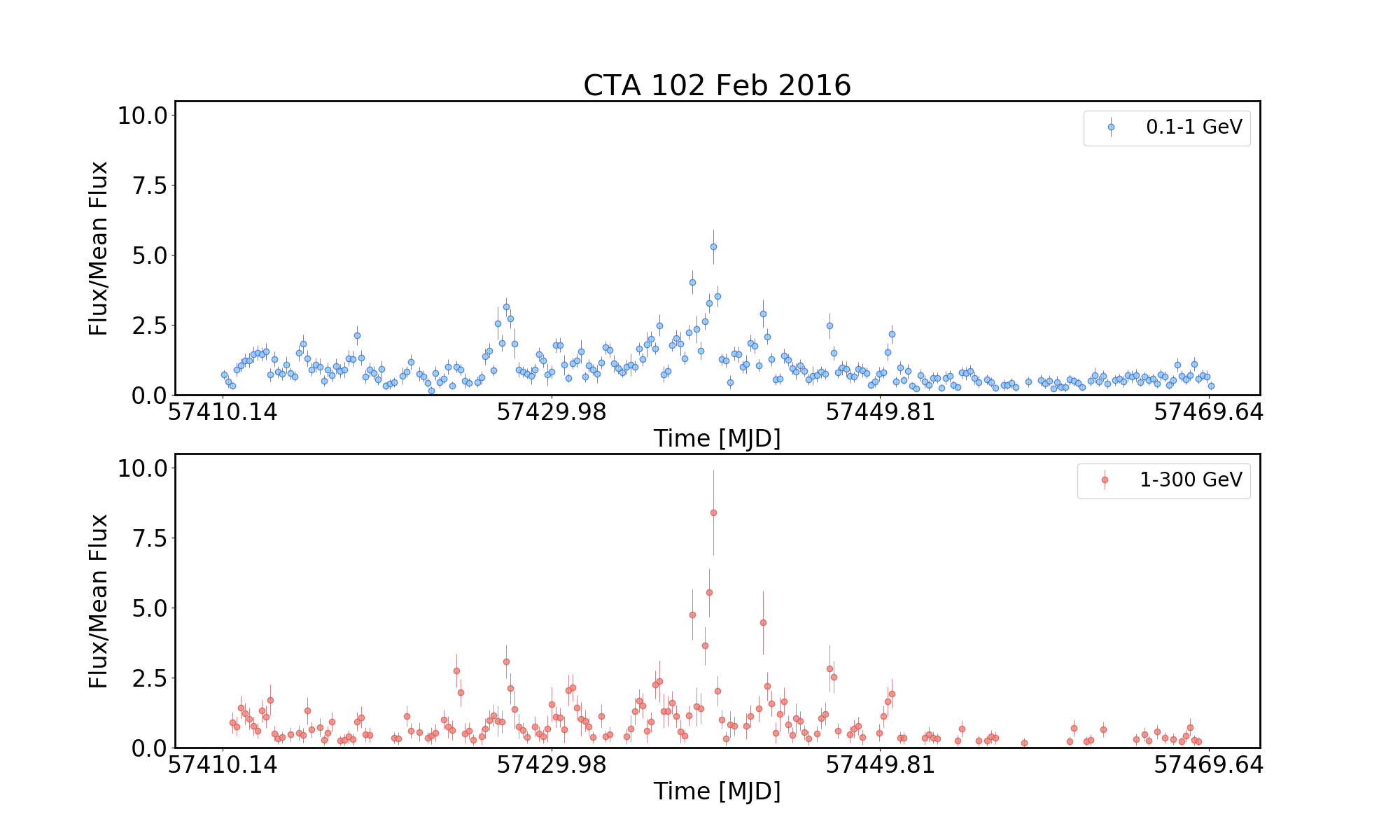}}
    
    \resizebox{\textwidth}{!}{
    \includegraphics{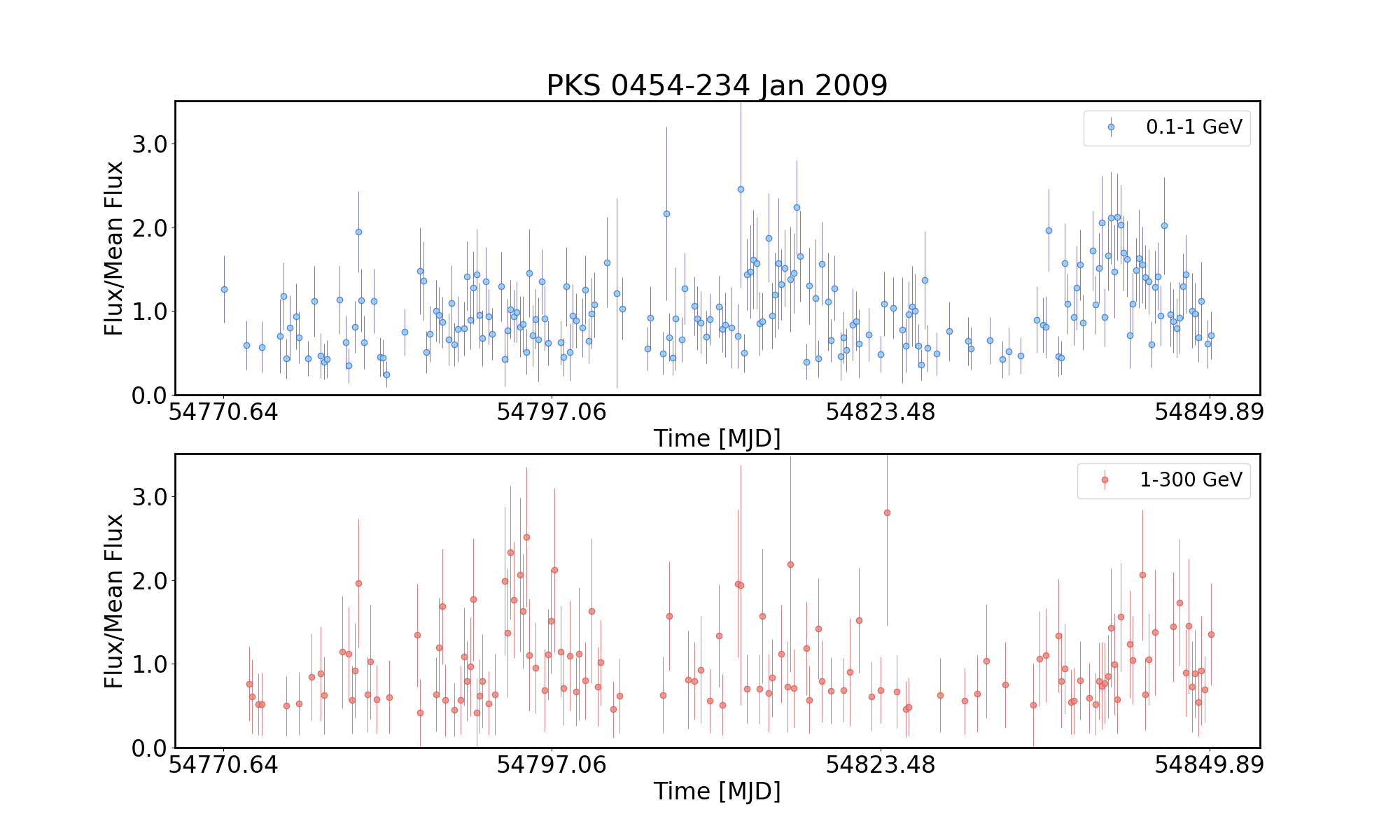}
    \includegraphics{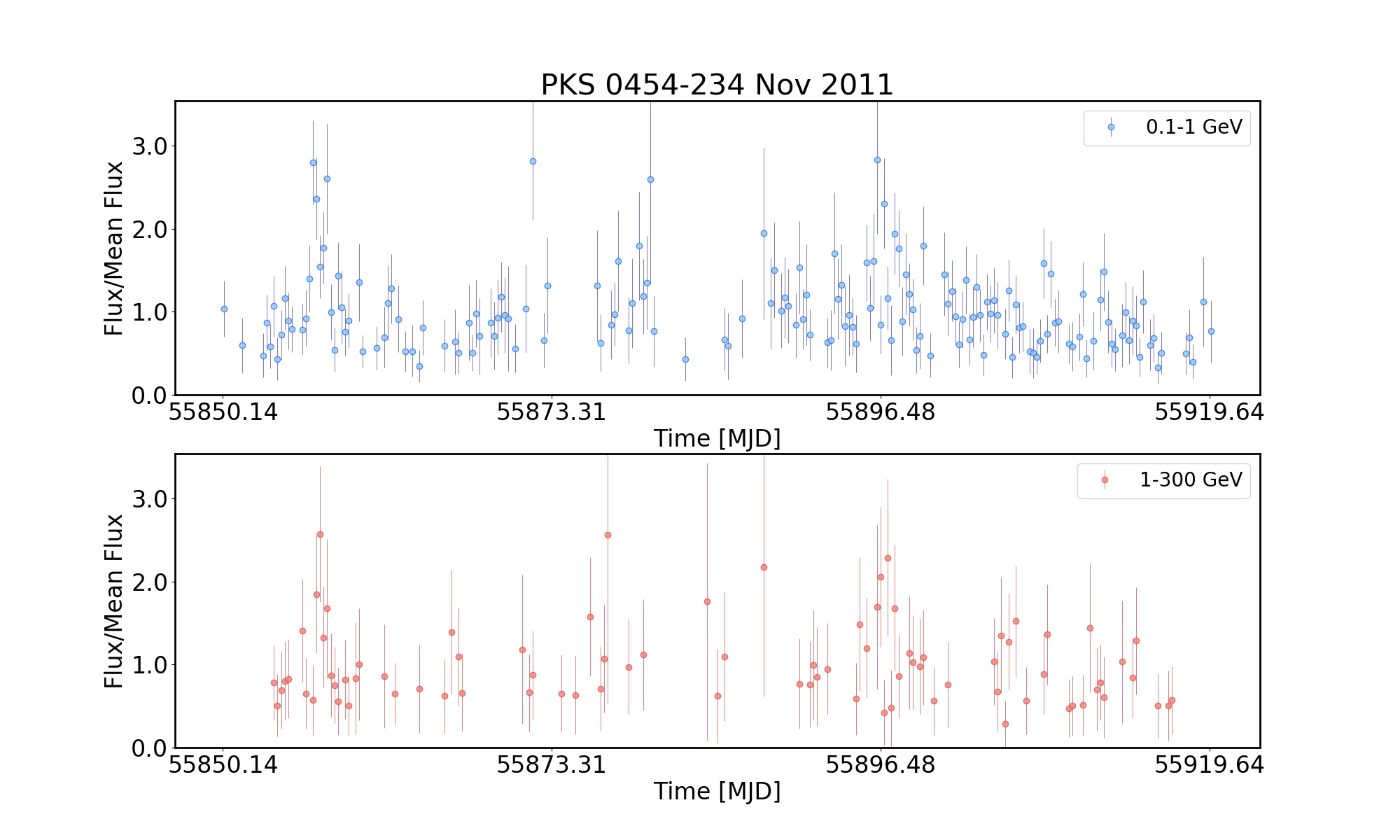}}
    
    \resizebox{\textwidth}{!}{
    \includegraphics{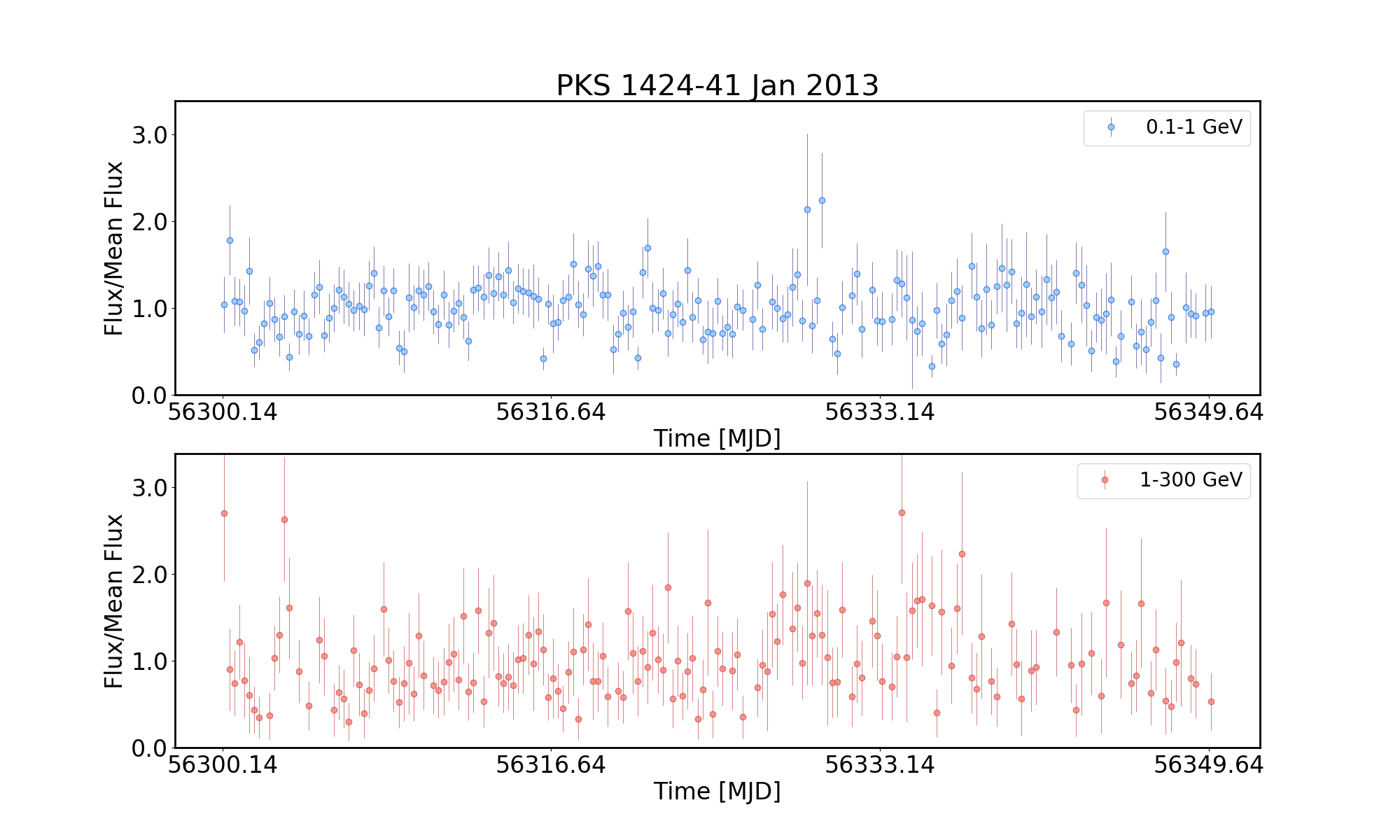}
    \includegraphics{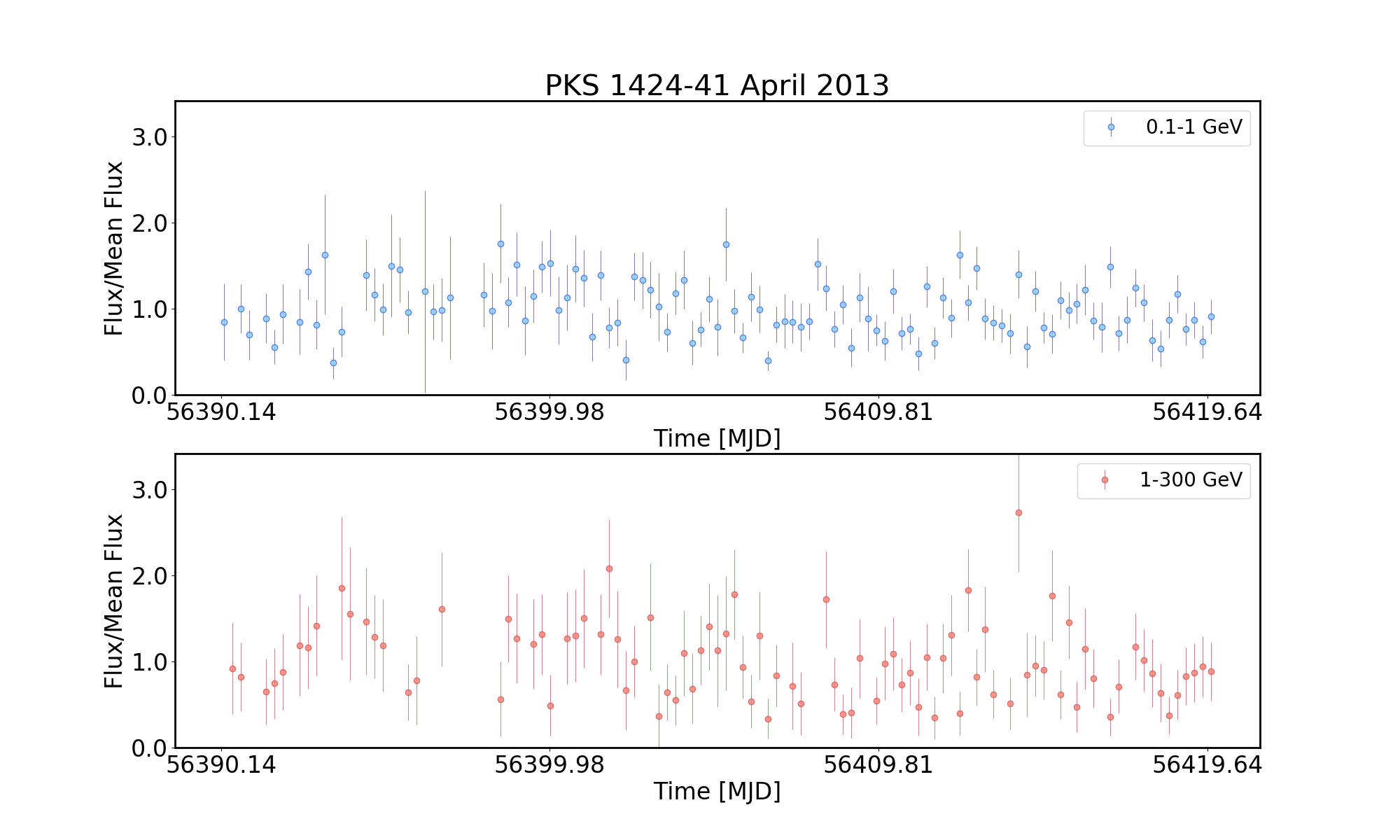}}
\end{figure*}

\begin{figure*}
    \centering
    \resizebox{\textwidth}{!}{
    \includegraphics{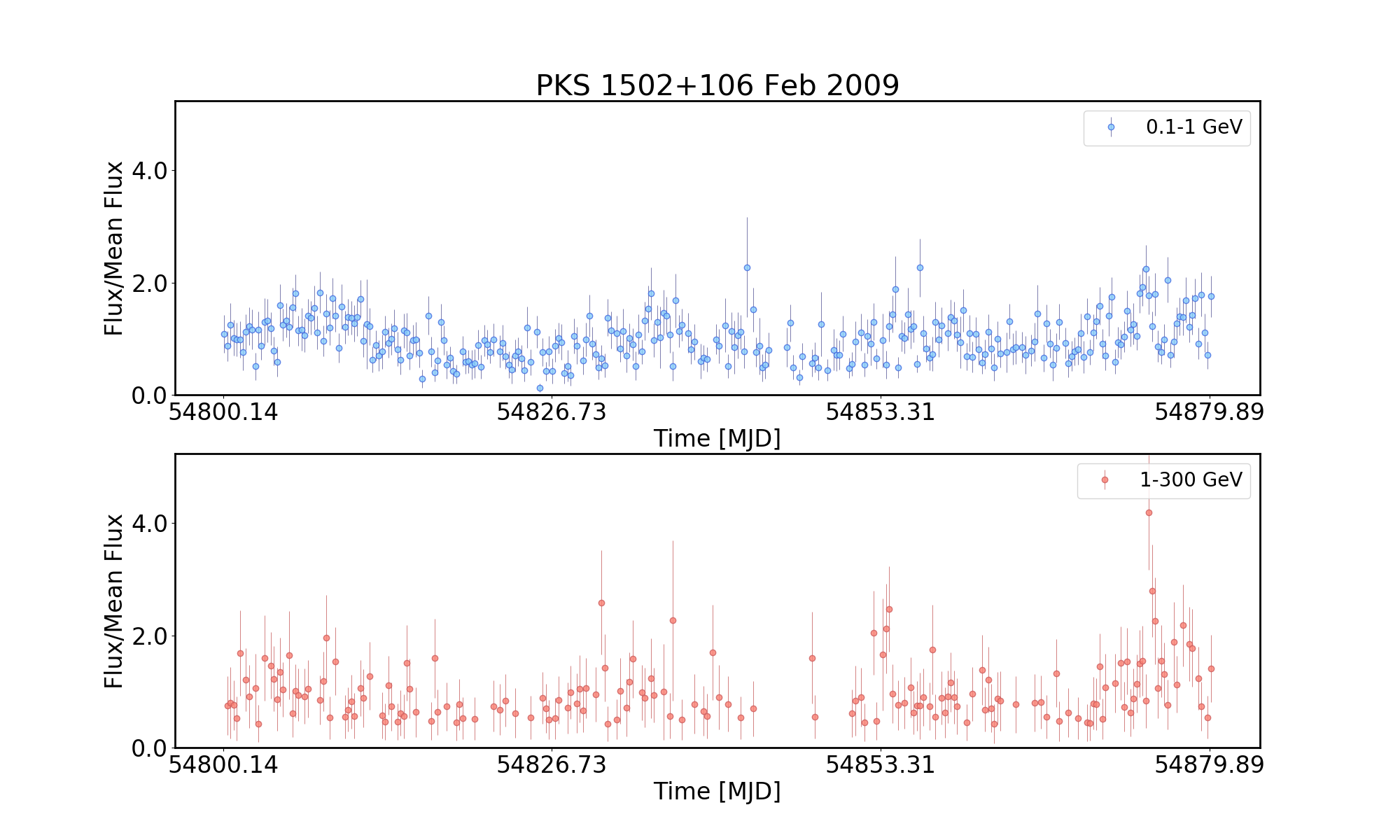}
    \includegraphics{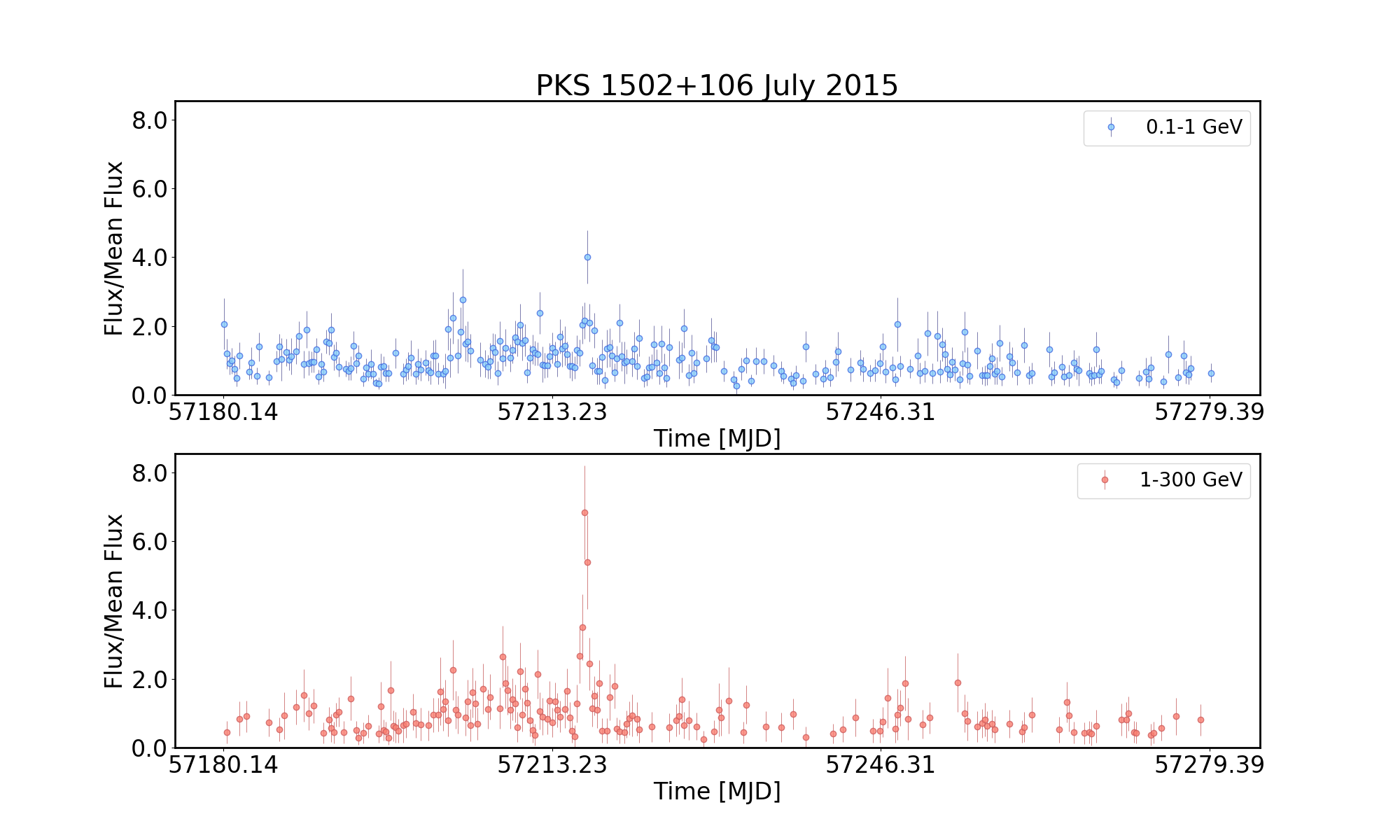}}
    \caption{Evolution of high- and low-energy flux in 6 hr bins during each flare period considered for B2 1520+31, CTA 102, PKS 0454-234, PKS 1424-41 and PKS 1502+106.}
    \label{fig: B2.}
\end{figure*}

\end{appendices}



\label{lastpage}
\end{document}